\documentclass[aps,superscriptaddress,showpacs,eqsecnum,twocolumn]{revtex4-1}
\usepackage[english]{babel}
\usepackage{graphicx}
\usepackage{stmaryrd}
\usepackage{amssymb}
\usepackage{amsfonts}
\usepackage{amsmath}
\usepackage{color}
\usepackage{xspace}
\usepackage{bm}
\definecolor{lightblue}{rgb}{0.13, 0.26, 0.99}
\usepackage{braket}
\def\be{\begin{equation}} \def\ee{\end{equation}}
\def\bea{\begin{eqnarray}} \def\eea{\end{eqnarray}}

\usepackage[
colorlinks=true,
urlcolor=lightblue,
citecolor=lightblue,
linkcolor=lightblue,
hyperfootnotes=false]{hyperref}

\allowdisplaybreaks

\newcommand{\e}{\varepsilon}
\newcommand{\Jl}{J_{\rm leg}}
\newcommand{\Jr}{J_{\rm rung}}

\begin{document}

\title{Dimensional modulation of spontaneous magnetic order in quasi-two-dimensional quantum antiferromagnets}
\author{Shunsuke C. Furuya}
\affiliation{Condensed Matter Theory Laboratory, RIKEN, Wako, Saitama 351-0198, Japan}
\author{Maxime Dupont}
\author{Sylvain Capponi}
\author{Nicolas Laflorencie}
\affiliation{Laboratoire de Physique Th\'eorique, Universit\'e de Toulouse, CNRS, UPS, France}
\author{Thierry Giamarchi}
\affiliation{Department of Quantum Matter Physics, University of Geneva, 24 Quai Ernest-Ansermet 1211 Geneva, Switzerland}
\date{\today}
\begin{abstract}
 Spontaneous symmetry breaking is deeply related to the dimensionality of a system.
 The N\'eel order going with spontaneous breaking of U(1) symmetry is
 safely allowed at any temperature for three-dimensional systems but
 allowed only at zero temperature for purely two-dimensional systems.
 We closely investigate how smoothly the ordering process of the three-dimensional system is modulated into that of the two-dimensional one
 with reduction of dimensionality,
 considering spatially anisotropic quantum antiferromagnets.
 We first show that the N\'eel temperature is kept finite even in the two-dimensional limit although the N\'eel order is greatly suppressed
 for low-dimensionality.
 This feature of the N\'eel temperature is highly nontrivial,
 which dictates how the order parameter is squashed under the reduction of dimensionality.
 Next we investigate this dimensional modulation of the order parameter.
 We develop our argument taking as an example a coupled spin-ladder system relevant for experimental studies.
 The ordering process is investigated multidirectionally using theoretical techniques of
 a mean-field method combined with analytical (exact solutions of quantum field theories)
 or numerial (density-matrix renormalization-group) methods, 
 a variational method, a renormalization-group study, 
 linear spin-wave theory, and quantum Monte Carlo simulations.
 We show that these methods independent of each other lead to the same conclusion about the dimensional modulation.
 \end{abstract}
\maketitle

\section{Introduction}\label{sec:intr}

\begin{figure}[b!]
 \centering
 \includegraphics[bb= 0 0 1200 600, width=\linewidth]{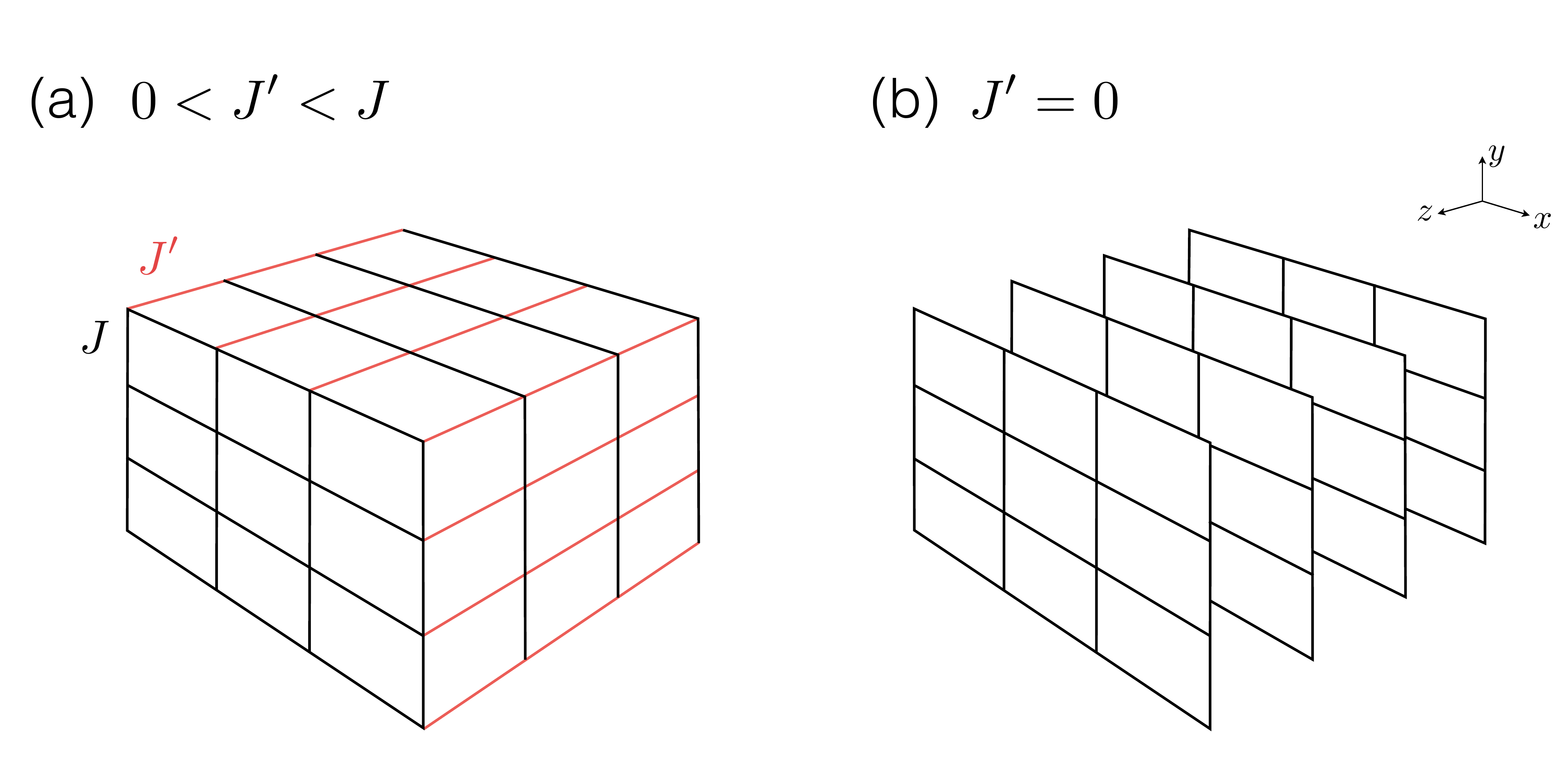}
 \caption{(Color online)
 (a) A 3D spin system on a cubic lattice.
 Black and red bonds represent the Heisenberg exchange interactions $J$ and $J'$ between neighboring spins respectively.
 When the exchange interaction $J'$ along the $z$ axis is gradually switched off,
 it converges to (b) a 2D spin system on layered square lattices.
 }
 \label{cube}
\end{figure}

Spatial dimension is an interesting parameter.
It dictates the fate of the wavefunction under a disordered potential~\cite{Abrahams_localization_1979, Giamarchi_localization_1d_1988},
restricts possible topological phases~\cite{Schnyder_classification_2008},
and affects spontaneous symmetry
breaking~\cite{MerminWagner_1966, Hohenberg_1967, Momoi_AF_1996, Furuya_unionjack_2014}.
Among various condensed-matter systems,
magnetic insulators are particularly interesting from the viewpoint of dimensionality.
In particular one can control with temperature their dimensionality.
Let us take as an example a spatially anisotropic quantum Heisenberg antiferromagnet on a three-dimensional cubic lattice
whose exchange interactions are $J$ in the $x$ and $y$ directions and $J'$ in the $z$ direction (Fig.~\ref{cube}).
For $J'\ll J$, this system is effectively identical to two-dimensional (2D) quantum spin systems [Fig.~\ref{cube}~(b)]
when the temperature $T$ is high enough to mask the interplane correlation due to $J'$.
On the other hand, when $T\ll J'$, the interplane coupling $J'$ is nonnegligible and leads to spontaneous N\'eel order.
In short, the Heisenberg antiferromagnet on  weakly coupled square lattices behaves
two-dimensionally for $T\gg J'$ and three-dimensionally for $T\ll J'$.
There must be a dimensional phase transition or crossover at a moderate temperature in between these two distinctive regions.
Dimensional phase transition and crossovers are common features of low-dimensional quantum spin systems realized in magnetic
insulators~\cite{Kohno_Cs2CuCl4_2007, Starykh_Cs2CuCl4_2010, Klanjsek_BPCB_2008, Schmidiger_DIMPY_2012, Garlea_dim_crossover_2009, Yamaguchi_3cl4fv_2013, Yamaguchi_3iv_2015, Klanjsek_bacovo_2015, Giamarchi_QPT_2010}. 
However, as detailed below, several important open questions remain.

\begin{figure}[b!]
 \centering
 \includegraphics[bb= 0 0 2100 700, width=\linewidth]{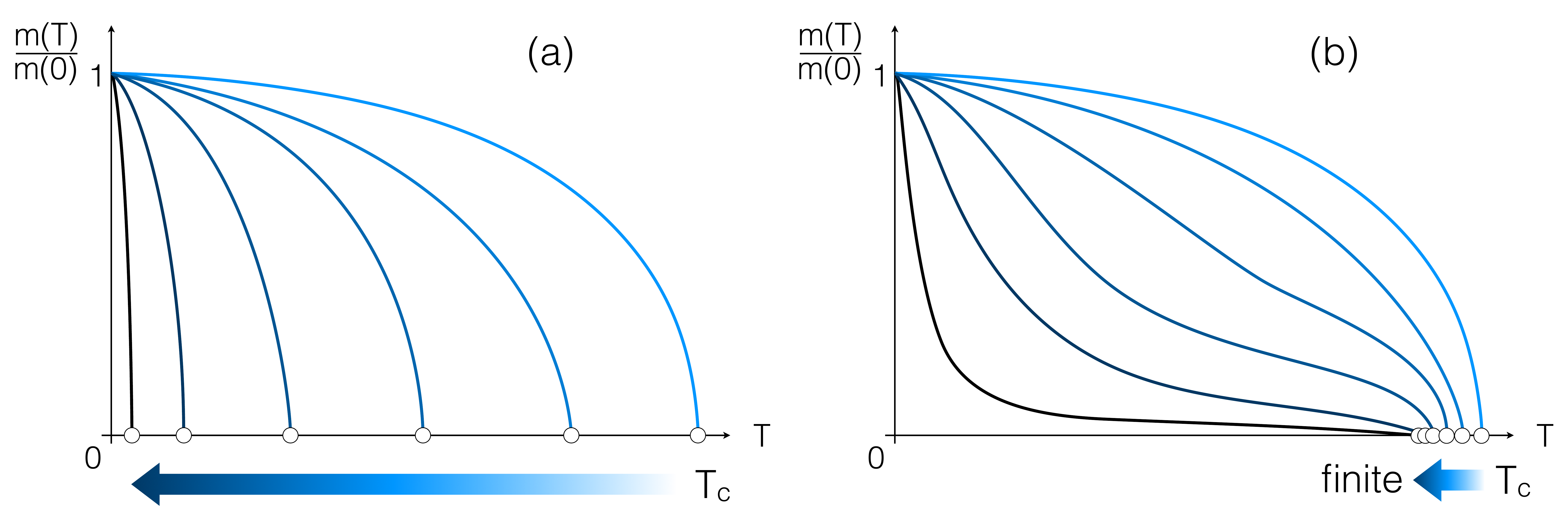}
 \caption{(Color online)
 Schematic figures to show the two possibilities of suppression of the N\'eel ordering by the reduction of the dimensionality
 from 3D (lighter blue) to quasi-2D (darker blue).
 (a) The naive possibility. The critical temperature  of the  3D ordered phase approaches zero as the interlayer interaction is reduced.
 (b) The other possibility, which turns out to be the case.
 The critical temperature remains finite even in the quasi-2D limit.
 As a consequence, the stepwise curve of the order parameter emerges.
 }
 \label{3Dto2D}
\end{figure}

\begin{figure}[b!]
 \centering
 \includegraphics[bb = 0 0 1050 700, width=0.9\linewidth]{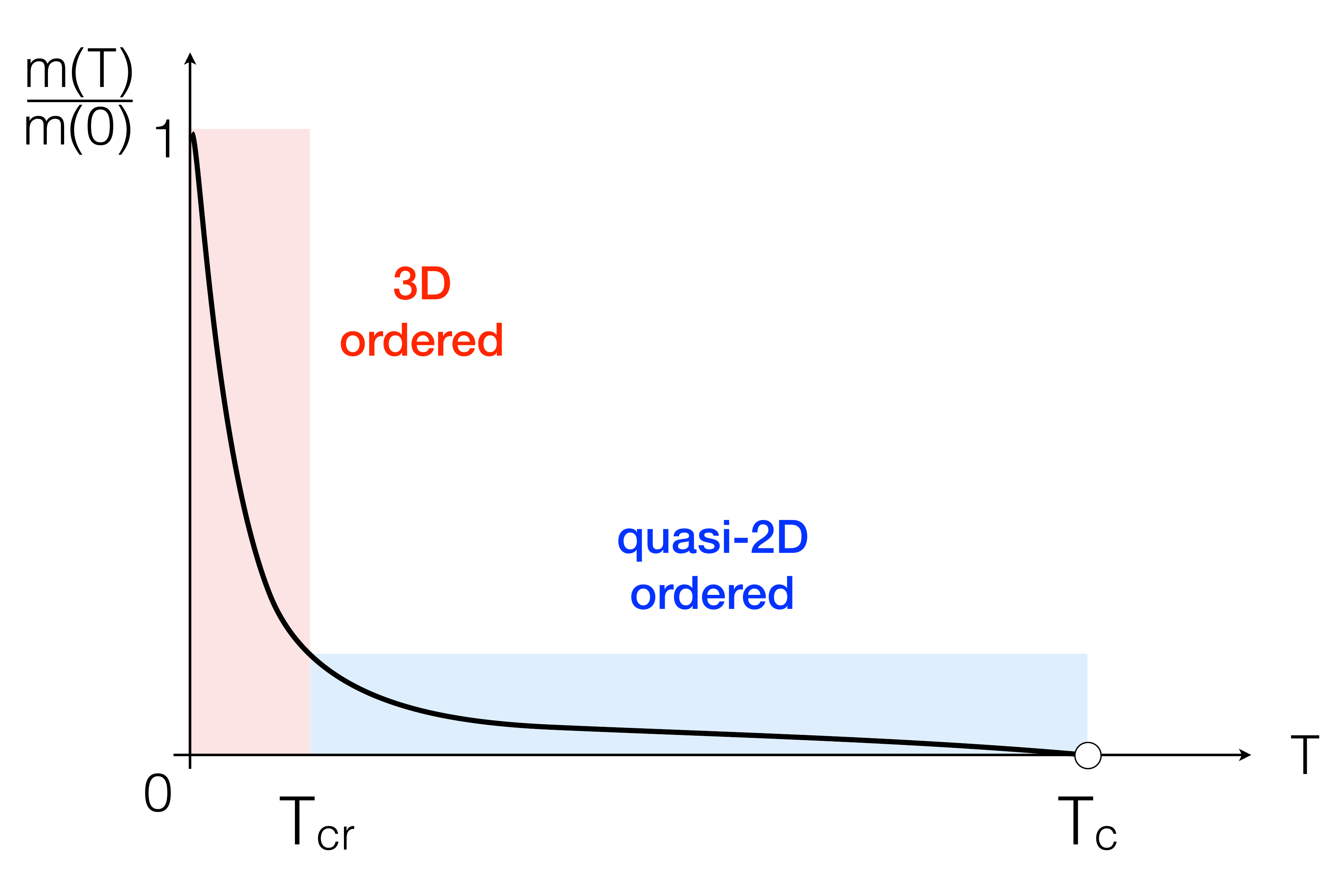}
 \caption{(Color online)
 A schematic distinction of the quasi-2D and 3D ordered phases.
 The quasi-2D ordered phase is spread near the critical temperature $T_c$ and associated by the suppressed but nonzero spontaneous order $m(T)$.
 The quasi-2D ordered phase inherits the critical behavior of the 2D Kosterlitz-Thouless critical phase.
 The 3D ordered phase is a rather conventional ordered phase with well developed interlayer correlation.
 Those phases are roughly separated at a crossover temperature $T_{\rm cr}$.
 }
 \label{quasi2D_def}
\end{figure}

The Heisenberg quantum antiferromagnet on the spatially anisotropic cubic lattice  [Fig.~\ref{cube}~(a)]
has a spontaneous N\'eel order $m(T)$ below a critical temperature $T_c$ for $J'/J\simeq 1$.
It is well known that $m(T)$ exhibits a domical temperature dependence (lightest-blue curves in Fig.~\ref{3Dto2D}).
On the other hand, it exhibits spontaneous N\'eel order only at zero temperature for $J'/J=0$.
Here we ask the question of how the $m(T)$ curve is modified when reducing $J'/J$.
There are \emph{a priori} two possibilities:  one with $T_c\to 0$ [Fig.~\ref{3Dto2D}~(a)]
and the other with $T_c$ kept finite [Fig.~\ref{3Dto2D}~(b)].
The former is the naive prediction while the latter is less expected.
Interestingly, we will argue in the following that the latter scenario is realized.
The $m(T)$ curve of Fig.~\ref{3Dto2D}~(b) dictates the existence of a quasi-2D ordered phase where
the system behaves two-dimensionally except for the suppressed but nonzero $m(T)$ breaking the U(1) symmetry (Fig.~\ref{quasi2D_def}).
In this paper we study the dimensional modulation of the $m(T)$ curve of Fig.~\ref{3Dto2D}~(b) 
on the experimentally relevant example of anisotropically coupled two-leg spin ladders.
We argue that this system shows a dimensional transition between 1D disordered and quasi-2D ordered phases and a dimensional crossover between
quasi-2D and 3D ordered phases.

This article is constructed as follows.
In Sec.~\ref{sec:model} we introduce a model of quantum antiferromagnet to describe the smooth reduction of dimensionality.
The ordering process is investigated in various complementary ways.
In Sec.~\ref{sec:rpa}, we adopt the random phase approximation (RPA) approach
in order to clarify the dimensionality dependence of the critical temperature.
While the RPA approach gives a strong evidence for Fig.~\ref{3Dto2D}~(b),
it is not useful to illustrate the ordering process.
To look into the temperature and dimensionality dependencies of the order parameter,
we employ a variational approach in Sec.~\ref{sec:variation}.
The variational analysis supports the result of the RPA analysis and
visualizes the ordering process with the decrease of temperature.
However, the variational analysis fails to provide clear characterization of a crossover temperature, which we denote as $T_{\rm cr}$,
between the quasi-2D ordered phase and 3D ordered phase (Fig.~\ref{quasi2D_def}).
To clarify the meaning of $T_{\rm cr}$,
we develop a renormalization-group (RG) argument with the aid of a classical approximation in Sec.~\ref{sec:rg}.
While all those theoretical techniques are independent of each other,
they are perfectly consistent in supporting the conclusion of Fig.~\ref{3Dto2D}~(b).
Among them, the classical approach of Sec.~\ref{sec:rg} is useful in comparison with numerical calculation as shown in Sec.~\ref{sec:classical_approx}.
In Sec.~\ref{sec:num}, numerical simulations are carried out for a system of weakly coupled spin chains in 2D, which is equivalent to our system under
the classical approximation,
and are successfully compared with analytical results.
In Sec.~\ref{sec:Z2} we look briefly at effects of the spin anisotropy on the ordering process.
All the discussions are summarized in Sec.~\ref{sec:summary}.
Some additional technical points can be found in the appendices.

\begin{figure}[b!]
 \includegraphics[bb= 0 0 1440 1440, width=\linewidth]{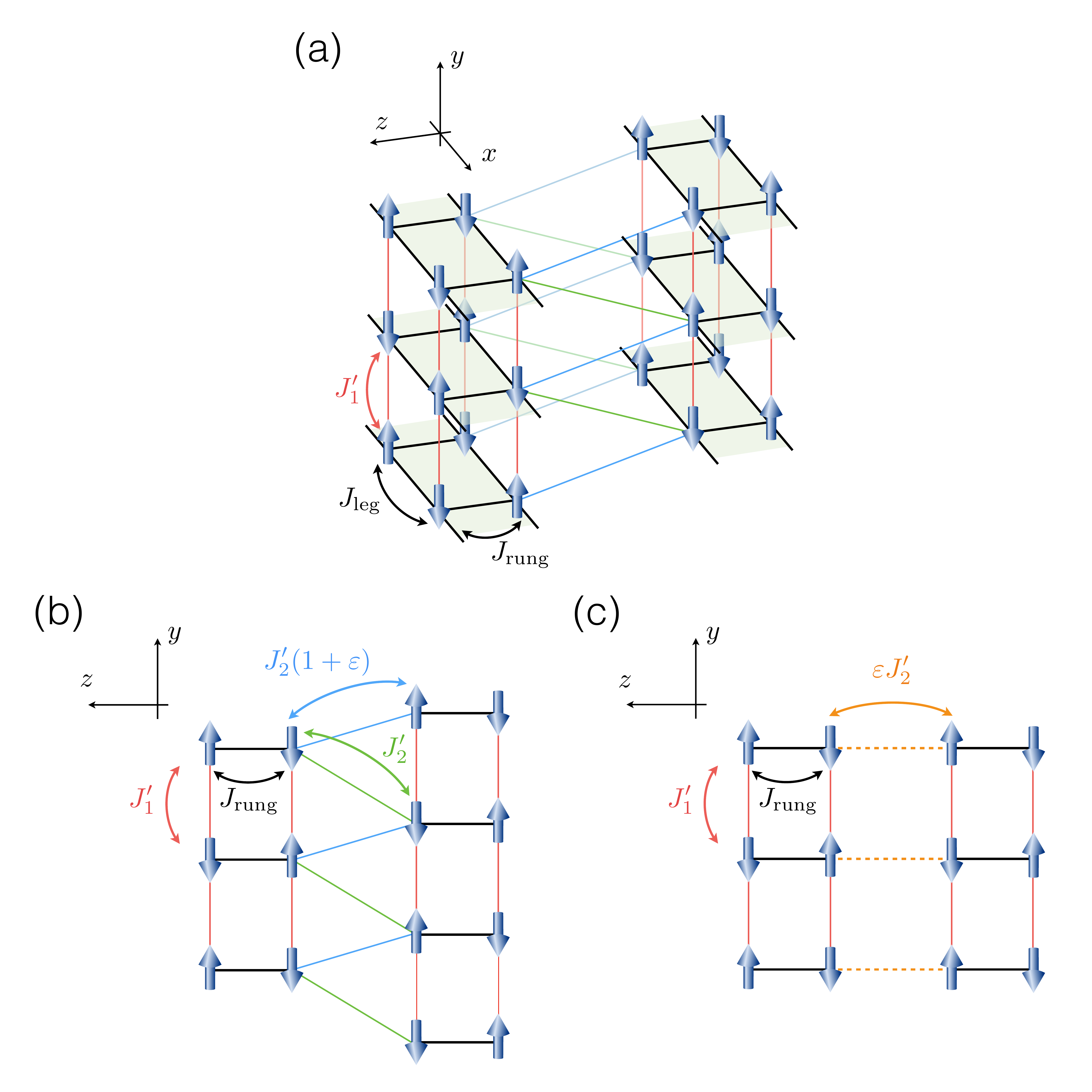}
 \caption{(Color online)
 The two-leg ladders are layered with an unfrustrated
 interladder  coupling $J'_1$ (red lines).
 The other frustrated interladder couplings have an imbalance,
 $(1+\e)J'_2$  (blue lines) and $J'_2$ (green lines).
 }
 \label{lattice}
\end{figure}

\section{Model and material}\label{sec:model}

In order to discuss the dimensional modulation of the magnetic ordering process,
we consider a system of anisotropically coupled $S=1/2$ two-leg spin ladders [Fig.~\ref{lattice}~(a)], 
which has the following Hamiltonian:
\begin{align}
 \mathcal H
 &= \sum_{\mu,\nu}\mathcal H_{\mu,\nu} + J'_1
 \sum_{j,l,\mu,\nu} \bm S_{j,l,\mu,\nu} \cdot \bm S_{j,l,\mu+1,\nu}
 \notag \\
 & \quad
 +J'_2 \sum_{j,\mu,\nu} \bm S_{j,2,\mu,\nu} \cdot \bigl\{
 (1+\e)\bm S_{j,1,\mu,\nu+1} + \bm S_{j,1,\mu-1,\nu+1}\bigr\},
 \label{H}
\end{align}
with an unfrustrated interladder interaction $J'_1$ along the $y$ axis
and imperfectly frustrated interactions $J'_2$ and $J'_2(1+\e)$ in the $z$ direction [Fig.~\ref{lattice}~(b)].
$\mathcal H_{\mu,\nu}$ represents the Hamiltonian of a two-leg spin ladder,
\begin{align}
 \mathcal H_{\mu,\nu}
 &= \Jl \sum_{j,l} \bm S_{j,l,\mu,\nu} \cdot \bm S_{j+1,l,\mu,\nu}
 \notag \\
 & \quad
 +\Jr \sum_{j} \bm S_{j,1,\mu,\nu} \cdot \bm S_{j,2,\mu,\nu}
 -h \sum_{j,l}S^z_{j,l,\mu,\nu}.
 \label{H_ladder}
\end{align}
Here, $h=g\mu_BH$ is proportional to the magnetic field $H$,
$g$ is the Land\'e factor and $\mu_B$ is the Bohr magneton.
We assume that the intraladder interactions $\Jl$ and $\Jr$ and the
interladder interactions $J'_1$, $J'_2$ and $J'_2(1+\e)$ satisfy
\begin{equation}
 \Jl > \Jr \gg J'_1 \gg \e J'_2>0.
  \label{hierarchy}
\end{equation}
In this paper we use the units $\hbar = k_B = a_0=1$ for simplicity.
Here, $a_0$ is the unit of lattice spacing.
In Eqs.~\eqref{H} and \eqref{H_ladder},
the latin indices $j$ and $l$ specify the position of the spin in the ladder and
the greek indices $\mu$ and $\nu$ specify the position of the ladder.
The parameter $\e>0$ represents the imbalance of
interladder interactions perpendicular to the $xy$ plane,
$J'_2$ and $J'_2(1+\e)$ [Fig.~\ref{lattice}~(b)].

There are three reasons to choose the model \eqref{H}.
First the spin-ladder structure of the model \eqref{H} makes it possible to predict the critical temperature precisely (Sec.~\ref{sec:rpa}).
The 1D nature of the underlying model is
masked by well developed interladder correlations in the quasi-2D or 3D ordered phases
and thus not essential for our claim of the dimensional modulation.
Still, the 1D structure of the lattice is technically convenient for theoretical analyses as we will see later.

Second the imperfect geometrical frustration $\e>0$ is an easy and realistic way to implement highly spatially anisotropic interactions.
When $\e=0$,  the situation can be completley different from the $\e\not=0$ case~\cite{Kohno_Cs2CuCl4_2007, Kohno_Cs2CuCl4_2009}.
The imperfectly frustrated interladder interactions $J'_2$ and $J'_2(1+\e)$ [Fig.~\ref{lattice}~(b)] in the $yz$ plane are replaceable by
a single unfrustrated interaction $\e J'_2$ [Fig.~\ref{lattice}~(c)] as far as the N\'eel order is concerned,
as we will confirm this point in Sec.~\ref{sec:rpa}.
Although the interladder interaction $J'_2$ is not necessarily much smaller than the other one $J'_1$,
the effective interaction $\e J'_2$ can become easily much smaller than $J'_1$.
In this paper, we discuss the quasi-2D ordered phase in terms of the parameter $\e$.

Last but not least the model \eqref{H} is useful for analyzing the
strong-leg spin-ladder compound $\mathrm{(C_7H_{10}N)_2CuBr_4}$ (also known as DIMPY).
DIMPY has recently been under active experimental and theoretical investigations~\cite{Schmidiger_DIMPY_2012, Hong_DIMPY_2010, Jeong_DIMPY_2013, Schmidiger_DIMPY_2013a, Schmidiger_DIMPY_2013b, Glazkov_DIMPY_2015, Ozerov_DIMPY_2015}
because it is the first spin-ladder compound with strong leg interactions, $\Jl>\Jr$.
In this paper we take DIMPY as an example and use parameters estimated numerically~\cite{Schmidiger_DIMPY_2012}.

\section{Random phase approximation}\label{sec:rpa}

\subsection{Susceptibility}

One of the simplest ways to deal with a phase transition is the mean-field (MF) method.
In our system, the N\'eel order $m(T)$ manifests itself under a high magnetic field $h$.
The order $m(T)$ grows perpendicularly to the direction of the magnetic field and
breaks the U(1) symmetry around the magnetic field.
Since the spontaneous N\'eel order breaking of a continuous symmetry is prohibited in purely 1D systems at finite temperatures~\cite{MerminWagner_1966, Hohenberg_1967} and even at zero temperature~\cite{Momoi_AF_1996},
the MF approximation is effectively applicable only to the interladder interactions.
This interladder MF approximation is often precise enough
to determine the phase boundary of the N\'eel ordered phase~\cite{Klanjsek_BPCB_2008, Bouillot_BPCB_2011, Schmidiger_DIMPY_2012}.

The interladder correlation of the N\'eel order can be easily included at the level of RPA~\cite{Scalapino_RPA_1975, Schulz_CoupledChains_1996}.
RPA leads to the susceptibility
$\chi^{xx}(\bm q)=\int d\bm r e^{-i\bm q\cdot \bm r}\langle S^xS^x\rangle (\bm r)$~\cite{Scalapino_RPA_1975}:
\begin{equation}
 \chi^{xx}(\bm q) = \frac{\chi^{xx}_{\mathrm{1D}}(q_x)}{1+\mathcal J(\bm q)\chi^{xx}_{\mathrm{1D}}(q_x)}.
  \label{chi^xx}
\end{equation}
Here $\chi^{xx}_{\mathrm{1D}}$ represents the susceptibility of the single ladder, which we call the 1D susceptibility.
$\mathcal J(\bm q)$ is the Fourier transform of the interladder interactions,
\begin{equation}
  \mathcal J(\bm q) = J'_1\cos q_y +  J'_2(1+\e)\cos q_z +J'_2 \cos(-q_y +q_z).
   \label{J_def}
\end{equation}
Equation \eqref{J_def} shows that both the imperfectly frustrated coupling [Fig.~\ref{lattice}~(b)] and the highly anisotropic unfrustrated couplings [Fig.~\ref{lattice}~(c)]
are equivalent as far as the N\'eel order developed at the wave vector $(q_y,q_z)=(\pi, \pi)$ is concerned:
\begin{equation}
 \mathcal J(q_x, \pi, \pi) = -J'_1 -\e J'_2.
\end{equation}

The RPA formula \eqref{chi^xx} tells us that the phase transition occurs at a temperature where the following equation,
\begin{equation}
 1+\mathcal J(\bm q)\chi^{xx}_{\mathrm{1D}}(q_x) = 0,
  \label{Tc_eq}
\end{equation}
is satisfied.
Equation~\eqref{Tc_eq} also gives the wavevector $\bm q=(q_x,q_y,q_z)$ of the order parameter $\langle e^{i\bm q\cdot \bm r}S^x_{j,l,\mu,\nu}\rangle$.
As we see below, the 1D susceptibility  has the largest contribution from $q_x\simeq \pi$.
Besides $q_y=q_z=\pi$ follows from the fact that the 1D susceptibility is positive.

\subsection{1D susceptibility in gapped phases}\label{subsec:h=0}

Here and in the next section, we deal with the 1D susceptibility of a single spin ladder at finite temperatures.
At zero magnetic field, the spin ladder is in a gapped nonmagnetic phase at low temperatures $T\ll \Jr$,
where the susceptibility is exponentially suppressed $\chi^{xx}_{\mathrm{1D}}(q_x)\propto e^{-\Delta/T}$, $\Delta$ being the finite spin gap.
Because of the exponential suppression, Eq.~\eqref{Tc_eq} has no solution for small interladder
interactions~\cite{Sakai_HaldanePhase_RPA_1989, Wierschem_quasi1dHaldane_2014}.
This is also the case under nonzero magnetic fields as far as the spin ladder is in the gapped phase, that is, $h<h_{c1}\equiv \Delta$
The susceptibility is also exponentially small in the saturated phase under an extremely high field $h>h_{c2}$,
when Eq.~\eqref{Tc_eq} becomes solutionless again.

\subsection{1D susceptibility in the Tomonaga-Luttinger liquid phase}\label{subsec:TLL}

\begin{figure}[t!]
 \centering
 \includegraphics[bb= 0 0 800 400, width=\linewidth]{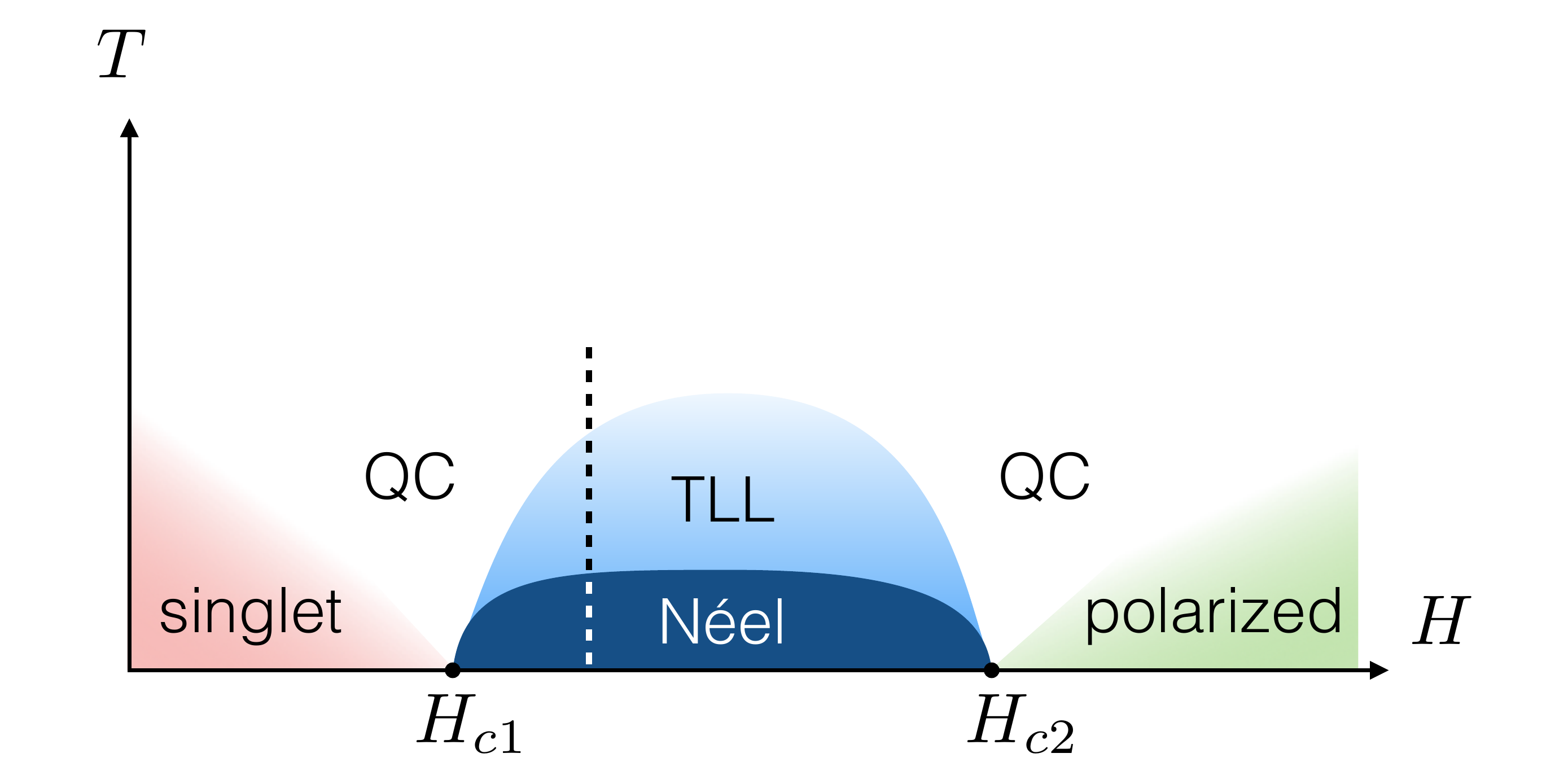}
 \caption{(Color online)
 A schematic phase diagram of the coupled spin ladders is presented in the $T$-$H$ 
 plane~\cite{Giamarchi_rev_BEC, Bouillot_BPCB_2011, Schmidiger_DIMPY_2012}.
 The two-leg spin ladder \eqref{H_ladder} has a gapped ``singlet''
 phase at low field where a nonzero excitation gap exists.
 The strong magnetic field closes the excitation gap and allows the
 field-induced TLL phase.
 As we further increase the magnetic field, all spins are polarized
 along the field direction.
 $H_{c1}\equiv h_{c1}/g\mu_B$ and $H_{c2}\equiv h_{c2}/g\mu_B$ denote the critical fields at zero temperature
 of the single ladder system that separate the field-induced TLL phase
 from the singlet and polarized phases.
 In the system of 3D coupled antiferromagnetic spin ladders, the N\'eel
 ordered  phase exists on the low-temperature side of the field-induced TLL phase.
 ``QC'' represents the quantum critical regions spread above the quantum
 critical points $H_{c1}$ and $H_{c2}$.
 In the rest of the paper, we discuss the temperature dependence of the
 N\'eel order along the dashed vertical line.
 }
 \label{PD_ladders}
\end{figure}

When the magnetic field $h$ is increased from zero, the  spin gap vanishes at $h_{c1}$.
That is, the magnetic field induces a quantum phase transition from the gapped phase into the Tomonaga-Luttinger liquid (TLL) phase.
The TLL phase extends up to the saturation field $h_{c2}$.
Note that the phase transitions at $h=h_{c1}, h_{c2}$ occur only at zero temperature.
No phase transition occurs at finite temperatures.
Instead quantum critical regions spread around $h_{c1}$ and $h_{c2}$ (QC in Fig.~\ref{PD_ladders}).

The 1D susceptibility in the field-induced TLL phase can be computed exactly~\cite{Giamarchi_book}.
Since the antiferromagnetic fluctuation of $S^x_{j,l,\mu,\nu}$ along the leg is developed more than any other fluctuations,
the $q_x=\pi$ component contributes most to the 1D susceptibility~\cite{Chitra_ladder_1997, Bouillot_BPCB_2011}.
Near $q_x=\pi$, it is given by
\begin{align}
 \chi^{xx}_{\mathrm{1D}}(q_x)
 & = \frac{A_x\sin(\frac{\pi}{4K})}u \biggl(\frac{2\pi T}u\biggr)^{\frac 1{2K}-2} \notag \\
 & \quad \times B\biggl(i\frac{ v(q_x-\pi)}{4\pi T} + \frac 1{8K}, 1-\frac 1{4K}\biggr)\notag \\
 & \quad \times B\biggl(-i\frac{ v(q_x-\pi)}{4\pi T} + \frac 1{8K}, 1-\frac 1{4K}\biggr).
 \label{1Dsus_TLL}
\end{align}
Here $A_x$ is a nonuniversal parameter which appears in the bosonization formulas of the spin operators (see Appendix~\ref{app:attraction}), 
$B(x,y)=\Gamma(x)\Gamma(y)/\Gamma(x+y)$ is the Beta function and
$\Gamma(z)$ is the Gamma function.
$K$ is the so-called TLL parameter that characterizes interaction strength and dictates the correlations of the TLL~\cite{Giamarchi_book}.

The TLL is attractive for $K>1$ and repulsive for $K<1$.
It is noninteracting at $K=1$.
Therefore the susceptibility depends crucially on $K$.
The TLL parameter of DIMPY is known to be $1<K<3/2$ depending on the magnetic field~\cite{Schmidiger_DIMPY_2012}.
This is in sharp contrast to a strong-\emph{rung} spin ladder compound $\mathrm{(C_5H_{12}N)_2CuBr_4}$ (also known as BPCB),
which is in the repulsive regime: $1/2<K<1$~\cite{Klanjsek_BPCB_2008}.
Actually, a spin ladder can yield in general an attractive field-induced TLL as long as the leg interaction is stronger than the rung.
A generic argument of the attractive field-induced TLL of the strong-leg spin ladder is briefly given in Appendix~\ref{app:attraction}.

The 1D susceptibility Eq.~\eqref{1Dsus_TLL} at $q_x=\pi$ satisfies $\chi^{xx}_{\mathrm{1D}}(\pi)>0$
and becomes arbitrarily large with a power law $T^{1/2K-2}$ as $T\to 0$.
These properties of the 1D susceptibility ensures the existence of  a solution of Eq.~\eqref{Tc_eq}.
Indeed, there exists  a unique solution,
\begin{equation}
 T_c = \frac v{2\pi}\biggl[\frac{2A_x (J'_1+\e J'_2)  \sin(\frac{\pi}{4K})B^2(\frac 1{8K}, 1-\frac  1{4K})}v \biggr]^{\frac{2K}{4K-1}},
  \label{T_c}
\end{equation}
where the wavevector of the order is $\bm q=(\pi,\pi,\pi)$.
For instance, using parameters of DIMPY~\cite{Schmidiger_DIMPY_2012} at finite magnetization $\langle S^z_{j,l}\rangle\simeq 0.2$ for reference, we get:
\begin{equation}
 K\simeq 1.3, \quad v\simeq 1.5\Jl, \quad A_x\simeq 0.18.
  \label{param_DIMPY}
\end{equation}
Given these parameters, the critical temperature \eqref{T_c} is estimated to be
\begin{equation}
 T_c \simeq 1.3 \Jl \biggl(\frac{J'_1+\e J'_2}{\Jl}\biggr)^{\frac{2K}{4K-1}}.
  \label{T_c_RPA}
\end{equation}
In the rest of this article, we will use the parameters \eqref{param_DIMPY} when needed.

\subsection{2D limit}

The RPA analysis tells us the fate of the critical temperature in the 2D limit $\e J'_2/J'_1\to 0$.
Even when $\e J'_2=0$, the susceptibility \eqref{chi^xx} diverges at a certain finite temperature.
This temperature is nothing but the Kosterlitz-Thouless (KT) transition temperature $T_{\rm KT}$.
For $T<T_{\rm KT}$ we have a critical phase (called KT phase) that preserves the U(1) symmetry.
Note that the susceptibility is divergent everywhere in the KT phase.

Let us see our system as a weakly coupled 2D systems with an infinitesimal interlayer coupling $\e J'_2$ (Fig.~\ref{cube}).
Performing the RPA calculation with respect to $\e J'_2$, we obtain the 3D susceptibility,
\begin{equation}
 \chi^{xx} (\bm q) = \frac{\chi^{xx}_{\rm 2D}(q_x,q_y)}{1+\mathcal J'(\bm q)\chi^{xx}_{\rm 2D}(q_x,q_y)},
  \label{chi^xx_q2D_RPA}
\end{equation}
with $\mathcal J'(\bm q)=J'_2(1+\e)\cos q_z+J'_2\cos (-q_y+q_z)$.
$\chi^{xx}_{\rm 2D}$ is the susceptibility in the 2D limit $\e J'_2/J'_1=0$.
The susceptibility \eqref{chi^xx_q2D_RPA} is divergent for $\bm q=(\pi,\pi,\pi)$  at a temperature $T_c$ which is a solution of
\begin{equation}
 \left.\chi^{xx}_{\rm 2D}(\pi,\pi)\right|_{T=T_c} = \frac 1{\e J'_2}.
  \label{Tc_2D_RPA}
\end{equation}
Since the right-hand side of Eq.~\eqref{Tc_2D_RPA} is large but finite,
the 2D susceptibility at $T_c$ on the left hand side must be finite.
Therefore it immediately follows that
\begin{equation}
 T_{\rm KT} < T_c
  \label{Tc_2D_bound}
\end{equation}
for any $\e J'_2/J'_1> 0$.
The relation \eqref{Tc_2D_bound} of the critical temperature and the KT temperature indicates that
the critical temperature converges to a finite value, which is $T_{\rm KT}$,  in the quasi 2D limit.
Therefore, our RPA analysis supports the dimensional reduction scenario sketched in Fig.~\ref{3Dto2D}~(b).

The phase diagram of the model \eqref{H} is schematically drawn in Fig.~\ref{PD_ladders}.
As long as $\e J'_2/J'_1>0$, the global structure of the phase diagram of Fig.~\ref{PD_ladders} is kept unchanged.
Nevertheless, the parameter $\e J'_2/J'_1$  affects the nature of the N\'eel ordered phase seriously.
To see that effect, we use the variational approach in the next section and the RG approach in Sec.~\ref{sec:rg}.

\section{Variational method}\label{sec:variation}

\subsection{Self-consistent equations}\label{sec:sce}

The variational method is designed to build a quadratic action that best approximates the nonlinear action of the original theory.
Using the bosonization representation of the spins~\cite{Giamarchi_book} the Hamiltonian of the system \eqref{H} leads to the 
Euclidean action:
\begin{align}
 \mathcal S
 &= \sum_{\mu,\nu}\frac{K}{2\pi v} \int d\tau dx \bigl\{
 (\partial_\tau\theta_{\mu,\nu})^2 + v^2 (\partial_x\theta_{\mu,\nu})^2
 \bigr\}
 \notag \\
 & \quad  +\frac{vg_u}{\pi}\sum_{\mu,\nu} \int d\tau dx \,
 \cos(2\phi_{\mu,\nu}+\pi (1-2M)x)
 \notag \\
 & \quad -J'_1A_x \sum_{\mu,\nu} \int d\tau dx \,
 \cos(\theta_{\mu,\nu}-\theta_{\mu+1,\nu}) \notag \\
 & \quad
 - \e J'_2A_x \sum_{\mu,\nu} \int d\tau dx \,
 \cos(\theta_{\mu,\nu}-\theta_{\mu,\nu+1}),
 \label{S_pre}
\end{align}
where  $M$ is the uniform magnetization density along the magnetic field and $g_u\propto \Jr$.
The first two lines of Eq.~\eqref{S_pre} represent the collection of the single-ladder interactions $\sum_{\mu,\nu}\mathcal H_{\mu,\nu}$~\cite{Giamarchi_book}.
The $\theta_{\mu,\nu}$ field describes the N\'eel order perpendicular to the magnetic field,
$ (-1)^{j+l+\mu+\nu}S^x_{j,l,\mu,\nu}\simeq\sqrt{A_x} \cos \theta_{\mu,\nu}(x)$.
The other field, $\phi_{\mu,\nu}$, is dual to $\theta_{\mu,\nu}$ and satisfies the commutation relation,
$[\phi_{\mu,\nu}(x), \partial_{x'}\theta_{\mu',\nu'}(x')]=i\pi\delta_{\mu,\nu'}\delta_{\nu,\nu'}\delta(x-x')$.
When $g_u=J'_1=J'_2=0$, the action \eqref{S_pre} represents a set of mutually independent TLL.
The cosine term of $\phi_{\mu,\nu}$ generates the spin gap of the low-field gapped phase with $M=0$.

The action \eqref{S_pre} contains both $\phi_{\mu,\nu}$ and $\theta_{\mu,\nu}$.
The presence of the N\'eel order $m(T)=\sqrt{A_x}\langle \cos\theta_{\mu,\nu}(x)\rangle$
allows us to discard $\phi_{\mu,\nu}$ from the action \eqref{S_pre} because of the uncertainty originating from the commutation relation.
That is, the action in the N\'eel ordered phase is effectively written in $\theta_{\mu,\nu}$ only:
\begin{align}
 \mathcal S
 &= \sum_{\mu,\nu}\frac{K}{2\pi v} \int d\tau dx \bigl\{
 (\partial_\tau\theta_{\mu,\nu})^2 + v^2 (\partial_x\theta_{\mu,\nu})^2
 \bigr\}
 \notag \\
 & \quad -J'_1A_x \sum_{\mu,\nu} \int d\tau dx \,
 \cos(\theta_{\mu,\nu}-\theta_{\mu+1,\nu}) \notag \\
 & \quad
 - \e J'_2A_x \sum_{\mu,\nu} \int d\tau dx \,
 \cos(\theta_{\mu,\nu}-\theta_{\mu,\nu+1}),
 \label{S}
\end{align}
Although the elimination of $\phi_{\mu,\nu}$ simplified the action \eqref{S},
it is still highly nonlinear and difficult to deal with.

The basic idea of the variational method is to search for a quadratic variational action,
\begin{equation}
 \mathcal S_{\rm v} = \frac T{2\Omega} \sum_{\omega_n, \bm k}G_{\rm v}^{-1}(i\omega_n,\bm k)|\theta (i\omega_n, \bm k)|^2,
  \label{Sv}
\end{equation}
that approximates the action \eqref{S} best in a sense stated in the next paragraph.
Here $\omega_n$ is the Matsubara frequency,
$\bm k=(k_x, k_y, k_z)=(q_x-\pi,q_y-\pi,q_z-\pi)$ is the wavevector shifted by $(\pi, \pi, \pi)$  for later convenience
and $\Omega$ is the volume of the system.
The field $\theta(i\omega_n,\bm k)$ is the Fourier transform of $\theta_{\mu,\nu}(\tau, x)$.

We determine the Green's function $G_{\rm v}(i\omega_n, \bm k)$ according to the variational principle~\cite{Feynman_book}.
The free energy $F$ is defined as $F=-T \ln Z$,
where $Z$ is the partition function $Z = \int\mathcal D
\theta_{\mu,\nu}(x) e^{-\mathcal S}$.
Likewise we can define another free energy,
\begin{align}
 F_{\rm v}
 &\equiv -T \ln \int \mathcal D\theta_{\mu,\nu}(x) \,e^{-\mathcal S_{\rm v}}.
\end{align}
The following relation is useful.
\begin{align}
 F
 &\le F_{\rm var},
 \label{F_ineq}
\end{align}
where $F_{\rm var}$ is the variational free energy defined as
\begin{equation}
 F_{\rm var} = F_{\rm v} + T \langle (\mathcal S - \mathcal S_{\rm
  v})\rangle_{\rm  v}.
  \label{F_var_def}
\end{equation}
The average $\langle \cdot \rangle_{\mathrm v}$ is taken with respect to
the variational action \eqref{Sv}.
The optimal variational action $\mathcal S_{\rm v}$ is determined so as to minimize the variational free energy \eqref{F_var_def}.
It is derived from the saddle-point equation,
\begin{equation}
 \frac{\delta F_{\mathrm{var}}}{\delta G_{\rm v}}= 0.
  \label{saddle}
\end{equation}
Straighforward calculations (Appendix \ref{app:F_var}) lead to
\begin{align}
 F_{\mathrm{var}}
 &= -\frac T2 \sum_{\omega_n, \bm k}\ln G_{\rm v}(i\omega_n, \bm k) \notag \\
 & \quad
 +\frac{KT}{2\pi v} \sum_{\omega_n, \bm k} (\omega_n^2+v^2k_x^2)
 G_{\rm v}(i\omega_n,\bm k) \notag \\
 &\quad -J'_1A_x \Omega \exp \biggl[-\frac
 T{2\Omega}\sum_{\omega_n, \bm k} F(k_y)G_{\rm v}(i\omega_n,\bm k)\biggr] \notag \\
 &\quad -\e J'_2A_x\Omega \exp\biggl[-\frac T{2 \Omega}
 \sum_{\omega_n,\bm k}F(k_z)G_{\rm v}(i\omega_n,\bm k)\biggr],
 \label{F_var}
\end{align}
with  $F(z)=2-2\cos z$.
Substituting the expression \eqref{F_var} into the saddle point equation \eqref{saddle},
we obtain the variational Green's function,
\begin{equation}
G_{\rm v}(i\omega_n,  \bm k)
  = \frac {\pi v}K
  \frac 1{\omega_n^2+v^2k_x^2+v_y^2F(k_y)+v_z^2F(k_z)}.
  \label{Gv}
\end{equation}
The saddle-point equation \eqref{saddle} requires
\begin{align}
 v_y
 &= \sqrt{\frac{\pi vJ'_1 A_x}{K}} \exp \biggl[-\frac
 T{4\Omega}\sum_{\omega_n,  \bm k} F(k_y)G_{\rm v}(i\omega_n,\bm k)\biggr],
 \label{vy_SCE} \\
 v_z
 &= \sqrt{\frac{\pi v\e J'_2 A_x}{K}} \exp\biggl[-\frac T{4 \Omega}
 \sum_{\omega_n, \bm k}F(k_z)G_{\rm v}(i\omega_n,\bm k)\biggr].
 \label{vz_SCE}
\end{align}
$v_y$ and $v_z$ are to be determined self-consistently.
The quadratic Green's function \eqref{Gv} leads to
\begin{equation}
 m(T)
 =\sqrt{A_x}\exp\biggl[-\frac T{2 \Omega}\sum_{\omega_n,\bm k}G_{\rm v}(i\omega_n,\bm k)\biggr].
 \label{mx_var}
\end{equation}
The right hand sides of Eqs.~\eqref{vy_SCE} and \eqref{vz_SCE} depend on the temperature.
$v_y(T)$ and $v_z(T)$ measure how much the interladder correlation in the $y$ and $z$ directions are developed
at a given temperature $T$.

$\bm v= (v, v_y, v_z)$ is the velocity of the Nambu-Goldstone mode.
Near $\bm k=0$ we can approximate $F(k_a)$ as
\begin{equation}
  F(k_a)\simeq k_a^2.
   \label{F_approx}
\end{equation}
and $G_{\rm v}(i\omega_n,\bm k)$ as
\begin{equation}
 G_{\rm v}(i\omega_n,  \bm k) \simeq \frac{\pi v}K \frac
  1{\omega_n^2 + v^2 k_x^2+v_y^2k_y^2 + v_z^2k_z^2}.
  \label{Gv_NG}
\end{equation}
Thus the variational action \eqref{Sv} describes a gapless excitation
with the anisotropic velocity $\bm v = (v,v_y,v_z)$, namely
the Nambu-Goldstone mode originating from the spontaneous breaking of the U(1) rotational symmetry.
The gapful amplitude mode can also be derived within the RPA scheme~\cite{Cazalilla_NJP_2006}.

The approximation \eqref{F_approx} is justified when $v_a(T)/T\gg 1$.
This can be interpreted as follows.
$v_aF(k_a)$ gives the dispersion relation of the Nambu-Goldstone mode in the $a$ direction.
The energy ``band'' $E=v_aF(k_a)$ is located within a range $-v_a\le E\le v_a$.
For $v_a(T) \lesssim T$, all excitations of the energy band $E=v_aF(k_a)$ must be taken into account.
On the other hand, a low temperature $T\ll v_a(T)$ suppresses high-energy excitations with
large $|k_a|$, justifying the quadratic approximation \eqref{F_approx}.
In other words, $T/v_a(T)$ represents an effective cutoff of $|k_a|$.
This relation of the effective cutoff and the quadratic approximation \eqref{F_approx}
is related to the RG argument of Sec.~\ref{sec:rg}.

\subsection{At zero temperature}\label{sec:zero}

Let us solve the self-consistent equations \eqref{vy_SCE} and \eqref{vz_SCE} first at $T=0$.
The hierarchy $\e J'_2/J'_1\ll 1$ motivates us to assume
\begin{equation}
 \frac{v_z(T=0)}{v_y(T=0)} \ll 1.
  \label{presume_ratio}
\end{equation}
We will solve the self-consistent equations under the assumption \eqref{presume_ratio} and then check that the solution indeed verifies this assumption.
We expand the integrand in Eq.~\eqref{vy_SCE} with small $v_z/v_y$:
\begin{align}
 &\frac T{ \Omega}\sum_{\omega_n,\bm k} F(k_y)
 G_{\rm v}(i\omega_n,\bm k) \notag \\
 &= \frac{\pi v}{2K}\int_{-\pi}^\pi \frac{d\bm k}{(2\pi)^3}
 \frac{F(k_y)}{\sqrt{v^2k_x^2+v_y^2F(k_y)+v_z^2F(k_z)}} \notag \\
 &\simeq \frac 1K \biggl[\ln\biggl(\frac{2\pi v}{v_y}\biggr) - \frac
 12\biggr].
\end{align}
Substituting this into the self-consistent equation \eqref{vy_SCE},
we obtain $v_y$ at $T=0$,
\begin{align}
 \frac{v_y(T=0)}v
 &= c_y\biggl(\frac{J'_1}{\Jl}\biggr)^{\frac{2K}{4K-1}},
  \label{vy_sol_T=0}
\end{align}
with $c_y=(\pi \Jl A_x/vK)^{2K/(4K-1)}(e^{1/2}/2\pi)^{1/2K}$.
The same procedure leads to
\begin{align}
 \frac{v_z(T=0)}v
 &= c_z\sqrt{\frac{\e J'_2}{J'_1}}
 \biggl(\frac{J'_1}{\Jl}\biggr)^{\frac{2K}{4K-1}},
 \label{vz_sol_T=0}
\end{align}
with  $c_z=(c_y/4\pi^2)^{1/4K}\sqrt{\pi \Jl A_x/vK}$.
Thus the self-consistent solutions \eqref{vy_sol_T=0} and \eqref{vz_sol_T=0} turns out to satisfy the assumption \eqref{presume_ratio}
because $v_z(T=0)/v_y(T=0)$ is of the order of $\sqrt{\e J'_2/J'_1}\ll 1$.

Using the velocities \eqref{vy_sol_T=0} and \eqref{vz_sol_T=0}, we can calculate the N\'eel order parameter
\eqref{mx_var} at $T=0$,
\begin{align}
 m(T=0)
 &=\sqrt{A_x}\biggl(\frac{c_y}{2\pi}\biggr)^{\frac 1{4K}}\biggl(\frac{J'_1}{\Jl}\biggr)^{\frac 1{8K-2}}.
 \label{m_T=0}
\end{align}
As it is expected, the small interladder coupling $\e J'_2$ has little influence on the order parameter \eqref{m_T=0} at zero temperature.

\subsection{Near critical temperature}

We move on to solving the self-consistent equations at finite temperatures.
In general it is challenging to solve analytically the self-consistent equations at finite temperatures.
Here we instead solve them graphically~\cite{Cazalilla_NJP_2006}.

The self-consistent equation for $v_y(T)/v_y(0)$ is then given by
\begin{widetext}
\begin{align}
 \frac{v_y(T)}{v_y(0)}
 &= \exp\left[-\frac T{4\Omega}\sum_{\omega, \bm k} F(k_y)G_{\rm v}(i\omega_n,\bm k)
 +\frac T{4\Omega}\sum_{\omega, \bm k} F(k_y)G_{\rm v}(i\omega_n,\bm k)\biggr|_{T=0}\right],
 \notag \\
 &=\exp\left[-\frac{\pi v}{8K}\int_{-\pi}^\pi \frac{dk_ydk_z}{(2\pi)^2}
 \int_{-T/v}^{T/v}
 \frac{dk_x}{2\pi}\frac{F(k_y)}{\sqrt{v^2k_x^2+v_y^2F(k_y)+v_z^2F(k_z)}}
 \biggl\{\coth\biggl(\frac 1{2T}\sqrt{v^2k_x^2+v_y^2F(k_y)
 +v_z^2F(k_z)}\biggr)-1\biggr\}
 \right].
 \label{vyT/vy0}
 \end{align}
The integration with $k_x$ is cut off at $T/v$ ($\ll 1$), as we argued in Sec.~\ref{sec:sce}.
To handle Eq.~\eqref{vyT/vy0}, we assume $v_y/T<1$ and expand the hyperbolic cotangent with respect to the small parameter $v_y/T$.
A complex but straightforward calculation~\cite{Cazalilla_NJP_2006} leads to
\begin{equation}
 \frac{v_y(T)}{v_y(0)}
  \simeq \biggl(\frac{2T}{v_y(T)}\biggr)^{\frac 1{4K}} \exp \biggl[-\frac
 1{4K}\biggl\{\frac{2T}{v_y(T)}- \frac 56
-\frac 4\pi \frac{v_z(T)}{v_y(T)}
 +2\biggl(\frac{v_y(T)}T\biggr)^2
 \biggr\}\biggr].
  \label{vy_SCE_expand}
\end{equation}
The other velocity $v_z(T)/v_z(0)$ is similarly derived:
\begin{equation}
 \frac{v_z(T)}{v_z(0)}\simeq
  \biggl(\frac{2T}{v_z(T)}\biggr)^{\frac 1{4K}}
  \exp \biggl[-\frac 1{4K}\biggl\{
  \frac {\pi T}{2v_y(T)} \biggl(1-\frac 2\pi \ln
  \biggl[\tan\biggl(\frac{v_z(T)}{4v_y(T)}\biggr)\biggr]\biggr)
  -\frac{11}6
  +\frac 23 \biggl(\frac{v_y(T)}T\biggr)^2 +\frac 23
  \biggl(\frac{v_z(T)}T\biggr)^2\biggr\}\biggr].
  \label{vz_SCE_expand}
\end{equation}
In addition, the order parameter $m(T)$ is also expanded as follows.
\begin{equation}
 \frac{m(T)}{m(0)}
  =
  \biggl(\frac{2T}{v_y(T)}\biggr)^{\frac 1{4K}}
  \exp\biggl[-\frac 1{4K}\biggl\{\frac{\pi T}{v_y(T)} \biggl(\frac
  23-\frac 1\pi \ln
  \biggl[\tan\biggl(\frac {v_z(T)}{4v_y(T)}\biggr)\biggr]\biggr)
  -\frac{11}6
  + \frac 23 \biggl(\frac{v_y(T)}T\biggr)^2 +\frac23
  \biggl(\frac{v_z(T)}T\biggr)^2 \biggr\}\biggr].
  \label{m_SCE_expand}
\end{equation}
\end{widetext}
Note that equations \eqref{vy_SCE_expand} and \eqref{vz_SCE_expand} have a trivial solution $v_y=v_z=0$.

\begin{figure}[b!]
 \centering
 \includegraphics[bb = 0 0 957 621, width=\linewidth]{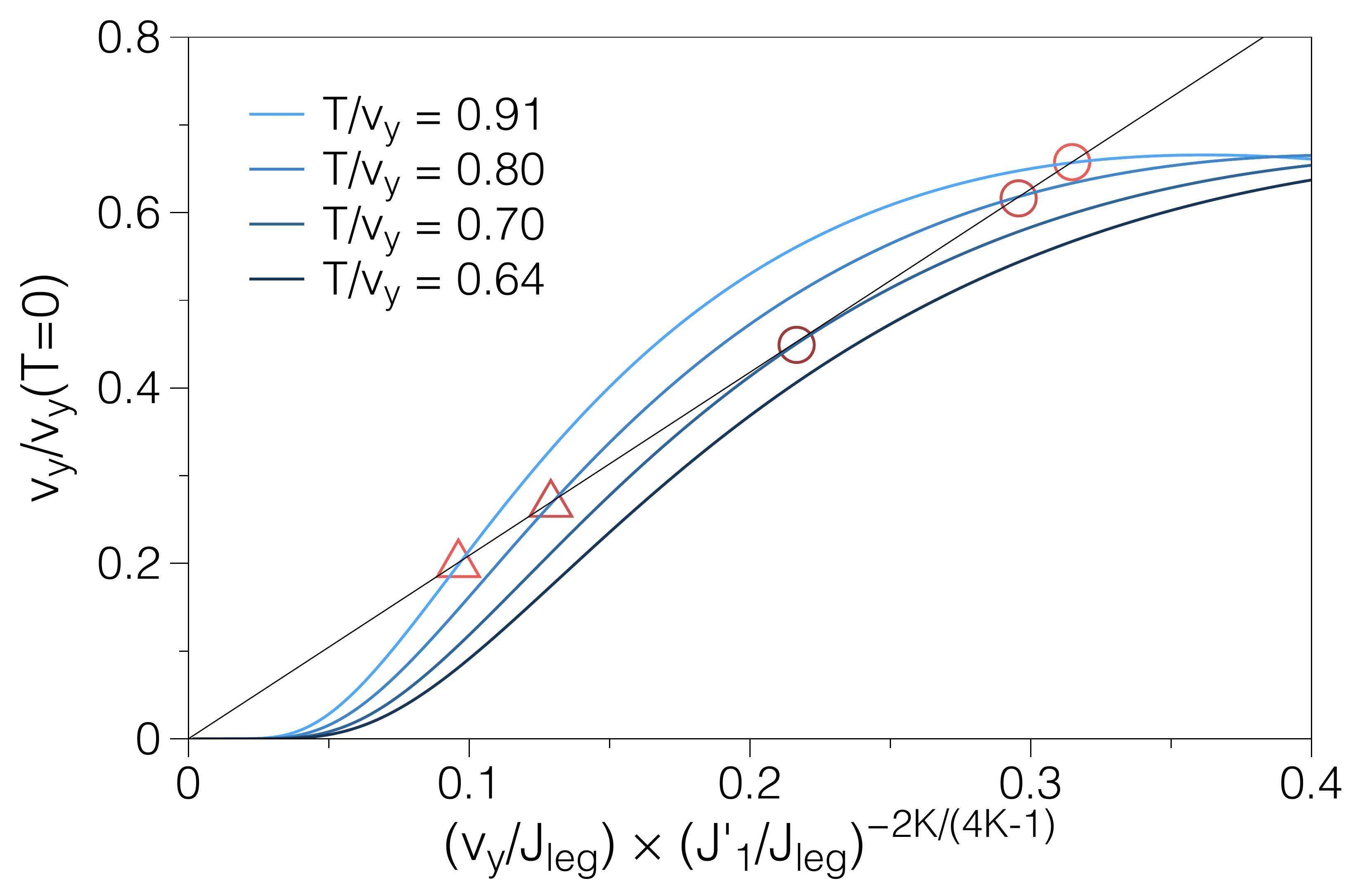}
 \caption{(Color online)
 The visulalization of the self-consistent equation
 \eqref{vy_SCE_expand} in the limit $v_z(T)\to 0$.
 The crossing points indicated by circles are the physical solutions
 $v_y(T)$ of Eq.~\eqref{vy_SCE_expand}.
 The other unphysical solutions indicated by triangles are rejected
 because they approach zero as the temperature goes to zero.
 }
 \label{SCE_vy}
\end{figure}

\begin{figure}[b!]
 \centering
 \includegraphics[bb = 0 0 956 623, width=\linewidth]{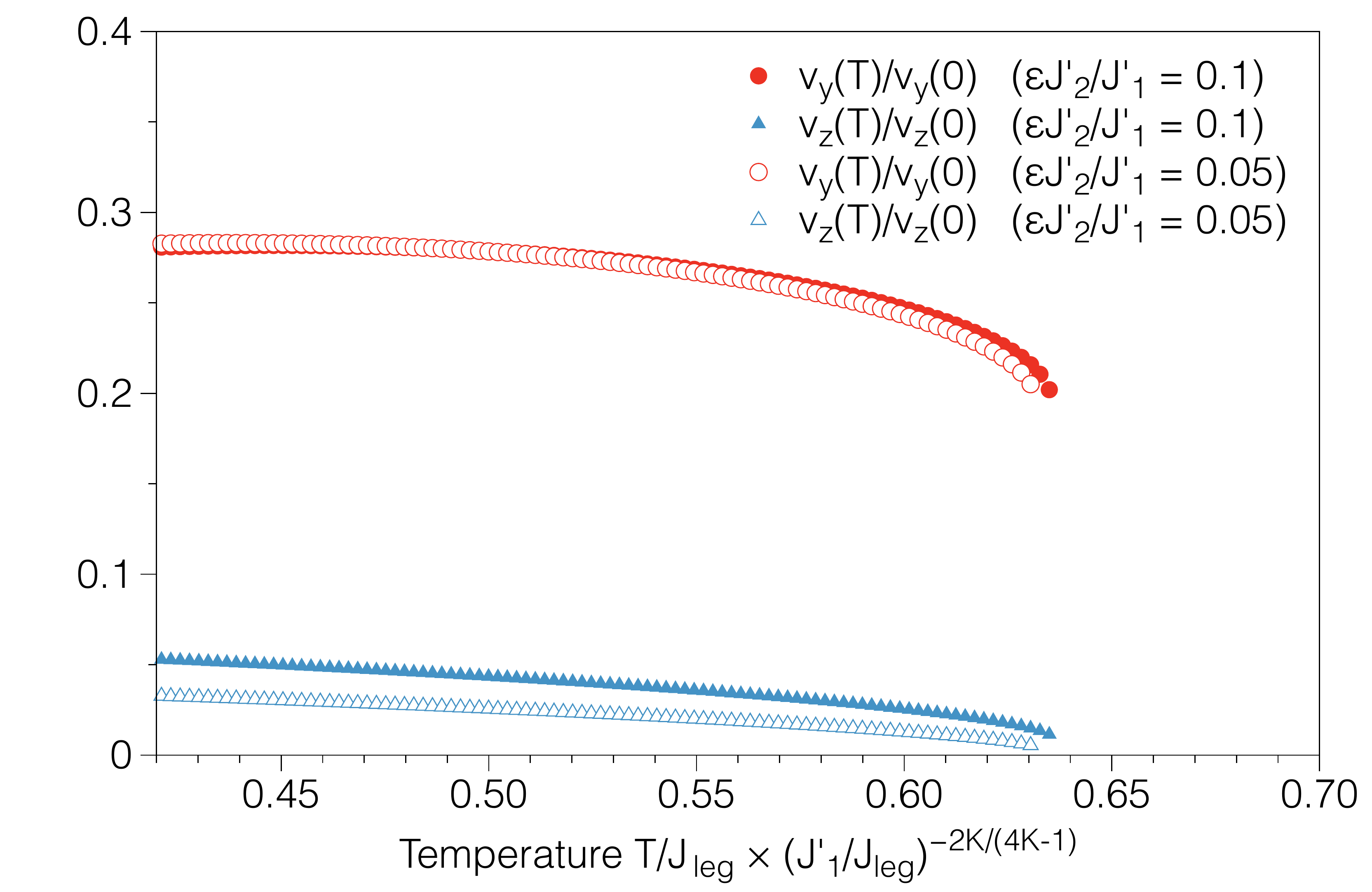}
 \caption{(Color online)
 Solutions of the self-consistent equations \eqref{vy_SCE_expand} and
 \eqref{vz_SCE_expand} for $\e J'_2/J'_1 = 0.1$ and $0.05$.
 While $v_y(T)$ is insensitive to the ratio $\e J'_2/J'_1$, the other
 velocity $v_z(T)$ is sensitive.
 }
 \label{vyvz_SCE}
\end{figure}

To get insight into the numerical procedure to solve them, we see the special case of the 2D limit.
In the purely 2D limit, the trivial solution $v_z(T)=0$ becomes the physical solution for Eq.~\eqref{vz_SCE_expand}.
Then Eq.~\eqref{vy_SCE_expand} becomes independent of $v_z(T)$
and allows us to draw the self-consistent equation \eqref{vy_SCE_expand} as a 2D graph.
Figure~\ref{SCE_vy} shows how $v_y(T)$ is determined from the graph.
$T_c$ is determined so that the curve on the right hand side of Eq.~\eqref{vy_SCE_expand} has a tangent point with
the line on the left hand side of Eq.~\eqref{vy_SCE_expand}.
Those curves have no nontrivial intersection for $T>T_c$ and have two nontrivial intersections for $T<T_c$.
We note that one of the two solutions, marked with triangles in Fig.~\ref{SCE_vy}, is unphysical because of its unphysical temperature dependence.
$v_y(T)$ gives the measure of the correlation in the $y$ direction.
Therefore we may expect $v_{y}(T)$ to be a  monotonically decreasing function of the temperature $T$.
We will confirm these expectations later.
On the other hand, the unphysical solution is increasing with increase of the temperature.
Discarding the unphysical one, we obtain the physical self-consistent solution.

\begin{figure}[b!]
 \centering
 \includegraphics[bb = 0 0 959 623, width=\linewidth]{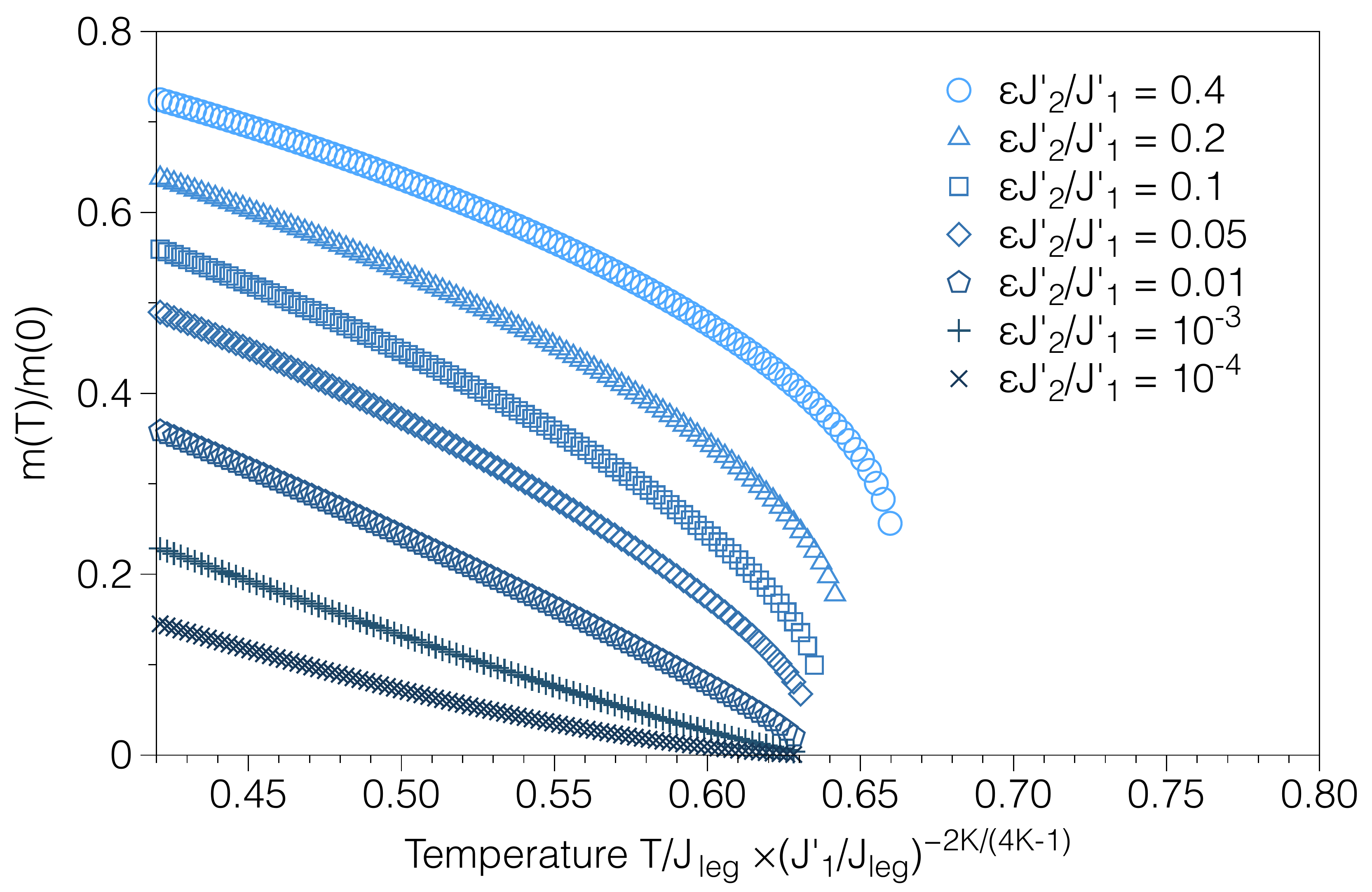}
 \caption{(Color online)
 The order parameter \eqref{m_SCE_expand} is plotted against  the temperature $T$.
When decreasing the ratio $\e J'_2/J'_1$,  the order parameter is
 more and more suppressed.
 On the other hand, the critical temperature is almost not affected as
  expected from the RPA result \eqref{T_c}.
 }
 \label{mag_q2d}
\end{figure}

For $\e J'_2/J'_1>0$ we solve Eqs.~\eqref{vy_SCE_expand} and \eqref{vz_SCE_expand} simultaneously.
The solving procedure is the same as in the 2D case.
Figure~\ref{vyvz_SCE} shows the self-consistent solutions $v_y(T)$ and $v_z(T)$ for $\e J'_2/J'_1 = 0.1$ and $0.05$.
We can see that $v_y(T)$ is indeed monotonically decreasing with increasing temperature $T$.
Thus we may rewrite the inequality $v_y(T)/T < 1$ as
\begin{equation}
 T_\ast < T
  \label{range_expand}
\end{equation}
where $T_\ast$ is defined as a solution of $T_\ast = v_y(T_\ast)$.
Using the parameters \eqref{param_DIMPY}, we obtain an approximate value $T_\ast\simeq 0.42\Jl$.

Thus far we have derived the critical temperature in two independent ways: RPA and the variational approach.
Given the parameters~\eqref{param_DIMPY},
the variational solution of $T_c\simeq 0.64\Jl (J'_1/\Jl)^{2K/(4K-1)}$ of Fig.~\ref{vyvz_SCE} is $50$~\% smaller  than the  RPA solution \eqref{T_c_RPA}.
The variational solution tends to underestimate the critical temperature
 because the variational method implicitly assumes a well developed order. On the contrary, RPA approach being a mean-field technique tends to overestimate the ordered phase.

We plotted the N\'eel order parameter \eqref{m_SCE_expand} for various values of $\e J'_2/J'_1$ in Fig.~\ref{mag_q2d},
where the order parameter is more suppressed with the reduction of the dimensionality (i.e. $\e J'_2/J'_1\to 0$).
Figure~\ref{mag_q2d} shows a jump of the order parameter at the critical temperature for various $\e J'_2/J'_1$.
The jump originates from the invalidity of the variational approach near $T_c$, as we explained.
Nevertheless Fig.~\ref{mag_q2d} strongly suggests that $T_c$ remains finite for infinitesimal $\e J'_2/J'_1$ in agreement with the RPA analysis.\\

\subsection{Deep in the ordered phase}

We could solve the self-consistent equations \eqref{vy_SCE_expand} and \eqref{vz_SCE_expand} thanks to the expansion with the small parameter $v_y(T)/T$.
This expansion is justified only for the range \eqref{range_expand}, near the critical temperature.
We note that another expansion with a small parameter is also possible deep in the ordered phase for $T<T_\ast$.
That is the expansion with $v_z(T)/T$.

When $T\ll T_\ast$, the velocity $v_y(T)$ is large enough to guarantee the quadratic approximation \eqref{F_approx}.
Let us expand the self-consistent equations with the small parameter $v_z(T)/T$ in the same way as in the previous subsection.
$v_y(T)/v_y(0)$ is represented as
\begin{widetext}
\begin{align}
 \frac{v_y(T)}{v_y(0)}
 &= \exp\biggl[ -\frac{\pi
 v}{8K}\int_{-\pi}^{\pi}\frac{dk_z}{2\pi}
 \int_{-T/v_y}^{T/v_y} \frac{dk_y}{2\pi}\int_{-T/v}^{T/v}\frac{dk_x}{2\pi}
 \frac{k_y^2}{\sqrt{v^2k_x^2+v_y^2k_y^2+v_z^2F(k_z)}}
 \biggl\{\coth\biggl(\frac 1{2T}
 \sqrt{v^2k_x^2+v_y^2k_y^2+v_z^2F(k_z)}\biggr)-1\biggr\}
 \biggr] \notag \\
 &\simeq \exp\biggl[-\frac{T^3}{96Kv_y(T)^3}\biggl\{
 \frac{17}{8}+9 \biggl(\frac{v_z(T)}T\biggr)^3
 \biggr\}\biggr].
 \label{vy_SCE_3d}
\end{align}
We used the effective cutoff $T/v_y$ in the integral of $k_y$.
The other velocity $v_z(T)$ obeys a self-consistent equation,
\begin{align}
 \frac{v_z(T)}{v_z(0)}
 &\simeq \biggl(\frac{v_z(T)}T\biggr)^{\frac T{4Kv_y(T)}}\exp\biggl[
 -\frac T{8Kv_y(T)}\biggl\{-2+\frac
 1{3\pi}
 +\frac{16}{3\pi}\frac{v_z(T)}T + \frac32
 \biggl(\frac{v_z(T)}T\biggr)^2\biggr\}\biggr].
 \label{vz_SCE_3d}
\end{align}
The order parameter $m(T)$ is
\begin{align}
 \frac{m(T)}{m(0)}
 &\simeq\biggl(\frac{v_z(T)}T\biggr)^{\frac T{4Kv_y(T)}}\exp\biggl[
 -\frac T{8Kv_y(T)}\biggl\{-1+\frac 1{3\pi}
 +\frac 4{\pi} \frac{v_z(T)}T
 +\biggl(\frac{v_z(T)}T\biggr)^2\biggr\}\biggr].
 \label{m_SCE_3d}
\end{align}
\end{widetext}
In analogy with the expansions \eqref{vy_SCE_expand}, \eqref{vz_SCE_expand} and \eqref{m_SCE_expand},
the expansions \eqref{vy_SCE_3d}, \eqref{vz_SCE_3d} and \eqref{m_SCE_3d} are valid in a temperature range,
\begin{equation}
 T_{\ast\ast} < T < T_\ast.
  \label{range_expand_3d}
\end{equation}
The lower bound $T_{\ast\ast}$ is a solution of an equation $T_{\ast\ast} = v_z(T_{\ast\ast})$.
The range \eqref{range_expand_3d} becomes wider as $\e J'_2/J'_1\to 0$ because
$T_\ast$ is kept finite while $T_{\ast\ast}$ goes to zero in that limit.

\begin{figure}[b!]
 \centering
 \includegraphics[bb = 0 0 1200 800, width=\linewidth]{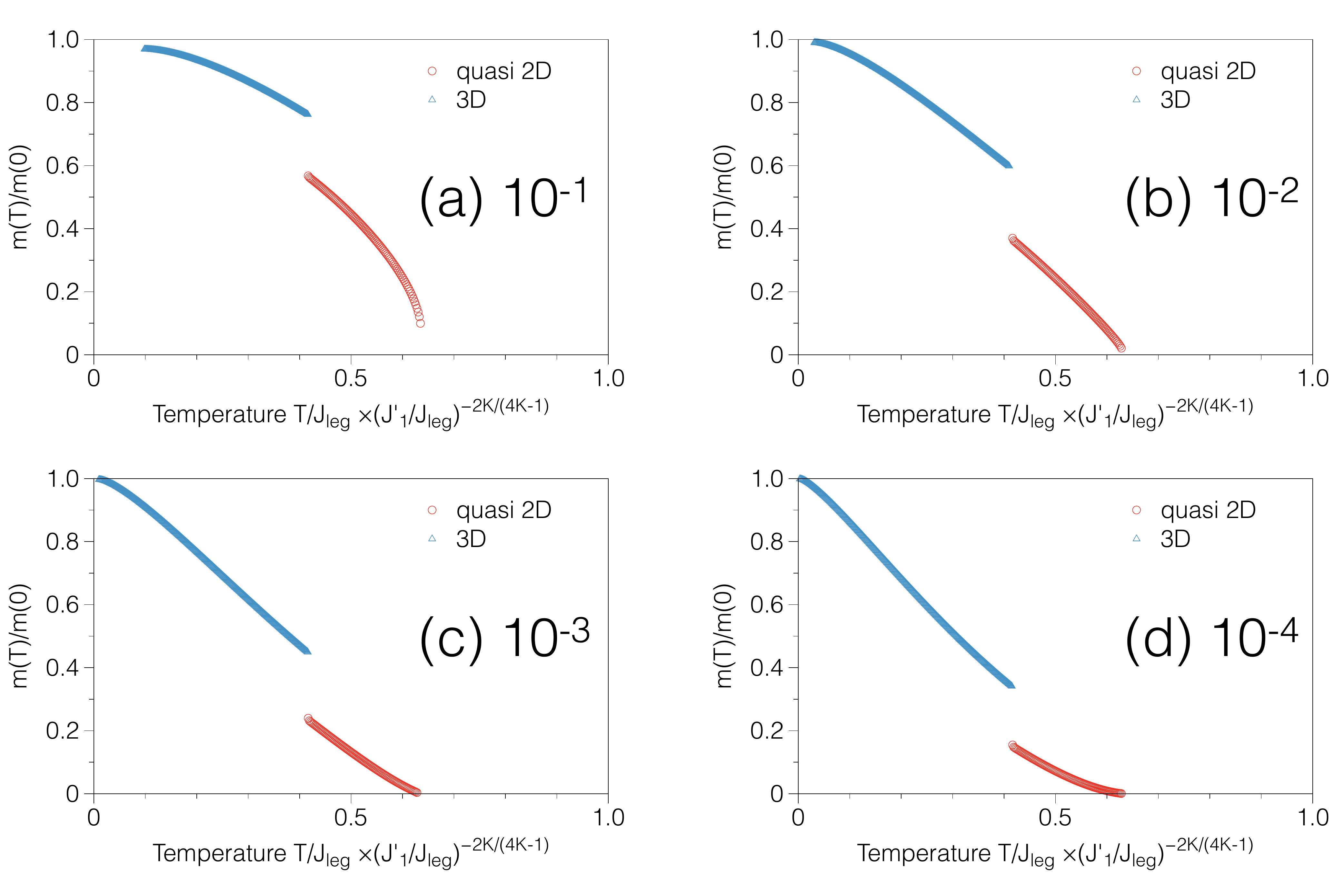}
 \caption{(Color online)
 The N\'eel order parameter $m(T)/m(0)$ is plotted against the
 temperature  $T$ for several ratios of  $\e J'_2/J'_1$: (a) $10^{-1}$, (b) $10^{-2}$, (c) $10^{-3}$ and $10^{-4}$.
 The red circles depict the order parameter in the temperature range
 $T_{\ast}< T < T_c$ of the quasi-2D phase.
 The blue triangles depict the order parameter in the range  \eqref{range_expand_3d}.
 }
 \label{stag_whole}
\end{figure}

Figure~\ref{stag_whole} shows the N\'eel order parameter in the whole temperature range for $\e J'_2/J'_1=10^{-1}, 10^{-2}, 10^{-3}$, and $10^{-4}$.
The red circles are the order parameter in the temperature range \eqref{range_expand}.
In addition, we plotted the order parameter \eqref{m_SCE_3d} in the range \eqref{range_expand_3d} with the blue triangles.
The order parameter \eqref{m_SCE_3d} turned out to converge very well to $m(0)$ as $T \to 0$.
This agreement shows the high accuracy of the approximation \eqref{vy_SCE_3d}, \eqref{vz_SCE_3d} and \eqref{m_SCE_3d}.

On the other hand, the curves of Eqs.~\eqref{m_SCE_3d} and \eqref{m_SCE_expand} disagree at $T=T_\ast$.
This is because the approximation \eqref{m_SCE_expand} with $v_y(T)/T\ll 1$ becomes less accurate as $T\to T_\ast +0$ and
the quadratic approximation \eqref{F_approx} becomes less reliable as $T\to T_\ast-0$.
Therefore the discontinuity of the order parameter at $T=T_\ast$ is a technical artifact.
Figure~\ref{stag_whole} implies a rapid growth of the N\'eel order below a certain temperature $T_{\rm cr}$.
The temperature $T_{\rm cr}$ is clearly important in order to distinguish two different regions of the ordered phases: the higher-temperature side
with the suppressed N\'eel order and the lower-temperature side with the full N\'eel order.
We call the former region as the quasi-2D phase and the latter as the 3D phase (Fig.~\ref{quasi2D_def}).
$T_{\rm cr}$ gives the crossover temperature between those phases.
In the next section, we discuss how to characterize the crossover temperature $T_{\rm cr}$.

\section{Characteristic temperatures: renormalization group analysis}\label{sec:rg}

The RPA analysis in Sec.~\ref{sec:rpa} uncovered that the critical temperature converges to the finite value $T_{\rm KT}$ in the 2D limit.
The variational analysis in Sec.~\ref{sec:variation} illustrated the temperature dependence of the N\'eel order parameter
in the quasi-2D and the 3D ordered phases.
The main claim of this section is that there is indeed a characteristic temperature $T_{\rm cr}$ which separates
the quasi-2D and 3D ordered phases.
Here we develop the RG analysis to clarify $T_{\rm cr}$.

\subsection{Critical temperature}

The RG transformation describes the dependence of couplings of the action and the Hamiltonian on the cutoff $E$ in energy.
In the course of the RG transformation the cutoff $E$ is reduced from its initial value $E=\Jl$.
As we already discussed, the effective cutoff of the wavenumber  in the $a$ direction is given by $T/v_a$
because the cutoff of the energy is given by $T$ itself.
Therefore the RG transformation is performed down to $E=T$.

There is a subtlety in the definition of the effective cutoff.
The effective cutoff in the energy is naively given by  $E=\max\{T, h\}$~\cite{Furuya_NLSM_2011}.
Then the RG flow seems to stop at $E=h$ at low temperatures $T<h$ of our interest (Fig.~\ref{PD_ladders}).
Actually, we can carry forward the RG transformation down to $E=T$ even when $T<h$ because of the following reason.
We need to recall that the Euclidean action \eqref{S} of our system is written  in terms of the $\theta_{\mu,\nu}$ field only
and that this field is related to the N\'eel order perpendicular to the magnetic field.
Therefore the magnetic field has no contribution to the action \eqref{S} except for determining the plane on which the N\'eel order lies.
This separation of the action \eqref{S} from the magnetic field enables us to set the effective cutoff to $E=T$ even under a strong magnetic field.

To perform the RG procedure, we rewrite  the action \eqref{S} as
\begin{align}
  \mathcal S
 &= \frac K{2\pi v}\sum_{\mu,\nu}\int d\tau dx \,
 \bigl\{(\partial_\tau\theta_{\mu,\nu})^2 +
 v^2(\partial_x\theta_{\mu,\nu})^2\bigr\} \notag \\
 & \quad -\frac{vg_1(T)}{\pi } \sum_{\mu,\nu} \int d\tau dx \,
 \cos(\theta_{\mu,\nu}-\theta_{\mu+1,\nu})   \notag \\
 & \quad
 - \frac{vg_2(T)}{\pi } \sum_{\mu,\nu} \int d\tau dx \,
 \cos(\theta_{\mu,\nu}-\theta_{\mu,\nu+1}).
 \label{H_E}
\end{align}
As well as the couplings $g_1(E)$ and $g_2(E)$, the TLL parameter $K$ also depends on the energy scale $E$ in principle.

Details of the RG equation depends on the dimensionality of the phase.
First we investigate the 1D phase of the field-induced TLL for $T>T_c$.
Adopting the bosonization formulas,  we obtain the bare values of the couplings,
\begin{equation}
 g_1(\Jl) = \frac{\pi J'_1A_x}v, \quad  g_2(\Jl) = \frac{\pi \e  J'_2A_x}v.
  \label{g_12_bare}
\end{equation}

As we lower the energy scale $E$, the couplings of Eq.~\eqref{H_E} are renormalized through the RG equations~\cite{Benfatto_SCfilm_2007},
\begin{align}
 \frac{dK}{d\ln E}
 &= -(g_1^2 + g_2^2)K^2,
 \label{K_RG} \\
 \frac{dg_n}{d\ln E}
 &= - \biggl(2-\frac 1{2K}\biggr)g_n.
  \label{g_12_RG}
\end{align}
The change \eqref{K_RG} of the TLL parameter $K$ during the RG transformation is exponentially small compared to those \eqref{g_12_RG} of $g_1$ and $g_2$.
In what follows we regard $K$ as a constant.
Given the TLL parameter $K>1/4$, the effective couplings $g_n(E)$ grow as follows.
\begin{equation}
 g_n(E) = g_n(\Jl) \biggl(\frac E{\Jl}\biggr)^{-\frac{4K-1}{2K}}.
  \label{g_n_eff}
\end{equation}

Let  $E_c$ be an energy scale at which the RG equation \eqref{g_12_RG} reaches the nonperturbative regime.
In other words $E_c$ represents a temperature below which the interladder correlation becomes nonnegligible.
Thus we may expect that the critical temperature \eqref{T_c} plays the role of $E_c$.
We can confirm this expectation as follows.
Solving Eq.~\eqref{g_12_RG} with the condition \eqref{g_12_bare} at $E=\Jl$ and $g_1(E_c)=1$, we obtain
\begin{equation}
 \frac{E_c}{\Jl} =\biggl(\frac{\pi A_x\Jl}{v}\biggr)^{\frac{2K}{4K-1}}
  \biggl(\frac{J'_1}{\Jl}\biggr)^{\frac{2K}{4K-1}}.
 \label{gap}
\end{equation}
$E_c$ obviously corresponds to the critical temperature \eqref{T_c}.
Note that  the other coupling $g_2(E)$ remains in the perturbative regime $g_2(E_c)\ll 1$ even for $E<E_c$ because of the anisotropy $\e J'_2/J'_1 < 1$.

\subsection{Crossover temperature}

Next we develop the RG analysis in the ordered phase for $T<T_c$ in the presence of a strong anisotropy $\e J'_2/J'_1\ll 1$.

In the 2D limit $\e J'_2/J'_1\to 0$, our system can be seen as a layered 2D superfluid system [Fig.~\ref{cube}~(b)],
where the interlayer coupling $\e J'_2$ plays the role of the Josephson coupling~\cite{Benfatto_SCfilm_2007}.
Let us investigate the RG equation of the interlayer coupling from that viewpoint.
We adopt the classical approximation by discarding the imaginary-time dependence: $\theta_{\mu,\nu}(\tau,x)\simeq \theta_{\mu,\nu}(x)$.
The classical approximation works well near the critical temperature~\cite{Sachdev_book}.
Although it is challenging to specify the precise range of validity of the classical approximation,
we can expect that the validity range is very wide in the 2D limit $\e J'_2/J'_1 \to 0$
because the system stays at the critical line at low temperatures $T<T_{\rm KT}$.

The classical approximation turns the action \eqref{H_E} into
\begin{align}
 \mathcal S_{\rm cl}
 &=\frac{K}{2\pi vT}\sum_{\nu}\int dxdy
 \bigl\{v^2(\partial_x\theta_\nu)^2 + v_y^2(\partial_y
 \theta_\nu)^2\bigr\}
 \notag \\
 & \quad -\frac{vg_2(T)}{\pi}\sum_{\nu}\int dxdy \cos(\theta_\nu-\theta_{\nu+1}).
 \label{S_cl_pre}
\end{align}
Here the interladder interaction $\cos(\theta_{\mu,\nu}-\theta_{\mu+1,\nu})$ is approximated as
\begin{equation}
 \cos(\theta_{\mu,\nu}-\theta_{\mu+1,\nu}) \simeq 1-\frac 12 (\partial_y \theta_{\nu})^2
  \label{quad_approx}
\end{equation}
and the classical field $\theta_{\mu,\nu}(x)$ is rewritten as $\theta_\nu(x,y)$.
Rescalings
\begin{align}
 y&=v_y\tau', \quad K' = \frac{v_y}TK, \quad
 g'_2 = \frac{v_y}T g_2,
 \label{rescaling}
\end{align}
replace the classical action \eqref{S_cl_pre} by
\begin{align}
 \mathcal S_{\rm cl}
 &=\frac{K'}{2\pi v}\sum_{\nu}\int dxd\tau'
 \bigl\{v^2(\partial_x\theta_\nu)^2 + (\partial_{\tau'}
 \theta_\nu)^2\bigr\} \notag \\
 & \quad -\frac{vg'_2(T)}{\pi}\sum_{\nu}\int dxd\tau' \cos(\theta_\nu-\theta_{\nu+1}).
 \label{S_cl}
\end{align}
Therefore, \emph{the action \eqref{S_cl} of the $(2+0)$-dimensional classical theory can also be seen as the one of a $(1+1)$-dimensional quantum theory at zero temperature}, where $\tau'$ is the imaginary time.

Taking advantage of this equivalence between the $(1+1)$-dimensional quantum theory and the $(2+0)$-dimensional classical theory,
we can derive the RG equation of $g'_2$ in the same manner as \eqref{g_12_RG}:
\begin{align}
 \frac{dg'_2}{d\ln E} &= - \biggl(2-\frac 1{2K'}\biggr)g'_2.
 \label{g_2_RG_cl}
\end{align}
Let $E_{\rm cr}$ be the energy scale below which $g'_2$ is nonperturbative.
Considering the analogy with the correspondence between $E_c$ and $T_c$,
we identify $E_{\rm cr}$ with the crossover temperature $T_{\rm cr}$.
The crossover temperature $T_{\rm cr}$ approaches zero in the 2D limit by definition.
Then $T_{\rm cr}\ll v_y$ follows  because $v_y$ is of the order of $J'_1$.
According to the rescaling \eqref{rescaling}, $K'\gg 1$ is valid for $T\lesssim T_{\rm cr}$ even though $K=O(1)$.

We identify $T_{\rm cr}$ with the energy scale $E_{\rm cr}$ at which the RG transformation \eqref{g_2_RG_cl} breaks down.
Then the crossover temperature satisfies the condition $g_2(T_{\rm cr})=1$.
Approximating the RG equation \eqref{g_2_RG_cl} as $dg'_2/d\ln E\simeq -2g'_2$ and integrating it over $E$,
we obtain
\begin{align}
 \frac{T_{\mathrm{cr}}}{T_c}
 & \simeq  \sqrt{\frac{v_y(T_c)}{T_c} \frac{\e J'_2}{J'_1}}.
 \label{T_cr}
\end{align}
In  particular, using parameters of Eq.~\eqref{param_DIMPY},
it equals to
\begin{equation}
 \frac{T_{\rm cr}}{T_c} \simeq  \sqrt{0.2 \frac{\e J'_2}{J'_1}}.
  \label{T_cr_num}
\end{equation}

\section{Classical approximation and order parameter}
\label{sec:classical_approx}

\begin{figure}[b!]
 \centering
 \includegraphics[bb = 0 0 959 623, width=\linewidth]{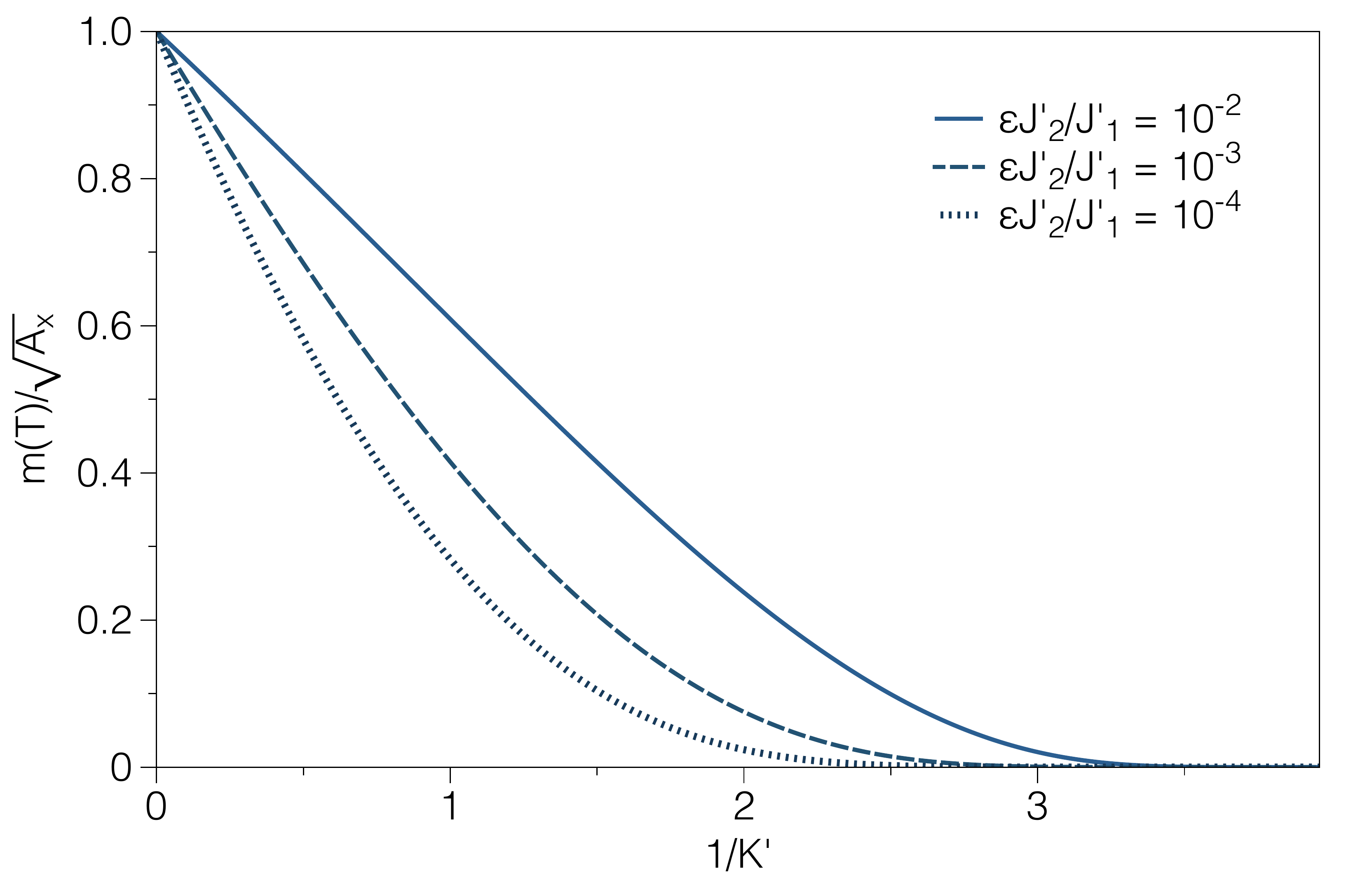}
 \caption{(Color online)
 The order parameter in the strongly 2D cases.
 The horizontal axis represents the temperature $1/K'\propto T$.
 }
 \label{mx_classical}
\end{figure}

In this Section, we develop further the analysis of the classical system and
describe the ordering process coherently in the entire region of the ordered phase.

Discarding the imaginary-time direction,
the classical approximation replaces the original $(3+1)$-dimensional quantum system to the $(3+0)$-dimensional classical one.
As we mentioned, the classical approximation works better for $\e J'_2/J'_1 \to 0$.
The small $\e J'_2/J'_1$ allows us to perform the MF approximation in order to replace the interplane interaction by an effective staggered field
$h_s(T)$.
Then the system is effectively $(2+0)$-dimensional and has an effective action,
\begin{align}
 \mathcal S_{\rm cl}
 &=\frac{K'}{2\pi v}\sum_{\nu}\int dxd\tau'
 \bigl\{v^2(\partial_x\theta_\nu)^2 + (\partial_{\tau'}
 \theta_\nu)^2\bigr\} \notag \\
 & \quad -2h_{\rm s}(T)\sum_{\nu}\int dxd\tau' \cos\theta_\nu,
 \label{S_cl_MF}
\end{align}
with
\begin{equation}
 h_s(T)\equiv \frac{vg'_2(T)m(T)}{\pi }=\frac{v \e J'_2 m(T)}{\pi}\frac{K'}K\biggl(\frac T{\Jl}\biggr)^{-2},
  \label{h_s}
\end{equation}
where the RG solution
$g'_2(T)=K'g_2(T)/K=K'\e J'_2 (T/\Jl)^{-2}/K$ was used.
The coupling $h_s(T)$ represents the mean field that depends on $m(T)$.
As a consequence of the MF approximation, the classical $(2+0)$-dimensional system \eqref{S_cl_MF} is equivalent to
the $(1+1)$-dimensional one.
Note that $h_s(T)$ depends on the N\'eel order $m(T)$ which is determined self-consistently from $h_s(T)$.
To do so, we compute the N\'eel order parameter $m(T)=\sqrt{A_x}\langle \cos\theta_\nu(x)\rangle$
using the equivalence to the $(1+1)$-dimensional system.
The field theory \eqref{S_cl_MF} is well known as the sine-Gordon theory.
Thanks to integrability of the sine-Gordon theory, we are able to calculate the order parameter exactly~\cite{Lukyanov_SG}:
\begin{align}
 m(T)
 &= \sqrt{A_x}\biggl(\frac{h_s}v\biggr)^{\frac1{8K'-1}}\frac{\frac{8K'}{8K'-1}\tan(\frac{\pi}{16K'-2})\Gamma^2(\frac 1{16K'-2})}{2\pi\Gamma^2(\frac{4K'}{8K'-1})}
 \notag \\
 & \quad \times
  \left[\frac{\pi\Gamma(\frac{8K'-1}{8K'})}{\Gamma(\frac 1{8K'})}\right]^{\frac{8K'}{8K'-1}}
 \label{m_SG_SCE}
\end{align}
Solving Eq.~\eqref{m_SG_SCE} with respect to $m(T)$, we obtain
\begin{align}
 m(T)
 &= \sqrt{A_x}\mathcal F(K')\biggl(\frac{\sqrt{A_x}\e J'_2}{\pi v}\frac{K'}K\biggr)^{\frac 1{8K'-2}}\biggl(\frac T{\Jl}\biggr)^{-\frac 1{4K'-1}},
 \label{m_SG}
\end{align}
with
\begin{align}
 \mathcal F(K')
 &= \left[\frac{\frac{8K'}{8K'-1} \tan(\frac{\pi}{16K'-2}) \Gamma^2(\frac 1{16K'-2})}{2\pi \Gamma^2(\frac{4K'}{8K'-1})}\right]^{\frac{8K'-1}{8K'-2}}
 \notag \\
 & \quad \times
  \left[\frac{\pi\Gamma(\frac{8K'-1}{8K'})}{\Gamma(\frac 1{8K'})}\right]^{\frac{8K'}{8K'-2}}
 \label{F}
\end{align}
The order parameter \eqref{m_SG} depends on the temperature via $g'_2(T)$ and the TLL parameter $K'\propto T^{-1}$.
Note that the variational calculation strongly suggests that
$v_y$ is constant at temperatures much lower than the critical temperature (Fig.~\ref{vyvz_SCE}).

Figure~\ref{mx_classical} shows the temperature dependence of the order parameter \eqref{m_SG}.
According to Eq.~\eqref{m_SG}, $m(T=0)$ equals to $\sqrt{A_x}$ regardless of the value of $J'_1$ and $J'_2$.
This is clearly incorrect because $m(T)$ must vanish in the limit of $J'_1, J'_2 \to 0$.
However the inconsistency of the classical approximation at $T=0$ is expected for the following reason.
The classical approximation relies on the quadratic expansion \eqref{quad_approx} of the interladder interaction.
This expansion is guaranteed by the large enough $g'_1(T)$, which breaks down obviously in the limit of $J'_1\to 0$.
On the other hand, the curves of Fig.~\ref{mx_classical} are qualitatively reasonable and
consistent with the result of the variational calculations (Fig.~\ref{stag_whole}).
Moreover, the $m(T)$ curves of Fig.~\ref{mx_classical} are free from the unphysical jump seen in Fig.~\ref{stag_whole}.

\section{Numerics for $T=0$ ordering of 2D weakly coupled TLL}
\label{sec:num}

We are at the stage of supporting our results obtained thus far with unbiased numerical calculations. While a full 3D computation is out-of-reach,~\footnote{On the one hand, frustration prohibits quantum Monte-Carlo simulations; on the other hand, even for a nonfrustrated model, 3D anisotropic systems would require very demanding simulations to be able to reach the ground-state properties.},
the classical approximation makes it possible.
Indeed, we have already found that the classical approximation leads to a reasonable result (Fig.~\ref{mx_classical}) consistent with the other analytical results.
The classical approximation has also a great advantage that it substantially reduces cost of numerical calculations.
In this section we provide numerical evidences to support previous analytical results by reproducing the $m(T)$ curve of Fig.~\ref{mx_classical}.
The numerical analysis allows us to go beyond the simple MF approximation with respect to $g'_2$ that we have made in Eqs.~\eqref{S_cl}
and \eqref{S_cl_MF} and test the validity of the MF approximation.

\subsection{Weakly coupled XXZ chains}

To do so, we replace the weakly coupled classical layer system \eqref{S_cl} at finite temperatures by
a weakly 2D coupled quantum chain model at zero temperature as follows.
This is possible because each classical layer is equivalent to a 1D quantum spin chain.
Particularly in the classical system \eqref{S_cl} for $h_{c1}<h<h_{c2}$,
each classical layer at a finite temperature  is equivalent to a weakly coupled TLL at zero temperature.
The interlayer coupling $g'_2$ in the former system is turned into a 2D interchain coupling, which we denote as $J'$,
of TLL in the latter system.
The TLL is a universal quantum state basically independent of underlying system to realize it.
Thus, in order to test the analytical result \eqref{m_SG_SCE}, we can use \emph{any 1D lattice} system  as long as it is described by TLL theory 
at zero temperature.

  Thus, to discuss the weakly 2D coupled TLL, we will use the simplest example, namely coupled XXZ spin-$1/2$ chains.
   The Hamiltonian is
\bea
  &\mathcal{H} &= \sum_{{i},n}\Bigl[S^x_{i,n}S^x_{i+1,n} + S^y_{i,n}S^y_{i+1,n}
   + \Delta S^z_{i,n}S^z_{i+1,n}\nonumber\\
  &+& J'\left(S^x_{i,n}S^x_{i,n+1} + S^y_{i,n}S^y_{i,n+1} + \Delta S^z_{i,n}S^z_{i,n+1}\right)\Bigr],
        \label{2d_xxz_ham}
\eea
 where $i$ and $n$ label the sites along and perpendicular
        to the chains. $\Delta\in[-1,1]$ is the Ising anisotropy
        governing the TLL parameter of isolated XXZ
        chains~\cite{luther_calculation_1975}, and $J'$ controls the
        strength of the transverse 2D interchain coupling between
        the chains. Note that the intrachain coupling along the
        chains is set to unity in this section.

The goal is to numerically investigate the ground-state order parameter as a
function of the TLL parameter $K$ in order to compare it
with various analytical approaches. We first discuss MF
approximations, both analytically (as previously presented in
Section~\ref{sec:classical_approx}), and within a numerical
scheme~\cite{Sakai_HaldanePhase_RPA_1989} based on the density-matrix renormalization group (DMRG)
technique~\cite{schollwock_density-matrix_2005}. We also compare
these results with a linear spin-wave (SW) theory.  Quantum
Monte Carlo (QMC) simulations are then used to deal with the exact quantum
mechanical problem and to compute the order parameter of weakly coupled
XXZ chains in the limit of small interchain coupling $J'\in
[0.005,0.1]$. To summarize our main result of the Section, in the strongly anisotropic regime, exact numerics agree
very well with MF theory, provided the transverse coupling is
renormalized $J'\to \alpha_{2\mathrm{D}} J'$, with $\alpha_{2\mathrm{D}} \approx 0.6$ found to be roughly independent of $K$.

\subsubsection{Mean-field theory}

\begin{figure}[!t]
    \centering
    \includegraphics[bb = 0 0 238 327, width=\columnwidth,clip=true]{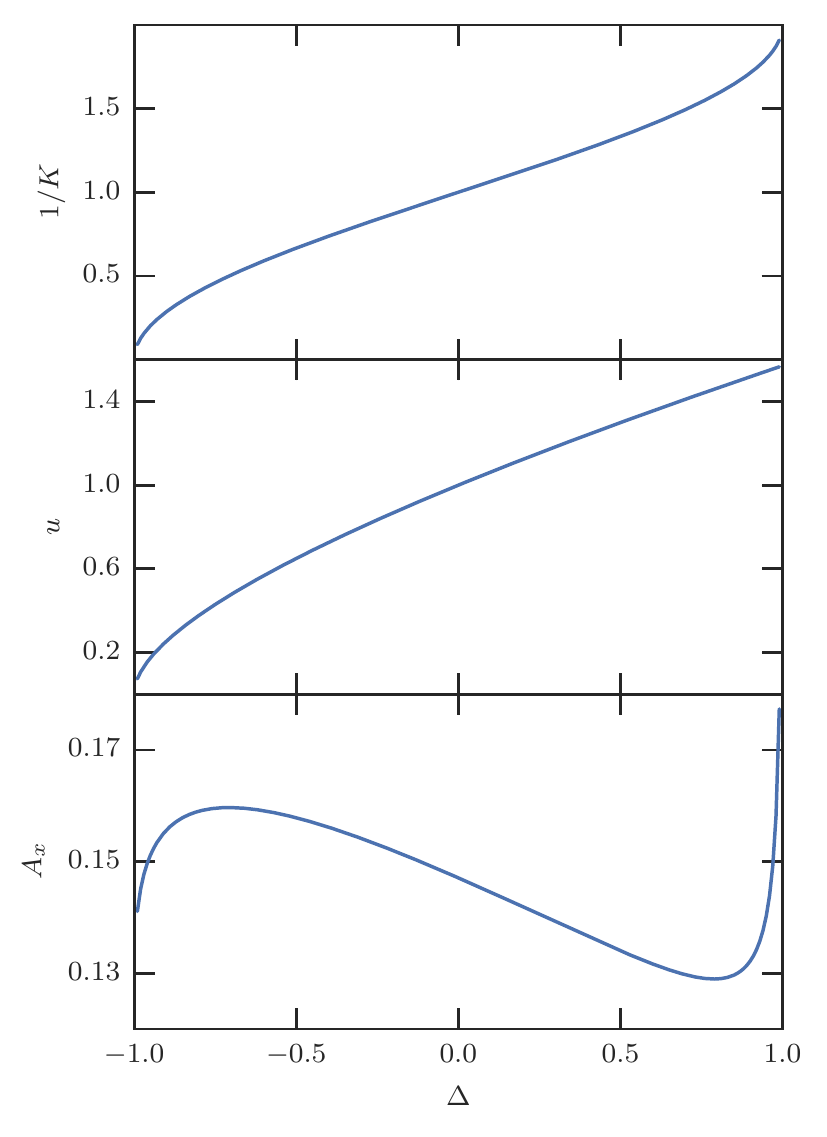}
    \caption{TLL parameter $1/K$, velocity of excitations $u$, and prefactor $A_x$ of the transverse correlations, all plotted as a function of the Ising anisotropy $\Delta$ of an isolated XXZ chain (Eq.~\eqref{2d_xxz_ham} with $J'=0$).}
    \label{fig:ll_params}
\end{figure}

As discussed in Refs.~\onlinecite{Klanjsek_BPCB_2008, Bouillot_BPCB_2011}, and also above, the transverse magnetization of an array of 2D coupled XXZ chains is given by
\begin{equation}
    m^x = \sqrt{2A_x}\mathcal F(K)\left(\frac{2\pi A_x J'}{u}\right)^{\frac{1}{8K-2}}
    \label{mx_ll}
\end{equation}
with $\mathcal F(K)$ [Eq.~\eqref{F}],
where $K$ is the TLL parameter, $u$ the velocity of excitations  and $A_x$ is the amplitude of the transverse correlation function~\cite{Giamarchi_book}. While $u$ and $K$ are know exactly from Bethe Ansatz for critical XXZ chains~\cite{luther_calculation_1975}:
\begin{equation}
    K=\frac{\pi}{2\arccos\left(-\Delta\right)}\quad\mathrm{and}\quad u=\frac{\pi\sqrt{1-\Delta^2}}{2\arccos{\Delta}},
    \label{ll_params}
\end{equation}
the amplitude $A_x$ is not known exactly.
Yet it can still be computed numerically using Lukyanov and Zamolodchikov~\cite{lukyanov_exact_1997,lukyanov_correlation_1999} analytical conjectured expression in the absence of magnetic field. These  parameters are plotted in Fig.~\ref{fig:ll_params} as a function of the Ising anisotropy $\Delta$.

Using these estimates of $K$, $u$ and $A_x$, one can  compute the $T=0$ order parameter $m^x$ in the MF approximation Eq.~\eqref{mx_ll} against $1/K$, displayed in Fig.~\ref{fig:qmc_final} where we also show estimates from a linear SW expansion, together with a more refined DMRG + MF procedure. Both approaches are discussed below.
\subsubsection{DMRG + Mean-Field}
In order to describe a 2D array of coupled chains, 
we rewrite the interchain coupling term of \eqref{2d_xxz_ham} in a standard MF way, neglecting quadratic quantum fluctuations,
\begin{equation}
    \mathcal{H}_{\mathrm{2D}}\longrightarrow\mathcal{H}_{\mathrm{MF}}=2J'm^x\sum_{i=1}^{L}\left(-1\right)^{i}S^x_i
    \label{ham_xxz_2d_mf}
\end{equation}
with $m^x=L^{-1}\sum_{i=1}^{L}\left(-1\right)^{i}\braket{S^x_i}$ the order parameter. The factor of $2$ stands for the number of neighboring chains coupled by $J'$. We obtain a model corresponding to a \emph{single} XXZ chain in an effective staggered (for $J'>0$) magnetic field in the $x$-direction. As $\braket{S^z_{\mathrm{tot}}}=0$ there is no $z$-oriented term in \eqref{ham_xxz_2d_mf}. While this could look like a MF artifact, we already pointed out that the $J'\Delta$ term in \eqref{2d_xxz_ham} has no effect at all on the value of the order parameter and can simply be dropped anyway, as we will show treating the $2$D coupling exactly using QMC [Fig.~\ref{fig:qmc_delta_temp}~(a)].

Numerical MF simulations can be performed in a self-consistent way
using the matrix-product state (MPS) formalism and the DMRG
algorithm~\cite{schollwock_density-matrix_2011,itensor_library}. To do so we start
with a nonzero initial guess for $m^x$ in the Hamiltonian (hence explicitly
breaking the U$(1)$ symmetry) and (tar)get the system ground
state. Once we have it, a new value of the order parameter is measured
and a new MF Hamiltonian is built accordingly. The procedure is
repeated until two consecutive measures of the transverse
magnetization appear to be within a given convergence criterion
($|m^x_{\mathrm{step}~j+1}-m^x_{\mathrm{step}~j}| < 10^{-4}$ in our
case).

The simulation takes longer as we decrease $J'$.
On top of that, the finite size effect becomes more severe as we approach the SU$(2)$ point at $\Delta=1$ (i.e. $1/K=2$).
In order to obtain reliable value of the order parameter in the thermodynamic limit $1/L\to0$,
careful extrapolation of $m^x$ is required, especially for smaller $J'$.
Here we perform the finite-size scaling using various polynomial fittings in $1/L$.

\begin{figure}[t]
    \centering
    \includegraphics[bb = 0 0 229 206, width=\columnwidth,clip=true]{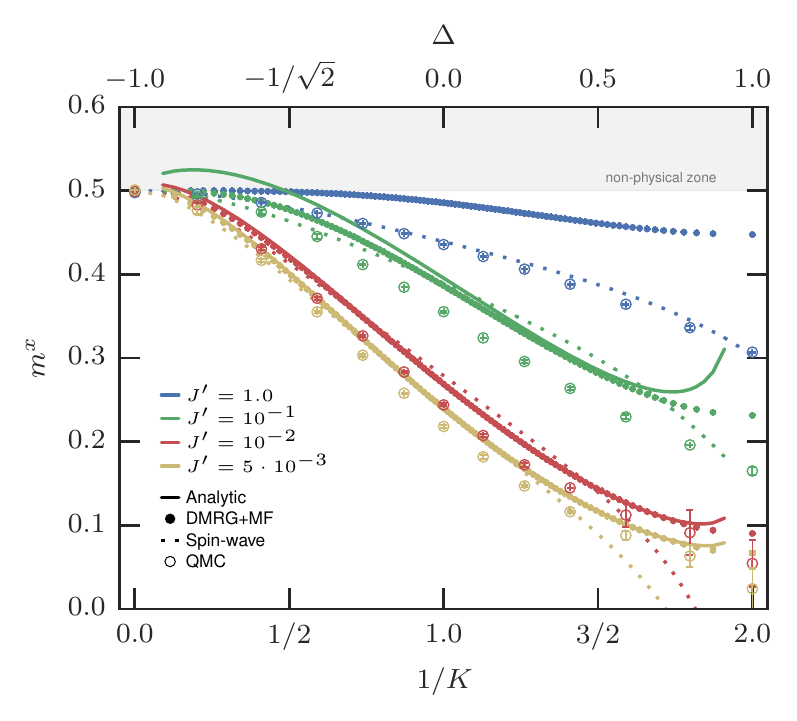}
    \caption{(Color online) Order parameter $m^x$ as a function of the TLL parameter $1/K$ plotted for various interchain couplings $J'$ for the different approaches developed in this section: Analytic (MF) Eq.~\eqref{mx_ll}, DMRG+MF, SW, and QMC (infinite size extrapolations).}
    \label{fig:qmc_final}
\end{figure}

As visible in Fig.~\ref{fig:qmc_final}, both analytical and numerical MF approaches agree better for smaller $J'$. The DMRG+MF is more controlled than the analytical approach when $J'$ is not very small, giving $m^x\le 0.5$, as it should be, in particular close to the ferromagnetic point $1/K\rightarrow 0$. Nor does it predict any divergence for $m^x$ close to $1/K\rightarrow 2$, attributed to the divergence the prefactor $A_x$ (Fig.~\ref{fig:ll_params}).
%

\subsubsection{Linear spin-wave}
As known for a long time, back to the seminal work by Anderson~\cite{anderson_approximate_1952}, linear SW theory gives excellent estimates at the $1/S$ order for the order parameter of $d\ge 2$ spin-$S$ quantum antiferromagnets, even for the most quantum case of $S=1/2$. The question of weakly coupled chains, where spatial anisotropy enhances quantum fluctuations is more delicate, as discussed in several works~\cite{Sakai_HaldanePhase_RPA_1989,azzouz_effect_1993,parola_realization_1993,affeck_plane_1994}. Using a standard treatment for computing the transverse order~\cite{coletta_semiclassical_2012}, one obtains the linear SW-corrected order parameter
\be
m^x=S
-\frac{1}{8\pi^2}\int{\rm{d}}k_x {\rm{d}}k_y\left(\frac{A(k_x,k_y)}{\Omega(k_x,k_y)}-1\right)
+\mathcal{O}(\frac{1}{S}),
\ee
where
\be
A(k_x,k_y)=\frac{1}{2}(\Delta-1)(\cos k_x+J'\cos k_y)+J'+1,
\ee
and the SW excitation spectrum
\bea
\Omega(k_x,k_y)&=&\sqrt{\Delta(\cos k_x+J'\cos k_y)+J'+1}\nonumber\\
&\times&\sqrt{J'+1-\cos k_x-J'\cos k_y}.
\eea
SW results are plotted together with MF estimates as well as with exact QMC results in Fig.~\ref{fig:qmc_final}.
In the repulsive TLL regime ($1/K>1$), the SW-corrected $m^x$ is strongly depleted for increasing anisotropy (decreasing $J'$), and deviates from MF results. On the other hand, for the attractive TLL regime ($1/K<1$), the agreement with MF is remarkable, in particular for smaller values of $J'$.
Nevertheless, we cannot expect the SW theory to be reliable for extremely small $J'$ for any $1/K>0$
because the SW expansion is not justified in the 1D limit $J' \to 0$.

\subsection{Quantum Monte Carlo study}

In order to go beyond the MF approximation and take exactly into account the 2D interchain coupling $J'$,
we use QMC through the stochastic series expansion (SSE) algorithm~\cite{syljuasen_quantum_2002,bauer_alps_2011}.
Since we are interested in ground state properties,
we need to perform QMC simulations at temperatures below the finite size gap of our finite size system, the lowest SW gap being dictated by the weak coupling $J'$.
Note also that one needs to perform a very careful finite size scaling analysis in order to reach the thermodynamic limit.

\subsubsection{Finite size effects and aspect ratio dependence}\label{sec:qmc_sizes}

\begin{figure*}[!t]
    \centering
    \includegraphics[bb = 0 0 512 184, width=\linewidth,clip=true]{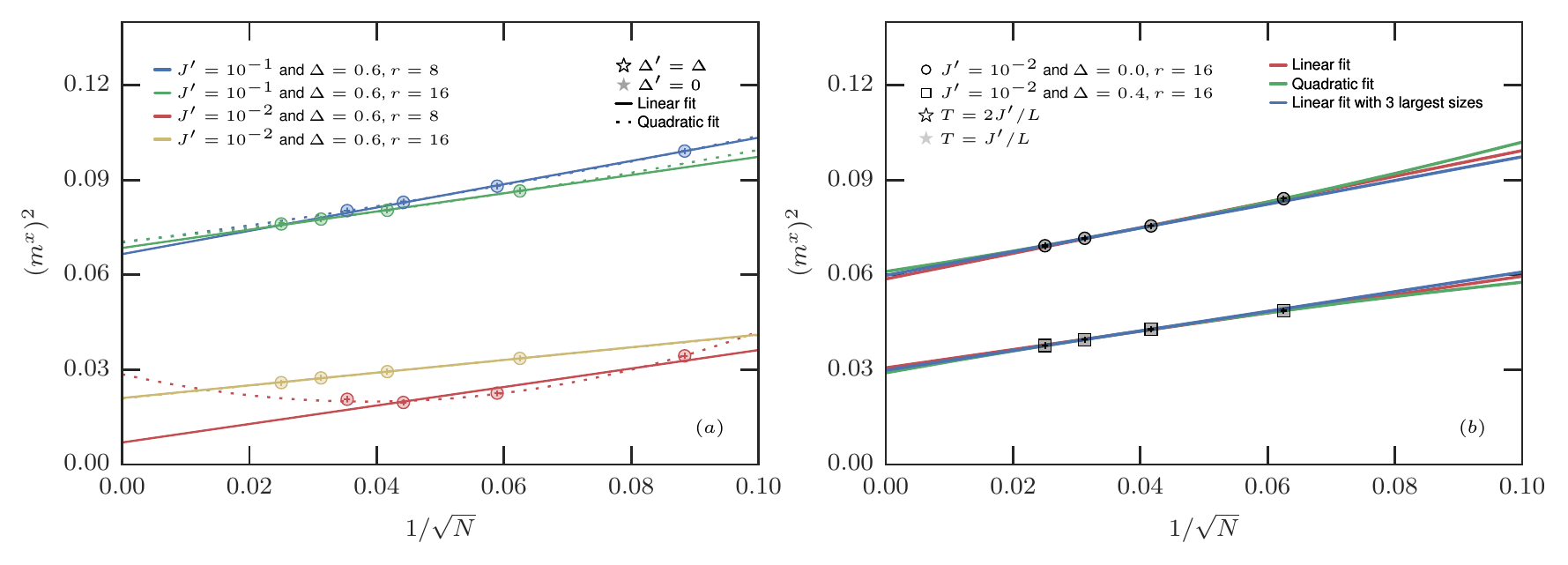}
    \caption{(Color online) The left panel (a) shows the thermodynamic limit extrapolation of the order parameter $m^x$ for $J'=10^{-1}$ and $J'=10^{-2}$ for $\Delta=0.6$ from finite samples of $N$ spins with different aspect ratios $r=8$ and $r=16$. Simulations with $J'\Delta\neq 0$ are plotted as empty symbols with an edge and $J'\Delta=0$ as filled translucent symbols without edge. The overlap of the two shows that $J'\Delta$ has no effect at all on the order parameter value.  The right panel (b) displays the thermodynamic limit extrapolation of the order parameter $m^x$ for $J'=10^{-2}$ and $\Delta=0.0$ and $\Delta=0.4$ from finite samples of $N$ spins with an aspect ratio $r=16$. The two temperatures are respectively plotted as empty symbols with an edge and filled translucent symbols without edge. An overlap of the two meaning that we are in the ground state. The QMC error bars are also shown but smaller than the symbol size. Various linear and quadratic fits are then performed, including more or less points to get the transverse magnetization value in the limit $N\rightarrow\infty$.}
    \label{fig:qmc_delta_temp}
\end{figure*}

We work on finite size systems with $N=L^2/r$ spins, where $L$ is the length of the $L/r$ coupled chains and $r$ is the aspect ratio of the 2D system.
Finite size systems with aspect ratio $r>1$ have been used before to reduce finite size effects for the anisotropic case~\cite{sandvik_multichain_1999} and surprisingly, also for isotropic case~\cite{white_neorder_2007}.
For the present study, we performed simulations for several system sizes with various aspect ratios $r=1,2,4,8,16,32$ and we found that $r=8$ (resp. $r=16$ and $r=32$) gave the best results for $J'=1$ (resp. $J'=10^{-1},10^{-2}$ and $J'=5\times 10^{-3}$), {\it{i.e.}} helps the convergence to the thermodynamic limit, although finite-size scaling analysis remains challenging for some parameters (see Fig.~\ref{fig:qmc_delta_temp} and discussion below).

We have performed QMC simulations both at $T=2J'/L$ and $T=J'/L$ in order to ensure that we are probing only the ground state,
which is verified in Fig.~\ref{fig:qmc_delta_temp}~(b).

Simulations have been performed for different $2$D couplings $J'=1,~10^{-1}$, $ 10^{-2}$ and $5\times10^{-3}$ for multiple Ising anisotropy values covering the whole range $\Delta\in[-1,1]$. The main difficulty is to extrapolate a reliable thermodynamic value of $m^x$ for small $J'$ couplings, especially close to $\Delta=1$. For each value of $\Delta$, we have performed various linear and quadratic fits of the QMC data as a function of $1/\sqrt{N}$: we show two cases in Fig.~\ref{fig:qmc_delta_temp}~(b).

As a result, the final $m^x$ value in the $N\rightarrow\infty$ limit is the mean value given by the different fits and its error bar is estimated as the standard deviation around this mean value. This leads to small error bars for the sets of parameters where all fits agree well. Note that these error bars do not reflect the QMC errors, which are much smaller, but rather gives an idea on the uncertainty due to  the infinite size extrapolation procedure.

Figure~\ref{fig:qmc_delta_temp} also that the interchain $S^z S^z$ interaction (with amplitude $J' \Delta$) has  no effect on the order parameter (within error bars), which confirms the assumption made at the MF level.

\subsubsection{Order parameter {\it{vs.}} TLL parameter}

We present in this subsection comparison of the QMC results with the various approaches described before, including analytical and numerical MF and SW expansion.
We compared them in Fig.~\ref{fig:qmc_final} for three values of $J'$.
We notice that the order parameter value given by QMC is always smaller than that by DMRG+MF,
which is expected as the MF approximation overestimates the order.

\begin{figure}[h!]
    \centering
    \includegraphics[bb = 0 0 229 206, width=\columnwidth,clip=true]{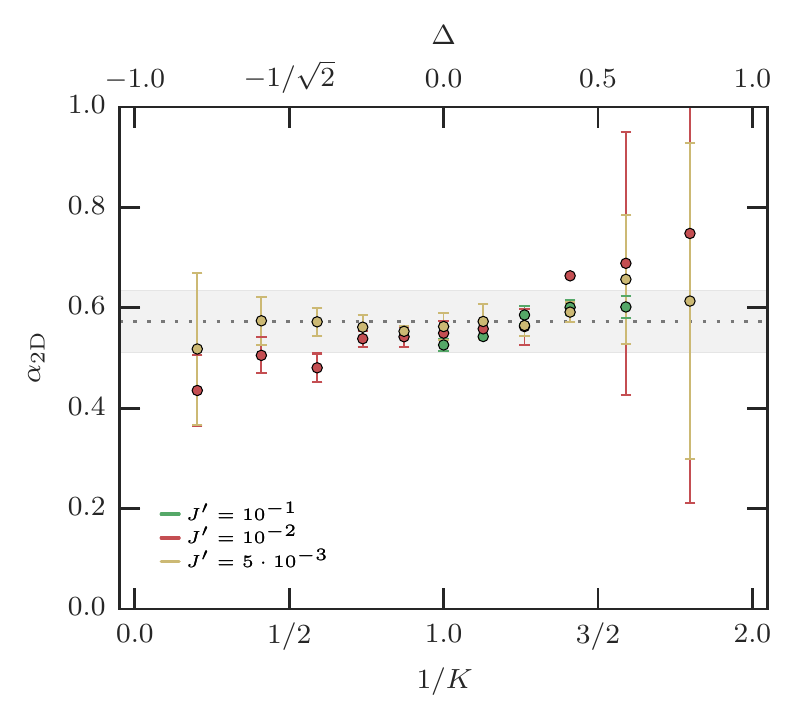}
    \caption{(Color online) Renormalization coefficient $\alpha_{\rm 2D}$ of the MF coupling $J'$ (see text). The only displayed points are those such that the absolute difference in $m^x$ between analytical MF and DMRG+MF is below $10^{-2}$. The displayed error bars are related to the extrapolated value of $m^x$ within QMC. Assuming a constant value, $\alpha_{2\mathrm{D}}=0.57~\pm~0.06$.}
    \label{fig:qmc_alpha}
\end{figure}

The MF approximation discarded the fluctuation, which causes the quantitative disagreement with the QMC data.
Now we ask ourselves whether an effective rescaling of $J'$ to $\alpha_{2\mathrm{D}}J'$ reconciles the MF and QMC results quantitatively so that
$m^x_{\mathrm{MF}}(\alpha_{2\mathrm{D}} J')=m^x_{\mathrm{QMC}}(J')$.
The renormalization factor $\alpha_{2\mathrm{D}}$ can be determined using the analytical expression \eqref{mx_ll}, although one needs to be careful because this analytical expression does not give exact MF results (compared to the numerical MF using DMRG considered as exact for reasons discussed above). This is why we limit ourselves to the sets of parameters $J'$ and $\Delta$ where analytical MF and DMRG+MF agree well. This renormalization factor $\alpha_{\rm 2D}$ of the MF coupling is shown against $1/K$ in Fig.~\ref{fig:qmc_alpha}~\footnote{We point out that the renormalization factor $\alpha_{2\mathrm{D}}$ is not defined at the ferromagnetic point $1/K=0$ where \textit{any value} of $J'$ leads to $m^x=0.5$ within both MF and QMC approaches.}, where it roughly remains constant over the full range of TLL parameter, with $\alpha_{2\mathrm{D}}=0.57~\pm~0.06$.

\section{Small spin anisotropy}\label{sec:Z2}

Thus far we assumed that there is no anisotropy in the spin space in contrast to the real space.
However the spin anisotropy exists in real materials to a greater or lesser extent.
In fact DIMPY has a weak uniform Dzyaloshinskii-Moriya interaction~\cite{Glazkov_DIMPY_2015, Ozerov_DIMPY_2015}.
In this section we see briefly how such a small spin anisotropy affects the physics discussed above.

\subsection{Longitudinal spin anisotropy}

The longitudinal spin anisotropy is easier to handle.
It  modulates the coupling of the $S^zS^z$ exchange interaction and changes only the TLL parameter $K$.
The small modulation of $K$ by the small anisotropy keeps all the qualitative results in this paper unchanged.
The gapless Nambu-Goldstone mode in the ordered phases is also unaffected by the spin anisotropy of the interladder interaction.
In fact the $S^zS^z$ interladder interaction is absent in the action \eqref{S}.

\subsection{Transverse spin anisotropy}

Compared to the longitudinal spin anisotropy, the transverse one can have more serious impact on the low-temperature physics of the ordered phase.

Let us consider a small transverse spin anisotropy $\Jl(\Delta_x-1)S^x_{j,l,\mu,\nu}S^x_{j+1,l,\mu,\nu}$ along the legs.
The intraladder transverse spin anisotropy  yields an additional term $\cos(2\theta_{\mu,\nu})$ to the single-spin-ladder Hamiltonian \eqref{H_ladder}.
Detailed analysis is given in the Appendix~\ref{app:XZZ}.
Although the cosine interaction gives rise to a finite excitation gap,
the gap is exponentially small and masked by the interladder interactions.
The small intraladder transverse spin anisotropy hardly affects the phase diagram of Fig.~\ref{PD_ladders}.

The transverse spin anisotropy lowers the symmetry of the system \eqref{H} in the spin space from U(1)  to $\mathbb Z_2$.
In other words, the transverse spin anisotropy chooses the special direction along which the N\'eel order grows.
To see this, we focus on the strongly 2D case and employ the classical approximation.
Including the intraladder transverse anisotropy, the action \eqref{H_E} becomes
\begin{align}
 \mathcal S_{\rm cl}
 &=\sum_{\nu} \frac{K'}{2\pi v}\int dxd\tau' \bigl\{v^2(\partial_x\theta_\nu)^2
 +(\partial_{\tau'}\theta_\nu)^2 \bigr\} \notag \\
 & \quad
 +\lambda'_x(T)\sum_{\nu} \int dx d\tau'\cos(2\theta_\nu)
 \notag \\
 & \quad
 - \frac{vg'_2(T)}{\pi} \sum_{\mu,\nu} \int  dxd\tau' \,
 \cos(\theta_{\mu,\nu}-\theta_{\mu,\nu+1}),
 \label{H_E_XZZ}
 \end{align}
with $\lambda'_x(T)=K'\lambda_x(T)/K = \lambda_x^0(T/\Jl)^{-2}$ and $\lambda_x^0\propto \Jl(\Delta_x-1)$.
To deal with the nonlinear action \eqref{H_E_XZZ}, we adopt the MF approximation
and the classical approximation as well as Eq.~\eqref{S_cl_MF}.
The resultant action is as follows.
\begin{align}
 \mathcal S_{\rm cl}
 &=\sum_{\nu} \frac{K'}{2\pi v}\int dxd\tau' \bigl\{v^2(\partial_x\theta_\nu)^2
 +(\partial_{\tau'}\theta_\nu)^2 \bigr\} \notag \\
 & -2h'_s(T)\sum_{\nu}\int dxd\tau' \, \cos \theta_\nu,
\end{align}
where the mean field $h'_s(T)$ is given by
\begin{align}
 h'_s(T) &\equiv h_s(T) + \frac{2K'\lambda_x(T)m(T)}K
 \notag \\
 &=  \frac{vK'm(T)}{\pi K} \biggl(\e J'_2 + \frac{\pi \lambda_x^0}v\biggr)\biggl(\frac T{\Jl}\biggr)^{-2}.
 \label{h'_s}
\end{align}
The transverse anisotropy shifted the factor $\e J'_2$ in the effective field \eqref{h_s}
by $\pi\lambda_x^0/v$ [Eq.~\eqref{h'_s}].
It immediately follows that the order parameter $m(T)$ has a nonzero value at finite temperatures even when $\e J'_2=0$.
This is the most important consequence of the transverse anisotropy.
In the absence of the transverse anisotropy, the 2D system cannot have the nonzero N\'eel order breaking the continuous rotational symmetry at finite temperatures.
Since the transverse anisotropy breaks the U(1) symmetry to the $\mathbb Z_2$ one,
nothing prevents the 2D system from having a nonzero order parameter.

The interladder transverse anisotropy merely generates additional interactions such as $\cos[2(\theta_{\mu,\nu}-\theta_{\mu+1,\nu})]$.
Thus the above discussion is directly applicable to the interladder transverse anisotropy
and that the same conclusion is derived.

\section{Summary and discussions}\label{sec:summary}

In this paper, we discussed the dimensional modulation of magnetic ordering process in spatially anisotropic quantum antiferromagnets.
Taking advantage of the small interladder and interlayer interactions,
we performed several complementary analyses: the RPA analysis (Sec.~\ref{sec:rpa}),
the variational method (Sec.~\ref{sec:variation}), the RG method (Sec.~\ref{sec:rg}) and the classical approximation analysis (Sec.~\ref{sec:classical_approx}).
All those analyses led to the dimensional reduction scenario sketched in Fig.~\ref{3Dto2D}~(b), rather than the naive expectation drawn in Fig.~\ref{3Dto2D}~(a).

The key observation here is the convergence of the critical temperature $T_c$ to the KT transition temperature $T_{\rm KT}$ in the 2D limit $\e J'_2/J'_1 \to 0$.
Thanks to this fact, the quasi-2D ordered phase emerge in the range $T_{\rm cr}<T<T_c$,
where $T_{\rm cr}$ represents the crossover temperature to the 3D phase at $T<T_{\rm cr}$
(Figs.~\ref{quasi2D_def}, \ref{stag_whole} and \ref{mx_classical}).
Since $T_c \to T_{\rm KT}$ and $T_{\rm cr} \to 0$ in the 2D limit, the quasi-2D ordered phase is smoothly connected to the KT phase of the 2D superfluid.
As we saw in the variational approach (Fig.~\ref{stag_whole}),
the N\'eel order $m(T)$ perpendicular to the magnetic field is strongly suppressed near the critical temperature when $\e J'_2/J'_1 \ll 1$.
We note that the same $m(T)$ curve is also derived with the aid of the classical approximation (Fig.~\ref{mx_classical}).
This agreement shows that our system in the quasi-2D ordered phase is well approximated by the classical system, for which we have provided a clear numerical confirmation based on quantum Monte Carlo simulations of an equivalent quantum system, compared to MF and SW approximations.
This characteristic of the quasi-2D ordered phase is inherited from the KT phase in the 2D limit.

We showed that our system (Fig.~\ref{lattice}) in the quasi-2D ordered phase can be seen as weakly coupled 2D critical systems.
The weak interplane interaction $\e J'_2$ originates from the imperfect geometrical frustration.
Geometrical frustration is a rich source of various unconventional quantum phases.
For example, the $S=1/2$ quantum Heisenberg antiferromagnet on the spatially anisotropic triangular lattice has an interesting quantum phase
with a characteristic triplon excitation going with the incommensurate N\'eel order along the direction of the magnetic field~\cite{Kohno_Cs2CuCl4_2007, Kohno_Cs2CuCl4_2009}.
One can find commonalities in our system and the spatially anisotropic triangular system.
However, the incommensurate order is not developed in our system
because the interplane geometrical frustration \eqref{J_def} is independent of the wavenumber $q_x$ along the leg.
In the theories of Refs.~\onlinecite{Kohno_Cs2CuCl4_2007, Kohno_Cs2CuCl4_2009},
the $q_x$ dependence of $\mathcal J(\bm q)$ is crucial.
Therefore it would be interesting to extend our theory in order to discuss the possibility of unconventional quantum phases that result from the interplane frustration
involved with the wavenumber of the order in the leg direction.
We leave it as an open problem.
We are convinced that our theory presented in this paper will be useful to investigate such quantum phases of matter.

\section*{Acknowledgment}

The authors are grateful to C. Berthier, M. Horvati\'c, M. Jeong and  M. Oshikawa for the valuable discussion.
This work was performed using HPC resources from GENCI (Grant Nos. x2015050225 and No. x2016050225), and is supported by
Swiss SNF under Division II, the French ANR program BOLODISS, R\'egion Midi-Pyr\'en\'ees, and JSPS KAKENHI Grant No. 16J04731.

\appendix

\section{Attractive field-induced TLL}\label{app:attraction}

\subsection{The TLL parameter}

Here we develop a low-energy effective theory of the attractive field-induced TLL of the strong-leg spin ladder.
The discussion in this appendix is useful to assure the inequality $K>1$ that  the TLL parameter $K$ of the field-induced TLL in our model \eqref{H} satisfies.
In addition it makes the article self-contained.
The low-energy theory is also beneficial to clarifying effects of the transverse spin anisotropy (Sec.~\ref{sec:Z2} and Appendix~\ref{app:XZZ}).

Since we are interested in the strong-leg spin ladder with $\Jr/\Jl<1$,
it is logical to start with the $\Jr=0$ case and then to include the rung interaction perturbatively.
The ratio  $\Jr/\Jl\simeq 0.58$ for DIMPY is not so small that the perturbation theory is naively justified.
On the other hand we have confirmed in Ref.~\onlinecite{Ozerov_DIMPY_2015} that
the perturbative approach succeeded in understanding the unconventional electron spin resonance of DIMPY.
We firmly believe that the perturbative approach describes physics of DIMPY at least qualitatively.

When $\Jr=0$, the spin ladder is composed of two independent chains.
On each leg the TLL fields $\phi_n$ and $\theta_n$ for $n=1,2$ are defined.
They satisfy the commutation relation
\begin{equation}
 [\phi_n(x), \partial_y\theta_n(y)]=i\pi \delta(x-y).
  \label{commutation}
\end{equation}
These fields are related to the spin on the $n$th leg as
\begin{align}
 S^z_{j,n} &= \frac 1\pi\partial_x\phi_n + (-1)^{j+n} \sqrt{A_z}\cos(2\phi_n),
 \label{Sz2phi} \\
 S^\pm_{j,n} &= e^{\mp i\theta_n} \bigl[ (-1)^{j+n}\sqrt{A_x} + \sqrt{B_x}\cos(2\phi_n)\bigr],
 \label{Spm2phi}
\end{align}
where $S^\pm_{j,n}=S^x_{j,n}\pm i S^y_{j,n}$ and $A_z$, $A_x$, and $B_z$ are constants.
We have omitted the greek indices to specify the position of the ladder.
The Hamiltonian is written as
\begin{align}
 \mathcal H_{\mu,\nu}
 &= \sum_{n=1,2}\frac{u_0}{2\pi}\int dx \, \biggl[ K_0(\partial_x\theta_n)^2 + \frac 1{K_0}(\partial_x\phi_n)^2 \biggr]
 \notag \\
 & \quad -\frac h{\pi} \sum_{n=1,2}\int dx \, \partial_x\phi_n,
 \label{H_chain}
\end{align}
with $u_0\propto \Jl$ and $K_0=(1+4/\pi)^{-1/2}$.
This value of $K_0$ is the bare value and the RG effect is not included.

At $h=0$, the SU(2) symmetry of Eq.~\eqref{H_chain} yields $K_0=1/2$.
The magnetic field affects the TLL parameter by terminating the RG flow at a cutoff specified by it.
Note that the magnetic field increases the TLL parameter $K_0$~\cite{Affleck_CuBenzoate_1999}.
At the leading order of $h/\Jl\ll 1$, the TLL parameter is given by
\begin{equation}
 K_0 = \frac 12 \biggl( 1- \frac 1{2\ln(h/\Jl)} + \cdots\biggr).
  \label{K_0_RG}
\end{equation}
One can obtain the exact value of $K_0$ as a function of $h$~\cite{Korepin_book}.

The TLL parameter controls the behavior of various physical quantities, for example, the susceptibility \eqref{1Dsus_TLL}.
However the TLL parameter has a subtle problem in its definition.
Rescalings of $\phi_n$ and $\theta_n$ actually change  the TLL parameter.
Let us consider a rescaling $\phi_n \to \alpha \phi_n$ and $\theta_n \to \alpha^{-1}\theta_n$ with a constant $\alpha$
so that the commutation relation \eqref{commutation} is kept intact.
Then the TLL parameter is subject to the rescaling $K_0 \to \alpha^{-2}K_0$.
In other words, the TLL parameter is uniquely determined only after specifying the parameter $\alpha$.

\subsection{Compactification relations}

 The compactification relations of $\phi_n$ and $\theta_n$ specify the parameter $\alpha$ uniquely.
Since $\phi_n$ and $\theta_n$ represent the phase degrees of freedom of the spin operator,
they are periodic functions.
In a standard notation of the bosonization~\cite{Giamarchi_book}, those periods are fixed with
the identification relations
\begin{align}
 \phi_n &\sim \phi_n + \pi N_nR,
 \label{compact_phi_n} \\
 \theta_n &\sim \theta_n + \frac{2\pi M_n}R,
 \label{compact_theta_n}
\end{align}
with $R=1$ and $N_n,M_n\in\mathbb Z$.
Here $\sim$ expresses the identification relation.
The parameter $R$ is called the compactification radius.

There are two ways to fix the parameter $\alpha$.
One is to have $R=1$~\cite{Chitra_ladder_1997, Bouillot_BPCB_2011}.
We employ this notation in this article.
The other is to have $K_0=1$~\cite{Hikihara_XXZ_2004, Furuya_boundary_2012}.
There is also an intermediate notation~\cite{Oshikawa_BCFT_2010}.
In what follows we emphasize that  fixing  $\alpha$ is important in defining the field-induced TLL in weakly coupled TLLs.

Let us now take the rung interaction $\Jr$ into account.
The Hamiltonian of the single spin ladder is composed of two parts:
\begin{equation}
 \mathcal H_{\mu,\nu}= \mathcal H_+ + \mathcal H_-,
\end{equation}
with
\begin{align}
 \mathcal H_+
 &= \frac{u_+}{2\pi}\int dx \, \biggl[ K_+ (\partial_x\theta_+)^2 + \frac 1{K_+}(\partial_x\phi_+)^2\biggr]
 \notag \\
 & \quad - \frac{\sqrt 2h}{\pi}\int dx \, \partial_x\phi_+
 +g_3 \int dx \, \cos(\sqrt 8\phi_+)
 \label{H_+}
\end{align}
and
\begin{align}
 \mathcal H_-
 &= \frac{u_-}{2\pi} \int dx \, \biggl[ K_-(\partial_x\theta_-)^2 + \frac 1{K_-}(\partial_x\phi_-)^2 \biggr]
 \notag \\
 & \quad
 +\int dx \, \bigl[ g_1\cos(\sqrt 2\theta_-) + g_2\cos(\sqrt 8\phi_-)\bigr],
 \label{H_-}
\end{align}
Here $\phi_\pm$ and $\theta_\pm$ are defined as
\begin{equation}
 \phi_\pm=\frac{\phi_1\pm\phi_2}{\sqrt 2}, \quad
  \theta_\pm = \frac{\theta_1\pm\theta_2}{\sqrt 2}
\end{equation}
and the couplings are given by $g_1, g_2, g_3\propto \Jr$, $u_\pm = u_0/K_\pm$ and
\begin{equation}
 K_\pm = K_0\biggl(1 \pm \frac{\Jr}{\pi \Jl}\biggr)^{-\frac 12}.
  \label{K_pm}
\end{equation}
Note that $\mathcal H_+$ and $\mathcal H_-$ are symmetric and antisymmetric with respect to the permutation of legs, respectively.
As well as $\phi_n$ and $\theta_n$, those symmetric and antisymmetric fields are compactified,
\begin{align}
 \phi_\pm \sim \phi_\pm + \pi N_\pm \frac 1{\sqrt 2}, \\
 \theta_\pm \sim \theta_\pm + 2\pi M_\pm \frac 1{\sqrt 2},
\end{align}
where $N_\pm=N_1\pm N_2$ and $M_\pm=M_1\pm M_2$.
Those integral parameters are subject to
\begin{align}
 N_+ &\equiv N_- \mod 2,
 \label{N_mod2} \\
 M_+ &\equiv M_- \mod 2.
 \label{M_mod2}
\end{align}
Importance of the relations \eqref{N_mod2} and \eqref{M_mod2}  is explained in depth in Ref.~\onlinecite{Oshikawa_BCFT_2010}.

The cosine interactions in Eq.~\eqref{H_+} and \eqref{H_-} can give rise to finite spin gaps both in the symmetric and antisymmetric sectors.
While the antisymmetric sector $\mathcal H_-$ is gapped for any magnetic field $h$,
the symmetric sector $\mathcal H_+$ is not because the magnetic field is even under the permutation of the legs.
Our aim here is to derive the low-energy effective field theory around the quantum critical point $h=h_{c1}$,
which is achieved by integrating out the gapped  antisymmetric sector $\mathcal H_-$.

The TLL parameters \eqref{K_pm} of those sectors represent bare values whose RG effects are not taken into account.
It is known that the renormalized TLL parameters  $K_\pm$ are increased in association with increase of the magnetic field $h$~\cite{Affleck_CuBenzoate_1999}.
Especially,  increase of $K_-$ makes the cosine $\cos(\sqrt 2\theta_-)$ more relevant than $\cos(\sqrt 8\phi_-)$.
Under the strong magnetic field, the excitation gap of  $\mathcal H_-$ is attributed mainly to $\cos(\sqrt 2\theta_-)$.
Let us denote the excitation gap of $\mathcal H_-$ as $\Delta_-$.
Given a temperature $T\ll \Delta_-$, the cosine potential $\cos(\sqrt 2\theta_-)$ strongly locks $\theta_-$  to one of its minima.
The strong locking allows us to set $M_-=0$, which affects the compactification of $\theta_+$ through Eq.~\eqref{M_mod2}.
When $M_-=0$, the integer $M_+$ must be an even number $2M'_+$  ($M'_+\in\mathbb Z$).
The compactification of $\phi_+$ and $\theta_+$ are replaced to
\begin{align}
 \phi_+ &\sim \phi_+ +\pi N_+ \frac 1{\sqrt 2},
 \label{compact_phi_+} \\
 \theta_- &\sim \theta_- + 2\pi M'_+ \sqrt{2}.
 \label{compact_theta_+}
\end{align}

The compactification relations \eqref{compact_phi_+} and \eqref{compact_theta_+} correspond to those [Eq.~\eqref{compact_phi_n} and \eqref{compact_theta_n}]
of $\phi_n$ and $\theta_n$ with $R=1/\sqrt 2$.
Therefore we need to rescale $\phi_+ \to\phi'_+\equiv  \sqrt 2\phi_+$ and $\theta_+ \to\theta'_+\equiv \theta_+/\sqrt 2$.
Integrating out the antisymmetric sector and rescaling the fields, we obtain the low-energy effective Hamiltonian,
\begin{align}
 \mathcal H_{\mu,\nu}
 &\simeq \frac{u_+}{2\pi} \int dx \biggl[ 2K_+(\partial_x\theta'_+)^2 + \frac 1{2K_+}(\partial_x\phi'_+)^2\biggr]
 \notag \\
 & \quad - \frac h\pi \int dx \, \partial_x\phi'_+ + g_3\int dx \, \cos(2\phi'_+).
 \label{H_C-IC}
\end{align}

The locking of $\theta_-$ affects the correspondence between the boson fields and the spin operators.
For example, the locking of $\theta_-$ leads to
\begin{equation}
 (-1)^{j+n}S^\pm_{j,n}=\sqrt{A_x}e^{\mp i\theta_+/\sqrt 2} = \sqrt{A_x}e^{\mp i \theta'_+}.
\end{equation}
If one uses $\phi_+$ and $\theta_+$, one needs to be careful about the difference in the correspondence between the spin and the boson operators.

\subsection{Commensurate-incommensurate transition}

Except for the factor $2$ in front of $K_+$, the model \eqref{H_C-IC} is nothing but the well-known sine-Gordon model in the presence of a chemical potential $h$~\cite{Giamarchi_book}.
The system \eqref{H_C-IC} is known to undergo a quantum phase transition from a commensurate gapped phase to an incommensurate gapless
one~\cite{Schulz_incommensurate_1980}.
At zero field $h=0$ the sine-Gordon model \eqref{H_C-IC} has an excitation gap $\Delta$.
The gapped phase extends for $h<\Delta$.
When $h>\Delta$, the system \eqref{H_C-IC} enters into a gapless phase, where the TLL excitation emerges around the new Fermi level specified by the chemical potential.
To obtain the Hamiltonian of the field-induced TLL, we need to linearize again the dispersion relation around the new Fermi level with the Fermi wavenumbers $\pm k_c$ ($k_c>0$).
A method for the relinearization is reviewed in depth in Ref.~\onlinecite{Giamarchi_book}
for the field theory \eqref{H_C-IC}.
Here we show its result only.
The Hamiltonian of the  field-induced TLL for $h>\Delta$ is given by
\begin{align}
 \mathcal H_{\mu,\nu}
 &= \frac v{2\pi}\int dx \, \biggl[ K(\partial_x\theta)^2 + \frac 1K(\partial_x\phi)^2\biggr]
 \notag \\
 &= \frac K{2\pi v}\int dx \, \bigl\{ (\partial_\tau \theta)^2 +v^2 (\partial_x\theta)^2\bigr\}.
 \label{H_FITLL}
\end{align}
where the fields $\phi$ and $\theta$ represent the field-induced TLL and equal to $\phi_{\mu,\nu}$ and $\theta_{\mu,\nu}$ in Eq.~\eqref{S}.
The TLL parameter $K$ of the field-induced TLL is given by
\begin{align}
 K \simeq 1 - \frac{u_+ k_c}{\Delta} \sinh (2\Theta),
 \label{K_FITLL}
\end{align}
with $k_c=\sqrt{(h^2-\Delta^2)/v^2}$ and
\begin{equation}
 e^{-2\Theta} = 2K_+.
  \label{Theta}
\end{equation}
The strong-leg spin ladder with $\Jr/\Jl\ll 1$ leads to
\begin{equation}
 \Theta<0
  \label{negative_Theta}
\end{equation}
as follows.
The small rung interaction has little impact on $K_\pm$ \eqref{K_pm}, which allows us to approximate $K_+\simeq K_0$.
From the fact $K_0> 1/2$ in the presence of the magnetic field~\cite{Affleck_CuBenzoate_1999},
Eq.~\eqref{negative_Theta} follows.
The negative $\Theta$ means that the TLL parameter \eqref{K_FITLL} of the field-induced TLL satisfies $K>1$ near the critical point $h=h_{c1}$.

\subsection{Attraction by back scattering}

We point out that the inequality \eqref{negative_Theta}, which is crucial to make the field-induced TLL attractive, is achieved only after
rescaling the compactification radius of Eqs.~\eqref{compact_phi_+} and \eqref{compact_theta_+}.
Since the rescaling of the compactification radius comes from the locking of the cosine interaction $\cos(\sqrt 2\theta_-)$,
the attraction originates from the back scattering term of $\theta_1$ and $\theta_2$.
This mechanism of the attraction by the back scattering was not pointed out before.

It is straightforward to generalize the above discussion to strong-leg spin ladders with $N$ legs.
Such an extension is beneficial to understanding systems of weakly coupled TLLs~\cite{Furuya_ESRwidth_2015}.
In the field-induced TLL phase of the $N$-leg ladder,
we have the TLL parameter \eqref{K_FITLL} with
\begin{equation}
 e^{-2\Theta} = NK_0,
\end{equation}
instead of Eq.~\eqref{Theta}.
$K_0$ is the TLL parameter of the ``center-of-mass'' field $\Phi_N=(\phi_1+\phi_2+\cdots +\phi_N)/\sqrt N$,
where $\phi_n$ ($n=1,2,\cdots, N$) represents the boson field on the $n$th leg.

\section{The variational free energy \eqref{F_var}}\label{app:F_var}

This Appendix is devoted to derivation of the variational free energy \eqref{F_var}.
The quadratic action \eqref{Gv} relates the partition function to
the Gaussian integral,
\begin{equation}
 \int \mathcal D \phi_{\mu,\nu}(x)e^{-\mathcal S_{\rm v}}
  =\mathrm{const.}\times \prod_{\omega_n, \bm k} G_{\rm v}^{\frac
  12}(i\omega_n, \bm k).
\end{equation}
It immediately follows that
\begin{equation}
 F_{\rm v} =\mathrm{const.} - \frac T2 \sum_{\omega_n, \bm k} \ln
  G_{\rm v} (i\omega_n, \bm k).
\end{equation}
The other term $\langle (\mathcal S-\mathcal S_{\rm v})\rangle_{\rm v}$ in
$F_{\rm var}$ [Eq.~\eqref{F_var_def}] is calculated as follows.
First $\langle \mathcal S_{\rm v}\rangle_{\rm v}$ is negligible because it is
independent of $G_{\rm v}$.
Second $\langle \mathcal S\rangle_{\rm v}$ is split into two terms: the
average of the kinetic term and the average of the cosine terms.
One needs the cumulant expansion of the Gaussian distribution
$\langle e^{s}\rangle_{\rm v}=\exp(\langle s^2\rangle_{\rm v}/2)$
to derive the average of the cosine term.
\begin{align}
 &\langle e^{i(\theta_{\mu,\nu}-\theta_{\mu+1,\nu})}\rangle_{\rm v}
 \notag \\
 &= \exp\biggl[-\frac 12 \langle
 (\theta_{\mu,\nu}-\theta_{\mu+1,\nu})^2\rangle_{\rm v}\biggr] \notag \\
 &= \exp\biggl[-\frac T{2\Omega}\sum_{\omega_n,\bm k} F(k_y)G_{\rm v}
 (i\omega_n, \bm k)\biggr].
 \label{cumulant}
\end{align}
 In the last line, we used the relation
\begin{align}
 \langle \theta_{\mu,\nu}(\tau,x)^2\rangle_{\rm v}=G_{\rm v}
 (\tau=0,\bm r=0),
\end{align}
where
$G_{\rm v}(\tau, \bm r)$ is expressed as  the Fourier transform of $G_{\rm v}(i\omega_n,\bm k)$:
\begin{equation}
 G_{\rm v} (\tau,\bm r)
 = \frac T{\Omega}\sum_{\omega_n,\bm k}
  e^{-i(\omega_n\tau-(\bm k+\bm \pi) \cdot \bm r)}
  G_{\rm v}(i\omega_n,\bm k).
\end{equation}
The real part of Eq.~\eqref{cumulant} leads to
\begin{align}
 &\int d\tau dx \sum_{\mu,\nu} \langle
  \cos(\theta_{\mu,\nu}-\theta_{\mu+1,\nu}) \rangle_{\rm v}
 \notag \\
 &= \frac{\Omega}T \exp\biggl[-\frac T{2\Omega}\sum_{\omega_n,\bm k} F(k_y)G_{\rm v}  (i\omega_n, \bm k)\biggr].
\end{align}
Combining these results, we obtain the variational free energy
\eqref{F_var}.

\section{Intraladder transverse spin anisotropy}\label{app:XZZ}

Here we discuss effects of a small transverse spin anisotropy.
Let us introduce an additional interaction $\Jl(\Delta_x-1) S^x_{j,l,\mu,\nu}S^x_{j+1,l,\mu,\nu}$ to the spin ladder Hamiltonian \eqref{H_ladder}.
As we saw in Appendix~\ref{app:attraction},  the rung interaction opens the spin gap.
The transverse spin anisotropy can open the spin gap even for $\Jr=0$.
The spin ladder Hamiltonian  for $\Jr=0$ is given by
\begin{align}
 \mathcal H_{\mu,\nu}
 &= \sum_{n=1,2}\frac{u_0}{2\pi} \int dx \, \biggl[ K_0(\partial_x\theta_n)^2 + \frac 1{K_0}(\partial_x\phi_n)^2\biggr]
 \notag \\
 & \quad - \sum_{n=1,2}\frac h{\pi}\int dx \, \partial_x\phi_n +\sum_{n=1,2}\lambda_x\int dx \, \cos(2\theta_n),
 \label{H_ladder_XZZ_h=0}
\end{align}
with $\lambda_x\propto \Jl(\Delta_x-1)$.
The cosine interaction $\cos(2\theta_n)$ has the scaling dimension $1/K_0$,
which means that it is marginal at $h=0$ since $K_0=1/2$.
The marginal interaction can generate an excitation gap depending on the sign of the coupling.
For $\lambda_x>0$ the cosine is marginally relevant and yields an exponentially small excitation gap.
On the other hand, for $\lambda_x<0$, the cosine is marginally irrelevant and keeps the spin ladder \eqref{H_ladder_XZZ_h=0} gapless.
Since the TLL parameter $K_0$ increases with $h$ [Eq.~\eqref{K_0_RG}], the excitation gap grows with $h$~\cite{Hikihara_XXZ_2004}.

Let us add the rung interaction to Eq.~\eqref{H_ladder_XZZ_h=0}.
The rung interaction generates three cosine interactions in Eqs.~\eqref{H_+} and \eqref{H_-}.
Having the scaling dimension $2K_0\simeq 1$ for $h\ll h_{c1}$,
all  those cosine interactions are relevant enough in the RG sense to generate a larger excitation gap
than the one generated by the transverse anisotropy $\lambda_x\sum_{n=1,2}\cos(2\theta_n)$.
Thus the excitation gap of the spin ladder for $h\ll h_{c1}$ is mostly governed by the rung interaction
and the low-energy theory in Appendix~\ref{app:attraction} works with a slight modification of parameters only.

In contrast the low-energy theory is seriously affected by the transverse anisotropy for $h\simeq h_{c1}$.
According to Eq.~\eqref{H_C-IC}, the rung interaction $\cos(2\phi'_+)$ and the Zeeman energy $h\partial_x\phi'_+/\pi$ compete with each other.
As a result of the competition, the excitation gap vanishes at $h=h_{c1}$.
The transverse anisotropy $\lambda_x\sum_{n=1,2}\cos(2\theta_n)$ is unaffected by the magnetic field
except for the renormalization effect of its scaling dimension.
Therefore,  under the magnetic field $h\simeq h_{c1}$, the spin gap is mostly dominated by the transverse anisotropy.
This concludes that the spin ladder is well approximated as two independent spin chains each of which has the transverse anisotropy.
According to Ref.~\onlinecite{Hikihara_XXZ_2004}, the anisotropy $\Delta_x=0.95$ only gives rise to a tiny excitation gap smaller than $4\times 10^{-4}\Jl$ at maximum.
In the coupled spin ladder system, the tiny excitation gap will be invisible because of the interladder interactions.
For example,  DIMPY has the interladder interaction $2J'_1+\e J'_2 \simeq 4.4 \times 10^{-3}\Jl$~\cite{Schmidiger_DIMPY_2012},
which is large enough to mask the transverse anisotropy even if it exists.


\begin{thebibliography}{62}%
\makeatletter
\providecommand \@ifxundefined [1]{%
 \@ifx{#1\undefined}
}%
\providecommand \@ifnum [1]{%
 \ifnum #1\expandafter \@firstoftwo
 \else \expandafter \@secondoftwo
 \fi
}%
\providecommand \@ifx [1]{%
 \ifx #1\expandafter \@firstoftwo
 \else \expandafter \@secondoftwo
 \fi
}%
\providecommand \natexlab [1]{#1}%
\providecommand \enquote  [1]{``#1''}%
\providecommand \bibnamefont  [1]{#1}%
\providecommand \bibfnamefont [1]{#1}%
\providecommand \citenamefont [1]{#1}%
\providecommand \href@noop [0]{\@secondoftwo}%
\providecommand \href [0]{\begingroup \@sanitize@url \@href}%
\providecommand \@href[1]{\@@startlink{#1}\@@href}%
\providecommand \@@href[1]{\endgroup#1\@@endlink}%
\providecommand \@sanitize@url [0]{\catcode `\\12\catcode `\$12\catcode
  `\&12\catcode `\#12\catcode `\^12\catcode `\_12\catcode `\%12\relax}%
\providecommand \@@startlink[1]{}%
\providecommand \@@endlink[0]{}%
\providecommand \url  [0]{\begingroup\@sanitize@url \@url }%
\providecommand \@url [1]{\endgroup\@href {#1}{\urlprefix }}%
\providecommand \urlprefix  [0]{URL }%
\providecommand \Eprint [0]{\href }%
\providecommand \doibase [0]{http://dx.doi.org/}%
\providecommand \selectlanguage [0]{\@gobble}%
\providecommand \bibinfo  [0]{\@secondoftwo}%
\providecommand \bibfield  [0]{\@secondoftwo}%
\providecommand \translation [1]{[#1]}%
\providecommand \BibitemOpen [0]{}%
\providecommand \bibitemStop [0]{}%
\providecommand \bibitemNoStop [0]{.\EOS\space}%
\providecommand \EOS [0]{\spacefactor3000\relax}%
\providecommand \BibitemShut  [1]{\csname bibitem#1\endcsname}%
\let\auto@bib@innerbib\@empty
\bibitem [{\citenamefont {Abrahams}\ \emph {et~al.}(1979)\citenamefont
  {Abrahams}, \citenamefont {Anderson}, \citenamefont {Licciardello},\ and\
  \citenamefont {Ramakrishnan}}]{Abrahams_localization_1979}%
  \BibitemOpen
  \bibfield  {author} {\bibinfo {author} {\bibfnamefont {E.}~\bibnamefont
  {Abrahams}}, \bibinfo {author} {\bibfnamefont {P.~W.}\ \bibnamefont
  {Anderson}}, \bibinfo {author} {\bibfnamefont {D.~C.}\ \bibnamefont
  {Licciardello}}, \ and\ \bibinfo {author} {\bibfnamefont {T.~V.}\
  \bibnamefont {Ramakrishnan}},\ }\href {\doibase 10.1103/PhysRevLett.42.673}
  {\bibfield  {journal} {\bibinfo  {journal} {Phys. Rev. Lett.}\ }\textbf
  {\bibinfo {volume} {42}},\ \bibinfo {pages} {673} (\bibinfo {year}
  {1979})}\BibitemShut {NoStop}%
\bibitem [{\citenamefont {Giamarchi}\ and\ \citenamefont
  {Schulz}(1988)}]{Giamarchi_localization_1d_1988}%
  \BibitemOpen
  \bibfield  {author} {\bibinfo {author} {\bibfnamefont {T.}~\bibnamefont
  {Giamarchi}}\ and\ \bibinfo {author} {\bibfnamefont {H.~J.}\ \bibnamefont
  {Schulz}},\ }\href {\doibase 10.1103/PhysRevB.37.325} {\bibfield  {journal}
  {\bibinfo  {journal} {Phys. Rev. B}\ }\textbf {\bibinfo {volume} {37}},\
  \bibinfo {pages} {325} (\bibinfo {year} {1988})}\BibitemShut {NoStop}%
\bibitem [{\citenamefont {Schnyder}\ \emph {et~al.}(2008)\citenamefont
  {Schnyder}, \citenamefont {Ryu}, \citenamefont {Furusaki},\ and\
  \citenamefont {Ludwig}}]{Schnyder_classification_2008}%
  \BibitemOpen
  \bibfield  {author} {\bibinfo {author} {\bibfnamefont {A.~P.}\ \bibnamefont
  {Schnyder}}, \bibinfo {author} {\bibfnamefont {S.}~\bibnamefont {Ryu}},
  \bibinfo {author} {\bibfnamefont {A.}~\bibnamefont {Furusaki}}, \ and\
  \bibinfo {author} {\bibfnamefont {A.~W.~W.}\ \bibnamefont {Ludwig}},\ }\href
  {\doibase 10.1103/PhysRevB.78.195125} {\bibfield  {journal} {\bibinfo
  {journal} {Phys. Rev. B}\ }\textbf {\bibinfo {volume} {78}},\ \bibinfo
  {pages} {195125} (\bibinfo {year} {2008})}\BibitemShut {NoStop}%
\bibitem [{\citenamefont {Mermin}\ and\ \citenamefont
  {Wagner}(1966)}]{MerminWagner_1966}%
  \BibitemOpen
  \bibfield  {author} {\bibinfo {author} {\bibfnamefont {N.~D.}\ \bibnamefont
  {Mermin}}\ and\ \bibinfo {author} {\bibfnamefont {H.}~\bibnamefont
  {Wagner}},\ }\href {\doibase 10.1103/PhysRevLett.17.1133} {\bibfield
  {journal} {\bibinfo  {journal} {Phys. Rev. Lett.}\ }\textbf {\bibinfo
  {volume} {17}},\ \bibinfo {pages} {1133} (\bibinfo {year}
  {1966})}\BibitemShut {NoStop}%
\bibitem [{\citenamefont {Hohenberg}(1967)}]{Hohenberg_1967}%
  \BibitemOpen
  \bibfield  {author} {\bibinfo {author} {\bibfnamefont {P.~C.}\ \bibnamefont
  {Hohenberg}},\ }\href {\doibase 10.1103/PhysRev.158.383} {\bibfield
  {journal} {\bibinfo  {journal} {Phys. Rev.}\ }\textbf {\bibinfo {volume}
  {158}},\ \bibinfo {pages} {383} (\bibinfo {year} {1967})}\BibitemShut
  {NoStop}%
\bibitem [{\citenamefont {Momoi}(1996)}]{Momoi_AF_1996}%
  \BibitemOpen
  \bibfield  {author} {\bibinfo {author} {\bibfnamefont {T.}~\bibnamefont
  {Momoi}},\ }\href {\doibase 10.1007/BF02175562} {\bibfield  {journal}
  {\bibinfo  {journal} {J. Stat. Phys.}\ }\textbf {\bibinfo {volume} {85}},\
  \bibinfo {pages} {193} (\bibinfo {year} {1996})}\BibitemShut {NoStop}%
\bibitem [{\citenamefont {Furuya}\ and\ \citenamefont
  {Giamarchi}(2014)}]{Furuya_unionjack_2014}%
  \BibitemOpen
  \bibfield  {author} {\bibinfo {author} {\bibfnamefont {S.~C.}\ \bibnamefont
  {Furuya}}\ and\ \bibinfo {author} {\bibfnamefont {T.}~\bibnamefont
  {Giamarchi}},\ }\href {\doibase 10.1103/PhysRevB.89.205131} {\bibfield
  {journal} {\bibinfo  {journal} {Phys. Rev. B}\ }\textbf {\bibinfo {volume}
  {89}},\ \bibinfo {pages} {205131} (\bibinfo {year} {2014})}\BibitemShut
  {NoStop}%
\bibitem [{\citenamefont {Kohno}\ \emph {et~al.}(2007)\citenamefont {Kohno},
  \citenamefont {Starykh},\ and\ \citenamefont
  {Balents}}]{Kohno_Cs2CuCl4_2007}%
  \BibitemOpen
  \bibfield  {author} {\bibinfo {author} {\bibfnamefont {M.}~\bibnamefont
  {Kohno}}, \bibinfo {author} {\bibfnamefont {O.~A.}\ \bibnamefont {Starykh}},
  \ and\ \bibinfo {author} {\bibfnamefont {L.}~\bibnamefont {Balents}},\ }\href
  {ttp://dx.doi.org/10.1038/nphys749} {\bibfield  {journal} {\bibinfo
  {journal} {Nature Phys.}\ }\textbf {\bibinfo {volume} {3}},\ \bibinfo {pages}
  {790} (\bibinfo {year} {2007})}\BibitemShut {NoStop}%
\bibitem [{\citenamefont {Starykh}\ \emph {et~al.}(2010)\citenamefont
  {Starykh}, \citenamefont {Katsura},\ and\ \citenamefont
  {Balents}}]{Starykh_Cs2CuCl4_2010}%
  \BibitemOpen
  \bibfield  {author} {\bibinfo {author} {\bibfnamefont {O.~A.}\ \bibnamefont
  {Starykh}}, \bibinfo {author} {\bibfnamefont {H.}~\bibnamefont {Katsura}}, \
  and\ \bibinfo {author} {\bibfnamefont {L.}~\bibnamefont {Balents}},\ }\href
  {\doibase 10.1103/PhysRevB.82.014421} {\bibfield  {journal} {\bibinfo
  {journal} {Phys. Rev. B}\ }\textbf {\bibinfo {volume} {82}},\ \bibinfo
  {pages} {014421} (\bibinfo {year} {2010})}\BibitemShut {NoStop}%
\bibitem [{\citenamefont {Klanj\v{s}ek}\ \emph {et~al.}(2008)\citenamefont
  {Klanj\v{s}ek}, \citenamefont {Mayaffre}, \citenamefont {Berthier},
  \citenamefont {Horvati\'{c}}, \citenamefont {Chiari}, \citenamefont
  {Piovesana}, \citenamefont {Bouillot}, \citenamefont {Kollath}, \citenamefont
  {Orignac}, \citenamefont {Citro},\ and\ \citenamefont
  {Giamarchi}}]{Klanjsek_BPCB_2008}%
  \BibitemOpen
  \bibfield  {author} {\bibinfo {author} {\bibfnamefont {M.}~\bibnamefont
  {Klanj\v{s}ek}}, \bibinfo {author} {\bibfnamefont {H.}~\bibnamefont
  {Mayaffre}}, \bibinfo {author} {\bibfnamefont {C.}~\bibnamefont {Berthier}},
  \bibinfo {author} {\bibfnamefont {M.}~\bibnamefont {Horvati\'{c}}}, \bibinfo
  {author} {\bibfnamefont {B.}~\bibnamefont {Chiari}}, \bibinfo {author}
  {\bibfnamefont {O.}~\bibnamefont {Piovesana}}, \bibinfo {author}
  {\bibfnamefont {P.}~\bibnamefont {Bouillot}}, \bibinfo {author}
  {\bibfnamefont {C.}~\bibnamefont {Kollath}}, \bibinfo {author} {\bibfnamefont
  {E.}~\bibnamefont {Orignac}}, \bibinfo {author} {\bibfnamefont
  {R.}~\bibnamefont {Citro}}, \ and\ \bibinfo {author} {\bibfnamefont
  {T.}~\bibnamefont {Giamarchi}},\ }\href {\doibase
  10.1103/PhysRevLett.101.137207} {\bibfield  {journal} {\bibinfo  {journal}
  {Phys. Rev. Lett.}\ }\textbf {\bibinfo {volume} {101}},\ \bibinfo {pages}
  {137207} (\bibinfo {year} {2008})}\BibitemShut {NoStop}%
\bibitem [{\citenamefont {Schmidiger}\ \emph {et~al.}(2012)\citenamefont
  {Schmidiger}, \citenamefont {Bouillot}, \citenamefont {M\"uhlbauer},
  \citenamefont {Gvasaliya}, \citenamefont {Kollath}, \citenamefont
  {Giamarchi},\ and\ \citenamefont {Zheludev}}]{Schmidiger_DIMPY_2012}%
  \BibitemOpen
  \bibfield  {author} {\bibinfo {author} {\bibfnamefont {D.}~\bibnamefont
  {Schmidiger}}, \bibinfo {author} {\bibfnamefont {P.}~\bibnamefont
  {Bouillot}}, \bibinfo {author} {\bibfnamefont {S.}~\bibnamefont
  {M\"uhlbauer}}, \bibinfo {author} {\bibfnamefont {S.}~\bibnamefont
  {Gvasaliya}}, \bibinfo {author} {\bibfnamefont {C.}~\bibnamefont {Kollath}},
  \bibinfo {author} {\bibfnamefont {T.}~\bibnamefont {Giamarchi}}, \ and\
  \bibinfo {author} {\bibfnamefont {A.}~\bibnamefont {Zheludev}},\ }\href
  {\doibase 10.1103/PhysRevLett.108.167201} {\bibfield  {journal} {\bibinfo
  {journal} {Phys. Rev. Lett.}\ }\textbf {\bibinfo {volume} {108}},\ \bibinfo
  {pages} {167201} (\bibinfo {year} {2012})}\BibitemShut {NoStop}%
\bibitem [{\citenamefont {Garlea}\ \emph {et~al.}(2009)\citenamefont {Garlea},
  \citenamefont {Zheludev}, \citenamefont {Habicht}, \citenamefont {Meissner},
  \citenamefont {Grenier}, \citenamefont {Regnault},\ and\ \citenamefont
  {Ressouche}}]{Garlea_dim_crossover_2009}%
  \BibitemOpen
  \bibfield  {author} {\bibinfo {author} {\bibfnamefont {V.~O.}\ \bibnamefont
  {Garlea}}, \bibinfo {author} {\bibfnamefont {A.}~\bibnamefont {Zheludev}},
  \bibinfo {author} {\bibfnamefont {K.}~\bibnamefont {Habicht}}, \bibinfo
  {author} {\bibfnamefont {M.}~\bibnamefont {Meissner}}, \bibinfo {author}
  {\bibfnamefont {B.}~\bibnamefont {Grenier}}, \bibinfo {author} {\bibfnamefont
  {L.-P.}\ \bibnamefont {Regnault}}, \ and\ \bibinfo {author} {\bibfnamefont
  {E.}~\bibnamefont {Ressouche}},\ }\href {\doibase 10.1103/PhysRevB.79.060404}
  {\bibfield  {journal} {\bibinfo  {journal} {Phys. Rev. B}\ }\textbf {\bibinfo
  {volume} {79}},\ \bibinfo {pages} {060404} (\bibinfo {year}
  {2009})}\BibitemShut {NoStop}%
\bibitem [{\citenamefont {Yamaguchi}\ \emph {et~al.}(2013)\citenamefont
  {Yamaguchi}, \citenamefont {Iwase}, \citenamefont {Ono}, \citenamefont
  {Shimokawa}, \citenamefont {Nakano}, \citenamefont {Shimura}, \citenamefont
  {Kase}, \citenamefont {Kittaka}, \citenamefont {Sakakibara}, \citenamefont
  {Kawakami},\ and\ \citenamefont {Hosokoshi}}]{Yamaguchi_3cl4fv_2013}%
  \BibitemOpen
  \bibfield  {author} {\bibinfo {author} {\bibfnamefont {H.}~\bibnamefont
  {Yamaguchi}}, \bibinfo {author} {\bibfnamefont {K.}~\bibnamefont {Iwase}},
  \bibinfo {author} {\bibfnamefont {T.}~\bibnamefont {Ono}}, \bibinfo {author}
  {\bibfnamefont {T.}~\bibnamefont {Shimokawa}}, \bibinfo {author}
  {\bibfnamefont {H.}~\bibnamefont {Nakano}}, \bibinfo {author} {\bibfnamefont
  {Y.}~\bibnamefont {Shimura}}, \bibinfo {author} {\bibfnamefont
  {N.}~\bibnamefont {Kase}}, \bibinfo {author} {\bibfnamefont {S.}~\bibnamefont
  {Kittaka}}, \bibinfo {author} {\bibfnamefont {T.}~\bibnamefont {Sakakibara}},
  \bibinfo {author} {\bibfnamefont {T.}~\bibnamefont {Kawakami}}, \ and\
  \bibinfo {author} {\bibfnamefont {Y.}~\bibnamefont {Hosokoshi}},\ }\href
  {\doibase 10.1103/PhysRevLett.110.157205} {\bibfield  {journal} {\bibinfo
  {journal} {Phys. Rev. Lett.}\ }\textbf {\bibinfo {volume} {110}},\ \bibinfo
  {pages} {157205} (\bibinfo {year} {2013})}\BibitemShut {NoStop}%
\bibitem [{\citenamefont {Yamaguchi}\ \emph {et~al.}(2015)\citenamefont
  {Yamaguchi}, \citenamefont {Miyagai}, \citenamefont {Kono}, \citenamefont
  {Kittaka}, \citenamefont {Sakakibara}, \citenamefont {Iwase}, \citenamefont
  {Ono}, \citenamefont {Shimokawa},\ and\ \citenamefont
  {Hosokoshi}}]{Yamaguchi_3iv_2015}%
  \BibitemOpen
  \bibfield  {author} {\bibinfo {author} {\bibfnamefont {H.}~\bibnamefont
  {Yamaguchi}}, \bibinfo {author} {\bibfnamefont {H.}~\bibnamefont {Miyagai}},
  \bibinfo {author} {\bibfnamefont {Y.}~\bibnamefont {Kono}}, \bibinfo {author}
  {\bibfnamefont {S.}~\bibnamefont {Kittaka}}, \bibinfo {author} {\bibfnamefont
  {T.}~\bibnamefont {Sakakibara}}, \bibinfo {author} {\bibfnamefont
  {K.}~\bibnamefont {Iwase}}, \bibinfo {author} {\bibfnamefont
  {T.}~\bibnamefont {Ono}}, \bibinfo {author} {\bibfnamefont {T.}~\bibnamefont
  {Shimokawa}}, \ and\ \bibinfo {author} {\bibfnamefont {Y.}~\bibnamefont
  {Hosokoshi}},\ }\href {\doibase 10.1103/PhysRevB.91.125104} {\bibfield
  {journal} {\bibinfo  {journal} {Phys. Rev. B}\ }\textbf {\bibinfo {volume}
  {91}},\ \bibinfo {pages} {125104} (\bibinfo {year} {2015})}\BibitemShut
  {NoStop}%
\bibitem [{\citenamefont {Klanj\v{s}ek}\ \emph {et~al.}(2015)\citenamefont
  {Klanj\v{s}ek}, \citenamefont {Horvati\'{c}}, \citenamefont {Kr\"amer},
  \citenamefont {Mukhopadhyay}, \citenamefont {Mayaffre}, \citenamefont
  {Berthier}, \citenamefont {Can\'evet}, \citenamefont {Grenier}, \citenamefont
  {Lejay},\ and\ \citenamefont {Orignac}}]{Klanjsek_bacovo_2015}%
  \BibitemOpen
  \bibfield  {author} {\bibinfo {author} {\bibfnamefont {M.}~\bibnamefont
  {Klanj\v{s}ek}}, \bibinfo {author} {\bibfnamefont {M.}~\bibnamefont
  {Horvati\'{c}}}, \bibinfo {author} {\bibfnamefont {S.}~\bibnamefont
  {Kr\"amer}}, \bibinfo {author} {\bibfnamefont {S.}~\bibnamefont
  {Mukhopadhyay}}, \bibinfo {author} {\bibfnamefont {H.}~\bibnamefont
  {Mayaffre}}, \bibinfo {author} {\bibfnamefont {C.}~\bibnamefont {Berthier}},
  \bibinfo {author} {\bibfnamefont {E.}~\bibnamefont {Can\'evet}}, \bibinfo
  {author} {\bibfnamefont {B.}~\bibnamefont {Grenier}}, \bibinfo {author}
  {\bibfnamefont {P.}~\bibnamefont {Lejay}}, \ and\ \bibinfo {author}
  {\bibfnamefont {E.}~\bibnamefont {Orignac}},\ }\href {\doibase
  10.1103/PhysRevB.92.060408} {\bibfield  {journal} {\bibinfo  {journal} {Phys.
  Rev. B}\ }\textbf {\bibinfo {volume} {92}},\ \bibinfo {pages} {060408}
  (\bibinfo {year} {2015})}\BibitemShut {NoStop}%
\bibitem [{\citenamefont {Giamarchi}(2010)}]{Giamarchi_QPT_2010}%
  \BibitemOpen
  \bibfield  {author} {\bibinfo {author} {\bibfnamefont {T.}~\bibnamefont
  {Giamarchi}},\ }\href@noop {} {\bibfield  {journal} {\bibinfo  {journal}
  {arXiv:1007.1029}\ } (\bibinfo {year} {2010})}\BibitemShut {NoStop}%
\bibitem [{\citenamefont {Kohno}(2009)}]{Kohno_Cs2CuCl4_2009}%
  \BibitemOpen
  \bibfield  {author} {\bibinfo {author} {\bibfnamefont {M.}~\bibnamefont
  {Kohno}},\ }\href {\doibase 10.1103/PhysRevLett.103.197203} {\bibfield
  {journal} {\bibinfo  {journal} {Phys. Rev. Lett.}\ }\textbf {\bibinfo
  {volume} {103}},\ \bibinfo {pages} {197203} (\bibinfo {year}
  {2009})}\BibitemShut {NoStop}%
\bibitem [{\citenamefont {Hong}\ \emph {et~al.}(2010)\citenamefont {Hong},
  \citenamefont {Kim}, \citenamefont {Hotta}, \citenamefont {Takano},
  \citenamefont {Tremelling}, \citenamefont {Turnbull}, \citenamefont {Landee},
  \citenamefont {Kang}, \citenamefont {Christensen}, \citenamefont {Lefmann},
  \citenamefont {Schmidt}, \citenamefont {Uhrig},\ and\ \citenamefont
  {Broholm}}]{Hong_DIMPY_2010}%
  \BibitemOpen
  \bibfield  {author} {\bibinfo {author} {\bibfnamefont {T.}~\bibnamefont
  {Hong}}, \bibinfo {author} {\bibfnamefont {Y.~H.}\ \bibnamefont {Kim}},
  \bibinfo {author} {\bibfnamefont {C.}~\bibnamefont {Hotta}}, \bibinfo
  {author} {\bibfnamefont {Y.}~\bibnamefont {Takano}}, \bibinfo {author}
  {\bibfnamefont {G.}~\bibnamefont {Tremelling}}, \bibinfo {author}
  {\bibfnamefont {M.~M.}\ \bibnamefont {Turnbull}}, \bibinfo {author}
  {\bibfnamefont {C.~P.}\ \bibnamefont {Landee}}, \bibinfo {author}
  {\bibfnamefont {H.-J.}\ \bibnamefont {Kang}}, \bibinfo {author}
  {\bibfnamefont {N.~B.}\ \bibnamefont {Christensen}}, \bibinfo {author}
  {\bibfnamefont {K.}~\bibnamefont {Lefmann}}, \bibinfo {author} {\bibfnamefont
  {K.~P.}\ \bibnamefont {Schmidt}}, \bibinfo {author} {\bibfnamefont {G.~S.}\
  \bibnamefont {Uhrig}}, \ and\ \bibinfo {author} {\bibfnamefont
  {C.}~\bibnamefont {Broholm}},\ }\href {\doibase
  10.1103/PhysRevLett.105.137207} {\bibfield  {journal} {\bibinfo  {journal}
  {Phys. Rev. Lett.}\ }\textbf {\bibinfo {volume} {105}},\ \bibinfo {pages}
  {137207} (\bibinfo {year} {2010})}\BibitemShut {NoStop}%
\bibitem [{\citenamefont {Jeong}\ \emph {et~al.}(2013)\citenamefont {Jeong},
  \citenamefont {Mayaffre}, \citenamefont {Berthier}, \citenamefont
  {Schmidiger}, \citenamefont {Zheludev},\ and\ \citenamefont
  {Horvati\ifmmode~\acute{c}\else \'{c}\fi{}}}]{Jeong_DIMPY_2013}%
  \BibitemOpen
  \bibfield  {author} {\bibinfo {author} {\bibfnamefont {M.}~\bibnamefont
  {Jeong}}, \bibinfo {author} {\bibfnamefont {H.}~\bibnamefont {Mayaffre}},
  \bibinfo {author} {\bibfnamefont {C.}~\bibnamefont {Berthier}}, \bibinfo
  {author} {\bibfnamefont {D.}~\bibnamefont {Schmidiger}}, \bibinfo {author}
  {\bibfnamefont {A.}~\bibnamefont {Zheludev}}, \ and\ \bibinfo {author}
  {\bibfnamefont {M.}~\bibnamefont {Horvati\ifmmode~\acute{c}\else
  \'{c}\fi{}}},\ }\href {\doibase 10.1103/PhysRevLett.111.106404} {\bibfield
  {journal} {\bibinfo  {journal} {Phys. Rev. Lett.}\ }\textbf {\bibinfo
  {volume} {111}},\ \bibinfo {pages} {106404} (\bibinfo {year}
  {2013})}\BibitemShut {NoStop}%
\bibitem [{\citenamefont {Schmidiger}\ \emph
  {et~al.}(2013{\natexlab{a}})\citenamefont {Schmidiger}, \citenamefont
  {Bouillot}, \citenamefont {Guidi}, \citenamefont {Bewley}, \citenamefont
  {Kollath}, \citenamefont {Giamarchi},\ and\ \citenamefont
  {Zheludev}}]{Schmidiger_DIMPY_2013a}%
  \BibitemOpen
  \bibfield  {author} {\bibinfo {author} {\bibfnamefont {D.}~\bibnamefont
  {Schmidiger}}, \bibinfo {author} {\bibfnamefont {P.}~\bibnamefont
  {Bouillot}}, \bibinfo {author} {\bibfnamefont {T.}~\bibnamefont {Guidi}},
  \bibinfo {author} {\bibfnamefont {R.}~\bibnamefont {Bewley}}, \bibinfo
  {author} {\bibfnamefont {C.}~\bibnamefont {Kollath}}, \bibinfo {author}
  {\bibfnamefont {T.}~\bibnamefont {Giamarchi}}, \ and\ \bibinfo {author}
  {\bibfnamefont {A.}~\bibnamefont {Zheludev}},\ }\href {\doibase
  10.1103/PhysRevLett.111.107202} {\bibfield  {journal} {\bibinfo  {journal}
  {Phys. Rev. Lett.}\ }\textbf {\bibinfo {volume} {111}},\ \bibinfo {pages}
  {107202} (\bibinfo {year} {2013}{\natexlab{a}})}\BibitemShut {NoStop}%
\bibitem [{\citenamefont {Schmidiger}\ \emph
  {et~al.}(2013{\natexlab{b}})\citenamefont {Schmidiger}, \citenamefont
  {M\"uhlbauer}, \citenamefont {Zheludev}, \citenamefont {Bouillot},
  \citenamefont {Giamarchi}, \citenamefont {Kollath}, \citenamefont {Ehlers},\
  and\ \citenamefont {Tsvelik}}]{Schmidiger_DIMPY_2013b}%
  \BibitemOpen
  \bibfield  {author} {\bibinfo {author} {\bibfnamefont {D.}~\bibnamefont
  {Schmidiger}}, \bibinfo {author} {\bibfnamefont {S.}~\bibnamefont
  {M\"uhlbauer}}, \bibinfo {author} {\bibfnamefont {A.}~\bibnamefont
  {Zheludev}}, \bibinfo {author} {\bibfnamefont {P.}~\bibnamefont {Bouillot}},
  \bibinfo {author} {\bibfnamefont {T.}~\bibnamefont {Giamarchi}}, \bibinfo
  {author} {\bibfnamefont {C.}~\bibnamefont {Kollath}}, \bibinfo {author}
  {\bibfnamefont {G.}~\bibnamefont {Ehlers}}, \ and\ \bibinfo {author}
  {\bibfnamefont {A.~M.}\ \bibnamefont {Tsvelik}},\ }\href {\doibase
  10.1103/PhysRevB.88.094411} {\bibfield  {journal} {\bibinfo  {journal} {Phys.
  Rev. B}\ }\textbf {\bibinfo {volume} {88}},\ \bibinfo {pages} {094411}
  (\bibinfo {year} {2013}{\natexlab{b}})}\BibitemShut {NoStop}%
\bibitem [{\citenamefont {Glazkov}\ \emph {et~al.}()\citenamefont {Glazkov},
  \citenamefont {Fayzullin}, \citenamefont {Krasnikova}, \citenamefont
  {Skoblin}, \citenamefont {Schmidiger}, \citenamefont {M\"uhlbauer},\ and\
  \citenamefont {Zheludev}}]{Glazkov_DIMPY_2015}%
  \BibitemOpen
  \bibfield  {author} {\bibinfo {author} {\bibfnamefont {V.~N.}\ \bibnamefont
  {Glazkov}}, \bibinfo {author} {\bibfnamefont {M.}~\bibnamefont {Fayzullin}},
  \bibinfo {author} {\bibfnamefont {Y.}~\bibnamefont {Krasnikova}}, \bibinfo
  {author} {\bibfnamefont {G.}~\bibnamefont {Skoblin}}, \bibinfo {author}
  {\bibfnamefont {D.}~\bibnamefont {Schmidiger}}, \bibinfo {author}
  {\bibfnamefont {S.}~\bibnamefont {M\"uhlbauer}}, \ and\ \bibinfo {author}
  {\bibfnamefont {A.}~\bibnamefont {Zheludev}},\ }\href@noop {} {\ }\bibinfo
  {note} {{}arXiv:1507.02503 (2015)}\BibitemShut {NoStop}%
\bibitem [{\citenamefont {Ozerov}\ \emph {et~al.}(2015)\citenamefont {Ozerov},
  \citenamefont {Maksymenko}, \citenamefont {Wosnitza}, \citenamefont
  {Honecker}, \citenamefont {Landee}, \citenamefont {Turnbull}, \citenamefont
  {Furuya}, \citenamefont {Giamarchi},\ and\ \citenamefont
  {Zvyagin}}]{Ozerov_DIMPY_2015}%
  \BibitemOpen
  \bibfield  {author} {\bibinfo {author} {\bibfnamefont {M.}~\bibnamefont
  {Ozerov}}, \bibinfo {author} {\bibfnamefont {M.}~\bibnamefont {Maksymenko}},
  \bibinfo {author} {\bibfnamefont {J.}~\bibnamefont {Wosnitza}}, \bibinfo
  {author} {\bibfnamefont {A.}~\bibnamefont {Honecker}}, \bibinfo {author}
  {\bibfnamefont {C.~P.}\ \bibnamefont {Landee}}, \bibinfo {author}
  {\bibfnamefont {M.~M.}\ \bibnamefont {Turnbull}}, \bibinfo {author}
  {\bibfnamefont {S.~C.}\ \bibnamefont {Furuya}}, \bibinfo {author}
  {\bibfnamefont {T.}~\bibnamefont {Giamarchi}}, \ and\ \bibinfo {author}
  {\bibfnamefont {S.~A.}\ \bibnamefont {Zvyagin}},\ }\href {\doibase
  10.1103/PhysRevB.92.241113} {\bibfield  {journal} {\bibinfo  {journal} {Phys.
  Rev. B}\ }\textbf {\bibinfo {volume} {92}},\ \bibinfo {pages} {241113}
  (\bibinfo {year} {2015})}\BibitemShut {NoStop}%
\bibitem [{\citenamefont {Bouillot}\ \emph {et~al.}(2011)\citenamefont
  {Bouillot}, \citenamefont {Kollath}, \citenamefont {L\"auchli}, \citenamefont
  {Zvonarev}, \citenamefont {Thielemann}, \citenamefont {R\"uegg},
  \citenamefont {Orignac}, \citenamefont {Citro}, \citenamefont {Klanj\v{s}ek},
  \citenamefont {Berthier}, \citenamefont {Horvati\'{c}},\ and\ \citenamefont
  {Giamarchi}}]{Bouillot_BPCB_2011}%
  \BibitemOpen
  \bibfield  {author} {\bibinfo {author} {\bibfnamefont {P.}~\bibnamefont
  {Bouillot}}, \bibinfo {author} {\bibfnamefont {C.}~\bibnamefont {Kollath}},
  \bibinfo {author} {\bibfnamefont {A.~M.}\ \bibnamefont {L\"auchli}}, \bibinfo
  {author} {\bibfnamefont {M.}~\bibnamefont {Zvonarev}}, \bibinfo {author}
  {\bibfnamefont {B.}~\bibnamefont {Thielemann}}, \bibinfo {author}
  {\bibfnamefont {C.}~\bibnamefont {R\"uegg}}, \bibinfo {author} {\bibfnamefont
  {E.}~\bibnamefont {Orignac}}, \bibinfo {author} {\bibfnamefont
  {R.}~\bibnamefont {Citro}}, \bibinfo {author} {\bibfnamefont
  {M.}~\bibnamefont {Klanj\v{s}ek}}, \bibinfo {author} {\bibfnamefont
  {C.}~\bibnamefont {Berthier}}, \bibinfo {author} {\bibfnamefont
  {M.}~\bibnamefont {Horvati\'{c}}}, \ and\ \bibinfo {author} {\bibfnamefont
  {T.}~\bibnamefont {Giamarchi}},\ }\href {\doibase 10.1103/PhysRevB.83.054407}
  {\bibfield  {journal} {\bibinfo  {journal} {Phys. Rev. B}\ }\textbf {\bibinfo
  {volume} {83}},\ \bibinfo {pages} {054407} (\bibinfo {year}
  {2011})}\BibitemShut {NoStop}%
\bibitem [{\citenamefont {Scalapino}\ \emph {et~al.}(1975)\citenamefont
  {Scalapino}, \citenamefont {Imry},\ and\ \citenamefont
  {Pincus}}]{Scalapino_RPA_1975}%
  \BibitemOpen
  \bibfield  {author} {\bibinfo {author} {\bibfnamefont {D.~J.}\ \bibnamefont
  {Scalapino}}, \bibinfo {author} {\bibfnamefont {Y.}~\bibnamefont {Imry}}, \
  and\ \bibinfo {author} {\bibfnamefont {P.}~\bibnamefont {Pincus}},\ }\href
  {\doibase 10.1103/PhysRevB.11.2042} {\bibfield  {journal} {\bibinfo
  {journal} {Phys. Rev. B}\ }\textbf {\bibinfo {volume} {11}},\ \bibinfo
  {pages} {2042} (\bibinfo {year} {1975})}\BibitemShut {NoStop}%
\bibitem [{\citenamefont {Schulz}(1996)}]{Schulz_CoupledChains_1996}%
  \BibitemOpen
  \bibfield  {author} {\bibinfo {author} {\bibfnamefont {H.~J.}\ \bibnamefont
  {Schulz}},\ }\href {\doibase 10.1103/PhysRevLett.77.2790} {\bibfield
  {journal} {\bibinfo  {journal} {Phys. Rev. Lett.}\ }\textbf {\bibinfo
  {volume} {77}},\ \bibinfo {pages} {2790} (\bibinfo {year}
  {1996})}\BibitemShut {NoStop}%
\bibitem [{\citenamefont {Sakai}\ and\ \citenamefont
  {Takahashi}(1989{\natexlab{a}})}]{Sakai_HaldanePhase_RPA_1989}%
  \BibitemOpen
  \bibfield  {author} {\bibinfo {author} {\bibfnamefont {T.}~\bibnamefont
  {Sakai}}\ and\ \bibinfo {author} {\bibfnamefont {M.}~\bibnamefont
  {Takahashi}},\ }\href {\doibase 10.1143/JPSJ.58.3131} {\bibfield  {journal}
  {\bibinfo  {journal} {J. Phys. Soc. Jpn.}\ }\textbf {\bibinfo {volume}
  {58}},\ \bibinfo {pages} {3131} (\bibinfo {year}
  {1989}{\natexlab{a}})}\BibitemShut {NoStop}%
\bibitem [{\citenamefont {Wierschem}\ and\ \citenamefont
  {Sengupta}(2014)}]{Wierschem_quasi1dHaldane_2014}%
  \BibitemOpen
  \bibfield  {author} {\bibinfo {author} {\bibfnamefont {K.}~\bibnamefont
  {Wierschem}}\ and\ \bibinfo {author} {\bibfnamefont {P.}~\bibnamefont
  {Sengupta}},\ }\href {\doibase 10.1103/PhysRevLett.112.247203} {\bibfield
  {journal} {\bibinfo  {journal} {Phys. Rev. Lett.}\ }\textbf {\bibinfo
  {volume} {112}},\ \bibinfo {pages} {247203} (\bibinfo {year}
  {2014})}\BibitemShut {NoStop}%
\bibitem [{\citenamefont {Giamarchi}(2004)}]{Giamarchi_book}%
  \BibitemOpen
  \bibfield  {author} {\bibinfo {author} {\bibfnamefont {T.}~\bibnamefont
  {Giamarchi}},\ }\href@noop {} {\emph {\bibinfo {title} {Quantum Physics in
  One Dimension}}}\ (\bibinfo  {publisher} {Oxford University Press},\ \bibinfo
  {address} {Oxford},\ \bibinfo {year} {2004})\BibitemShut {NoStop}%
\bibitem [{\citenamefont {Chitra}\ and\ \citenamefont
  {Giamarchi}(1997)}]{Chitra_ladder_1997}%
  \BibitemOpen
  \bibfield  {author} {\bibinfo {author} {\bibfnamefont {R.}~\bibnamefont
  {Chitra}}\ and\ \bibinfo {author} {\bibfnamefont {T.}~\bibnamefont
  {Giamarchi}},\ }\href {\doibase 10.1103/PhysRevB.55.5816} {\bibfield
  {journal} {\bibinfo  {journal} {Phys. Rev. B}\ }\textbf {\bibinfo {volume}
  {55}},\ \bibinfo {pages} {5816} (\bibinfo {year} {1997})}\BibitemShut
  {NoStop}%
\bibitem [{\citenamefont {Giamarchi}\ \emph {et~al.}(2008)\citenamefont
  {Giamarchi}, \citenamefont {R{\"u}egg},\ and\ \citenamefont
  {Tchernyshyov}}]{Giamarchi_rev_BEC}%
  \BibitemOpen
  \bibfield  {author} {\bibinfo {author} {\bibfnamefont {T.}~\bibnamefont
  {Giamarchi}}, \bibinfo {author} {\bibfnamefont {C.}~\bibnamefont
  {R{\"u}egg}}, \ and\ \bibinfo {author} {\bibfnamefont {O.}~\bibnamefont
  {Tchernyshyov}},\ }\href
  {http://www.nature.com/nphys/journal/v4/n3/abs/nphys893.html} {\bibfield
  {journal} {\bibinfo  {journal} {Nat. Phys.}\ }\textbf {\bibinfo {volume}
  {4}},\ \bibinfo {pages} {198} (\bibinfo {year} {2008})}\BibitemShut {NoStop}%
\bibitem [{\citenamefont {Feynman}(1998)}]{Feynman_book}%
  \BibitemOpen
  \bibfield  {author} {\bibinfo {author} {\bibfnamefont {R.~P.}\ \bibnamefont
  {Feynman}},\ }\href@noop {} {\emph {\bibinfo {title} {Statistical Mechanics:
  A Set of Lectures}}},\ \bibinfo {edition} {2nd}\ ed.\ (\bibinfo  {publisher}
  {Advanced Book Classics, Perseus},\ \bibinfo {address} {New York},\ \bibinfo
  {year} {1998})\BibitemShut {NoStop}%
\bibitem [{\citenamefont {Cazalilla}\ \emph {et~al.}(2006)\citenamefont
  {Cazalilla}, \citenamefont {Ho},\ and\ \citenamefont
  {Giamarchi}}]{Cazalilla_NJP_2006}%
  \BibitemOpen
  \bibfield  {author} {\bibinfo {author} {\bibfnamefont {M.~A.}\ \bibnamefont
  {Cazalilla}}, \bibinfo {author} {\bibfnamefont {A.~F.}\ \bibnamefont {Ho}}, \
  and\ \bibinfo {author} {\bibfnamefont {T.}~\bibnamefont {Giamarchi}},\ }\href
  {http://stacks.iop.org/1367-2630/8/i=8/a=158} {\bibfield  {journal} {\bibinfo
   {journal} {New J. Phys.}\ }\textbf {\bibinfo {volume} {8}},\ \bibinfo
  {pages} {158} (\bibinfo {year} {2006})}\BibitemShut {NoStop}%
\bibitem [{\citenamefont {Furuya}\ \emph {et~al.}(2011)\citenamefont {Furuya},
  \citenamefont {Oshikawa},\ and\ \citenamefont {Affleck}}]{Furuya_NLSM_2011}%
  \BibitemOpen
  \bibfield  {author} {\bibinfo {author} {\bibfnamefont {S.~C.}\ \bibnamefont
  {Furuya}}, \bibinfo {author} {\bibfnamefont {M.}~\bibnamefont {Oshikawa}}, \
  and\ \bibinfo {author} {\bibfnamefont {I.}~\bibnamefont {Affleck}},\ }\href
  {\doibase 10.1103/PhysRevB.83.224417} {\bibfield  {journal} {\bibinfo
  {journal} {Phys. Rev. B}\ }\textbf {\bibinfo {volume} {83}},\ \bibinfo
  {pages} {224417} (\bibinfo {year} {2011})}\BibitemShut {NoStop}%
\bibitem [{\citenamefont {Benfatto}\ \emph {et~al.}(2007)\citenamefont
  {Benfatto}, \citenamefont {Castellani},\ and\ \citenamefont
  {Giamarchi}}]{Benfatto_SCfilm_2007}%
  \BibitemOpen
  \bibfield  {author} {\bibinfo {author} {\bibfnamefont {L.}~\bibnamefont
  {Benfatto}}, \bibinfo {author} {\bibfnamefont {C.}~\bibnamefont
  {Castellani}}, \ and\ \bibinfo {author} {\bibfnamefont {T.}~\bibnamefont
  {Giamarchi}},\ }\href {\doibase 10.1103/PhysRevLett.98.117008} {\bibfield
  {journal} {\bibinfo  {journal} {Phys. Rev. Lett.}\ }\textbf {\bibinfo
  {volume} {98}},\ \bibinfo {pages} {117008} (\bibinfo {year}
  {2007})}\BibitemShut {NoStop}%
\bibitem [{\citenamefont {Sachdev}(2007)}]{Sachdev_book}%
  \BibitemOpen
  \bibfield  {author} {\bibinfo {author} {\bibfnamefont {S.}~\bibnamefont
  {Sachdev}},\ }\href@noop {} {\emph {\bibinfo {title} {Quantum phase
  transitions}}}\ (\bibinfo  {publisher} {Wiley Online Library},\ \bibinfo
  {year} {2007})\BibitemShut {NoStop}%
\bibitem [{\citenamefont {Lukyanov}(1997)}]{Lukyanov_SG}%
  \BibitemOpen
  \bibfield  {author} {\bibinfo {author} {\bibfnamefont {S.}~\bibnamefont
  {Lukyanov}},\ }\href@noop {} {\bibfield  {journal} {\bibinfo  {journal} {Mod.
  Phys. Lett. A}\ }\textbf {\bibinfo {volume} {12}},\ \bibinfo {pages} {2543}
  (\bibinfo {year} {1997})}\BibitemShut {NoStop}%
\bibitem [{Note1()}]{Note1}%
  \BibitemOpen
  \bibinfo {note} {On the one hand, frustration prohibits quantum Monte-Carlo
  simulations; on the other hand, even for a non-frustrated model, 3D
  anisotropic systems would require very demanding simulations to be able to
  reach the ground-state properties.}\BibitemShut {Stop}%
\bibitem [{\citenamefont {Luther}\ and\ \citenamefont
  {Peschel}(1975)}]{luther_calculation_1975}%
  \BibitemOpen
  \bibfield  {author} {\bibinfo {author} {\bibfnamefont {A.}~\bibnamefont
  {Luther}}\ and\ \bibinfo {author} {\bibfnamefont {I.}~\bibnamefont
  {Peschel}},\ }\href {\doibase 10.1103/PhysRevB.12.3908} {\bibfield  {journal}
  {\bibinfo  {journal} {Phys. Rev. B}\ }\textbf {\bibinfo {volume} {12}},\
  \bibinfo {pages} {3908} (\bibinfo {year} {1975})}\BibitemShut {NoStop}%
\bibitem [{\citenamefont
  {Schollw{\"o}ck}(2005)}]{schollwock_density-matrix_2005}%
  \BibitemOpen
  \bibfield  {author} {\bibinfo {author} {\bibfnamefont {U.}~\bibnamefont
  {Schollw{\"o}ck}},\ }\href {\doibase 10.1103/RevModPhys.77.259} {\bibfield
  {journal} {\bibinfo  {journal} {Rev. Mod. Phys.}\ }\textbf {\bibinfo {volume}
  {77}},\ \bibinfo {pages} {259} (\bibinfo {year} {2005})}\BibitemShut
  {NoStop}%
\bibitem [{\citenamefont {Lukyanov}\ and\ \citenamefont
  {Zamolodchikov}(1997)}]{lukyanov_exact_1997}%
  \BibitemOpen
  \bibfield  {author} {\bibinfo {author} {\bibfnamefont {S.}~\bibnamefont
  {Lukyanov}}\ and\ \bibinfo {author} {\bibfnamefont {A.}~\bibnamefont
  {Zamolodchikov}},\ }\href {\doibase 10.1016/S0550-3213(97)00123-5} {\bibfield
   {journal} {\bibinfo  {journal} {Nuclear Physics B}\ }\textbf {\bibinfo
  {volume} {493}},\ \bibinfo {pages} {571} (\bibinfo {year}
  {1997})}\BibitemShut {NoStop}%
\bibitem [{\citenamefont {Lukyanov}(1999)}]{lukyanov_correlation_1999}%
  \BibitemOpen
  \bibfield  {author} {\bibinfo {author} {\bibfnamefont {S.}~\bibnamefont
  {Lukyanov}},\ }\href {\doibase 10.1103/PhysRevB.59.11163} {\bibfield
  {journal} {\bibinfo  {journal} {Phys. Rev. B}\ }\textbf {\bibinfo {volume}
  {59}},\ \bibinfo {pages} {11163} (\bibinfo {year} {1999})}\BibitemShut
  {NoStop}%
\bibitem [{\citenamefont
  {Schollw{\"o}ck}(2011)}]{schollwock_density-matrix_2011}%
  \BibitemOpen
  \bibfield  {author} {\bibinfo {author} {\bibfnamefont {U.}~\bibnamefont
  {Schollw{\"o}ck}},\ }\href {\doibase 10.1016/j.aop.2010.09.012} {\bibfield
  {journal} {\bibinfo  {journal} {Annals of Physics}\ }\textbf {\bibinfo
  {volume} {326}},\ \bibinfo {pages} {96} (\bibinfo {year} {2011})}\BibitemShut
  {NoStop}%
\bibitem [{ite()}]{itensor_library}%
  \BibitemOpen
  \href {http://itensor.org} {}\bibinfo {note} {{MPS} calculations were
  performed using ITensor library (http://itensor.org)}\BibitemShut {NoStop}%
\bibitem [{\citenamefont {Anderson}(1952)}]{anderson_approximate_1952}%
  \BibitemOpen
  \bibfield  {author} {\bibinfo {author} {\bibfnamefont {P.~W.}\ \bibnamefont
  {Anderson}},\ }\href {\doibase 10.1103/PhysRev.86.694} {\bibfield  {journal}
  {\bibinfo  {journal} {Phys. Rev.}\ }\textbf {\bibinfo {volume} {86}},\
  \bibinfo {pages} {694} (\bibinfo {year} {1952})}\BibitemShut {NoStop}%
\bibitem [{\citenamefont {Azzouz}\ and\ \citenamefont {Dou{\c
  c}ot}(1993)}]{azzouz_effect_1993}%
  \BibitemOpen
  \bibfield  {author} {\bibinfo {author} {\bibfnamefont {M.}~\bibnamefont
  {Azzouz}}\ and\ \bibinfo {author} {\bibfnamefont {B.}~\bibnamefont {Dou{\c
  c}ot}},\ }\href {\doibase 10.1103/PhysRevB.47.8660} {\bibfield  {journal}
  {\bibinfo  {journal} {Phys. Rev. B}\ }\textbf {\bibinfo {volume} {47}},\
  \bibinfo {pages} {8660} (\bibinfo {year} {1993})}\BibitemShut {NoStop}%
\bibitem [{\citenamefont {Parola}\ \emph {et~al.}(1993)\citenamefont {Parola},
  \citenamefont {Sorella},\ and\ \citenamefont
  {Zhong}}]{parola_realization_1993}%
  \BibitemOpen
  \bibfield  {author} {\bibinfo {author} {\bibfnamefont {A.}~\bibnamefont
  {Parola}}, \bibinfo {author} {\bibfnamefont {S.}~\bibnamefont {Sorella}}, \
  and\ \bibinfo {author} {\bibfnamefont {Q.~F.}\ \bibnamefont {Zhong}},\ }\href
  {\doibase 10.1103/PhysRevLett.71.4393} {\bibfield  {journal} {\bibinfo
  {journal} {Phys. Rev. Lett.}\ }\textbf {\bibinfo {volume} {71}},\ \bibinfo
  {pages} {4393} (\bibinfo {year} {1993})}\BibitemShut {NoStop}%
\bibitem [{\citenamefont {Affeck}\ \emph {et~al.}(1994)\citenamefont {Affeck},
  \citenamefont {Gelfand},\ and\ \citenamefont {Singh}}]{affeck_plane_1994}%
  \BibitemOpen
  \bibfield  {author} {\bibinfo {author} {\bibfnamefont {I.}~\bibnamefont
  {Affeck}}, \bibinfo {author} {\bibfnamefont {M.~P.}\ \bibnamefont {Gelfand}},
  \ and\ \bibinfo {author} {\bibfnamefont {R.~R.~P.}\ \bibnamefont {Singh}},\
  }\href {\doibase 10.1088/0305-4470/27/22/009} {\bibfield  {journal} {\bibinfo
   {journal} {J. Phys. A: Math. Gen.}\ }\textbf {\bibinfo {volume} {27}},\
  \bibinfo {pages} {7313} (\bibinfo {year} {1994})}\BibitemShut {NoStop}%
\bibitem [{\citenamefont {Coletta}\ \emph {et~al.}(2012)\citenamefont
  {Coletta}, \citenamefont {Laflorencie},\ and\ \citenamefont
  {Mila}}]{coletta_semiclassical_2012}%
  \BibitemOpen
  \bibfield  {author} {\bibinfo {author} {\bibfnamefont {T.}~\bibnamefont
  {Coletta}}, \bibinfo {author} {\bibfnamefont {N.}~\bibnamefont
  {Laflorencie}}, \ and\ \bibinfo {author} {\bibfnamefont {F.}~\bibnamefont
  {Mila}},\ }\href {\doibase 10.1103/PhysRevB.85.104421} {\bibfield  {journal}
  {\bibinfo  {journal} {Phys. Rev. B}\ }\textbf {\bibinfo {volume} {85}},\
  \bibinfo {pages} {104421} (\bibinfo {year} {2012})}\BibitemShut {NoStop}%
\bibitem [{\citenamefont {Sylju{\r a}sen}\ and\ \citenamefont
  {Sandvik}(2002)}]{syljuasen_quantum_2002}%
  \BibitemOpen
  \bibfield  {author} {\bibinfo {author} {\bibfnamefont {O.~F.}\ \bibnamefont
  {Sylju{\r a}sen}}\ and\ \bibinfo {author} {\bibfnamefont {A.~W.}\
  \bibnamefont {Sandvik}},\ }\href {\doibase 10.1103/PhysRevE.66.046701}
  {\bibfield  {journal} {\bibinfo  {journal} {Phys. Rev. E}\ }\textbf {\bibinfo
  {volume} {66}},\ \bibinfo {pages} {046701} (\bibinfo {year}
  {2002})}\BibitemShut {NoStop}%
\bibitem [{\citenamefont {Bauer}\ \emph {et~al.}(2011)\citenamefont {Bauer},
  \citenamefont {Carr}, \citenamefont {Evertz}, \citenamefont {Feiguin},
  \citenamefont {Freire}, \citenamefont {Fuchs}, \citenamefont {Gamper},
  \citenamefont {Gukelberger}, \citenamefont {Gull}, \citenamefont {Guertler},
  \citenamefont {Hehn}, \citenamefont {Igarashi}, \citenamefont {Isakov},
  \citenamefont {Koop}, \citenamefont {Ma}, \citenamefont {Mates},
  \citenamefont {Matsuo}, \citenamefont {Parcollet}, \citenamefont {Paw{\l
  }owski}, \citenamefont {Picon}, \citenamefont {Pollet}, \citenamefont
  {Santos}, \citenamefont {Scarola}, \citenamefont {Schollw{\"o}ck},
  \citenamefont {Silva}, \citenamefont {Surer}, \citenamefont {Todo},
  \citenamefont {Trebst}, \citenamefont {Troyer}, \citenamefont {Wall},
  \citenamefont {Werner},\ and\ \citenamefont {Wessel}}]{bauer_alps_2011}%
  \BibitemOpen
  \bibfield  {author} {\bibinfo {author} {\bibfnamefont {B.}~\bibnamefont
  {Bauer}}, \bibinfo {author} {\bibfnamefont {L.~D.}\ \bibnamefont {Carr}},
  \bibinfo {author} {\bibfnamefont {H.~G.}\ \bibnamefont {Evertz}}, \bibinfo
  {author} {\bibfnamefont {A.}~\bibnamefont {Feiguin}}, \bibinfo {author}
  {\bibfnamefont {J.}~\bibnamefont {Freire}}, \bibinfo {author} {\bibfnamefont
  {S.}~\bibnamefont {Fuchs}}, \bibinfo {author} {\bibfnamefont
  {L.}~\bibnamefont {Gamper}}, \bibinfo {author} {\bibfnamefont
  {J.}~\bibnamefont {Gukelberger}}, \bibinfo {author} {\bibfnamefont
  {E.}~\bibnamefont {Gull}}, \bibinfo {author} {\bibfnamefont {S.}~\bibnamefont
  {Guertler}}, \bibinfo {author} {\bibfnamefont {A.}~\bibnamefont {Hehn}},
  \bibinfo {author} {\bibfnamefont {R.}~\bibnamefont {Igarashi}}, \bibinfo
  {author} {\bibfnamefont {S.~V.}\ \bibnamefont {Isakov}}, \bibinfo {author}
  {\bibfnamefont {D.}~\bibnamefont {Koop}}, \bibinfo {author} {\bibfnamefont
  {P.~N.}\ \bibnamefont {Ma}}, \bibinfo {author} {\bibfnamefont
  {P.}~\bibnamefont {Mates}}, \bibinfo {author} {\bibfnamefont
  {H.}~\bibnamefont {Matsuo}}, \bibinfo {author} {\bibfnamefont
  {O.}~\bibnamefont {Parcollet}}, \bibinfo {author} {\bibfnamefont
  {G.}~\bibnamefont {Paw{\l }owski}}, \bibinfo {author} {\bibfnamefont {J.~D.}\
  \bibnamefont {Picon}}, \bibinfo {author} {\bibfnamefont {L.}~\bibnamefont
  {Pollet}}, \bibinfo {author} {\bibfnamefont {E.}~\bibnamefont {Santos}},
  \bibinfo {author} {\bibfnamefont {V.~W.}\ \bibnamefont {Scarola}}, \bibinfo
  {author} {\bibfnamefont {U.}~\bibnamefont {Schollw{\"o}ck}}, \bibinfo
  {author} {\bibfnamefont {C.}~\bibnamefont {Silva}}, \bibinfo {author}
  {\bibfnamefont {B.}~\bibnamefont {Surer}}, \bibinfo {author} {\bibfnamefont
  {S.}~\bibnamefont {Todo}}, \bibinfo {author} {\bibfnamefont {S.}~\bibnamefont
  {Trebst}}, \bibinfo {author} {\bibfnamefont {M.}~\bibnamefont {Troyer}},
  \bibinfo {author} {\bibfnamefont {M.~L.}\ \bibnamefont {Wall}}, \bibinfo
  {author} {\bibfnamefont {P.}~\bibnamefont {Werner}}, \ and\ \bibinfo {author}
  {\bibfnamefont {S.}~\bibnamefont {Wessel}},\ }\href {\doibase
  10.1088/1742-5468/2011/05/P05001} {\bibfield  {journal} {\bibinfo  {journal}
  {Journal of Statistical Mechanics: Theory and Experiment}\ }\textbf {\bibinfo
  {volume} {2011}},\ \bibinfo {pages} {P05001} (\bibinfo {year}
  {2011})}\BibitemShut {NoStop}%
\bibitem [{\citenamefont {Sandvik}(1999)}]{sandvik_multichain_1999}%
  \BibitemOpen
  \bibfield  {author} {\bibinfo {author} {\bibfnamefont {A.~W.}\ \bibnamefont
  {Sandvik}},\ }\href {\doibase 10.1103/PhysRevLett.83.3069} {\bibfield
  {journal} {\bibinfo  {journal} {Phys. Rev. Lett.}\ }\textbf {\bibinfo
  {volume} {83}},\ \bibinfo {pages} {3069} (\bibinfo {year}
  {1999})}\BibitemShut {NoStop}%
\bibitem [{\citenamefont {White}\ and\ \citenamefont
  {Chernyshev}(2007)}]{white_neorder_2007}%
  \BibitemOpen
  \bibfield  {author} {\bibinfo {author} {\bibfnamefont {S.~R.}\ \bibnamefont
  {White}}\ and\ \bibinfo {author} {\bibfnamefont {A.~L.}\ \bibnamefont
  {Chernyshev}},\ }\href {\doibase 10.1103/PhysRevLett.99.127004} {\bibfield
  {journal} {\bibinfo  {journal} {Phys. Rev. Lett.}\ }\textbf {\bibinfo
  {volume} {99}},\ \bibinfo {pages} {127004} (\bibinfo {year}
  {2007})}\BibitemShut {NoStop}%
\bibitem [{Note2()}]{Note2}%
  \BibitemOpen
  \bibinfo {note} {We point out that the renormalization factor $\alpha
  _{2\protect \mathrm {D}}$ is not defined at the ferromagnetic point $1/K=0$
  where \protect \textit {any value} of $J'$ leads to $m^x=0.5$ within both MF
  and QMC approaches.}\BibitemShut {Stop}%
\bibitem [{\citenamefont {Affleck}\ and\ \citenamefont
  {Oshikawa}(1999)}]{Affleck_CuBenzoate_1999}%
  \BibitemOpen
  \bibfield  {author} {\bibinfo {author} {\bibfnamefont {I.}~\bibnamefont
  {Affleck}}\ and\ \bibinfo {author} {\bibfnamefont {M.}~\bibnamefont
  {Oshikawa}},\ }\href {\doibase 10.1103/PhysRevB.60.1038} {\bibfield
  {journal} {\bibinfo  {journal} {Phys. Rev. B}\ }\textbf {\bibinfo {volume}
  {60}},\ \bibinfo {pages} {1038} (\bibinfo {year} {1999})}\BibitemShut
  {NoStop}%
\bibitem [{\citenamefont {Korepin}\ \emph {et~al.}(1997)\citenamefont
  {Korepin}, \citenamefont {Bogoliubov},\ and\ \citenamefont
  {Izergin}}]{Korepin_book}%
  \BibitemOpen
  \bibfield  {author} {\bibinfo {author} {\bibfnamefont {V.~E.}\ \bibnamefont
  {Korepin}}, \bibinfo {author} {\bibfnamefont {N.~M.}\ \bibnamefont
  {Bogoliubov}}, \ and\ \bibinfo {author} {\bibfnamefont {A.~G.}\ \bibnamefont
  {Izergin}},\ }\href@noop {} {\emph {\bibinfo {title} {Quantum inverse
  scattering method and correlation functions}}}\ (\bibinfo  {publisher}
  {Cambridge university press},\ \bibinfo {year} {1997})\BibitemShut {NoStop}%
\bibitem [{\citenamefont {Hikihara}\ and\ \citenamefont
  {Furusaki}(2004)}]{Hikihara_XXZ_2004}%
  \BibitemOpen
  \bibfield  {author} {\bibinfo {author} {\bibfnamefont {T.}~\bibnamefont
  {Hikihara}}\ and\ \bibinfo {author} {\bibfnamefont {A.}~\bibnamefont
  {Furusaki}},\ }\href {\doibase 10.1103/PhysRevB.69.064427} {\bibfield
  {journal} {\bibinfo  {journal} {Phys. Rev. B}\ }\textbf {\bibinfo {volume}
  {69}},\ \bibinfo {pages} {064427} (\bibinfo {year} {2004})}\BibitemShut
  {NoStop}%
\bibitem [{\citenamefont {Furuya}\ and\ \citenamefont
  {Oshikawa}(2012)}]{Furuya_boundary_2012}%
  \BibitemOpen
  \bibfield  {author} {\bibinfo {author} {\bibfnamefont {S.~C.}\ \bibnamefont
  {Furuya}}\ and\ \bibinfo {author} {\bibfnamefont {M.}~\bibnamefont
  {Oshikawa}},\ }\href {\doibase 10.1103/PhysRevLett.109.247603} {\bibfield
  {journal} {\bibinfo  {journal} {Phys. Rev. Lett.}\ }\textbf {\bibinfo
  {volume} {109}},\ \bibinfo {pages} {247603} (\bibinfo {year}
  {2012})}\BibitemShut {NoStop}%
\bibitem [{\citenamefont {Oshikawa}(2010)}]{Oshikawa_BCFT_2010}%
  \BibitemOpen
  \bibfield  {author} {\bibinfo {author} {\bibfnamefont {M.}~\bibnamefont
  {Oshikawa}},\ }\href@noop {} {\bibfield  {journal} {\bibinfo  {journal}
  {arXiv:1007.3739}\ } (\bibinfo {year} {2010})}\BibitemShut {NoStop}%
\bibitem [{\citenamefont {Schulz}(1980)}]{Schulz_incommensurate_1980}%
  \BibitemOpen
  \bibfield  {author} {\bibinfo {author} {\bibfnamefont {H.~J.}\ \bibnamefont
  {Schulz}},\ }\href {\doibase 10.1103/PhysRevB.22.5274} {\bibfield  {journal}
  {\bibinfo  {journal} {Phys. Rev. B}\ }\textbf {\bibinfo {volume} {22}},\
  \bibinfo {pages} {5274} (\bibinfo {year} {1980})}\BibitemShut {NoStop}%
\bibitem [{\citenamefont {Furuya}\ and\ \citenamefont
  {Sato}(2015)}]{Furuya_ESRwidth_2015}%
  \BibitemOpen
  \bibfield  {author} {\bibinfo {author} {\bibfnamefont {S.~C.}\ \bibnamefont
  {Furuya}}\ and\ \bibinfo {author} {\bibfnamefont {M.}~\bibnamefont {Sato}},\
  }\href {\doibase 10.7566/JPSJ.84.033704} {\bibfield  {journal} {\bibinfo
  {journal} {J. Phys. Soc. Jpn.}\ }\textbf {\bibinfo {volume} {84}},\ \bibinfo
  {pages} {033704} (\bibinfo {year} {2015})}\BibitemShut {NoStop}%
\end{thebibliography}
\end{document}